\documentclass[prl,showkeys,twocolumn]{revtex4}
\usepackage{graphicx,amssymb,amsmath}
\usepackage{hyperref}

\begin{document}

\title{The fundamental diagram of urbanization}

\author{Giulia Carra}
\affiliation{Institut de Physique Th\'{e}orique, CEA, CNRS-URA 2306, F-91191,
Gif-sur-Yvette, France}
\author{Marc Barthelemy}
\email{marc.barthelemy@cea.fr}
\affiliation{Institut de Physique Th\'{e}orique, CEA, CNRS-URA 2306, F-91191,
Gif-sur-Yvette, France}
\affiliation{Centre d'Analyse et de Math\'ematique Sociales, (CNRS/EHESS) 190-198, avenue
de France, 75244 Paris Cedex 13, France}

\begin{abstract}

  The process of urbanization is one of the most important phenomenon
  of our societies and it is only recently that the availability of
  massive amounts of geolocalized historical data allows us to address
  quantitatively some of its features. Here, we discuss how the number
  of buildings evolves with population and we show on different
  datasets (Chicago, $1930-2010$; London, $1900-2015$; New York City,
  $1790-2013$; Paris, $1861-2011$) that this `fundamental diagram'
  evolves in a possibly universal way with three distinct
  phases. After an initial pre-urbanization phase, the first phase is
  a rapid growth of the number of buildings versus population. In a
  second regime, where residences are converted into another use (such
  as offices or stores for example), the population decreases while
  the number of buildings stays approximatively constant. In another
  subsequent phase, the number of buildings and the population grow
  again and correspond to a re-densification of cities. We propose a
  stochastic model based on these simple mechanisms to reproduce the
  first two regimes and show that it is in excellent agreement with
  empirical observations. These results bring evidences for the
  possibility of constructing a minimal model that could serve as a
  tool for understanding quantitatively urbanization and the future
  evolution of cities.

\end{abstract}


\maketitle
\section{Introduction}


Understanding urbanization and the evolution of urban system is a
long-standing problem tackled by geographers, historians, and
economists and has been abundantly discussed in the literature but
still represents a widely debated problem (see for example~\cite{Champion:2001}).
The term urbanization has been used in the literature with various
definitions, and depending has been considered as a continuous or an
intermittent process. In particular, urbanization measured by the
fraction of individuals (in a country for example) living in urban
areas describes a continuous process that gradually increased in many
countries with a quick growth since the middle of the 19$^{th}$
century until reaching values around $80\%$ in most european countries
(\cite{Antrop:2004}). Another definition has been introduced by
\cite{Fielding:1982} and presented by~\cite{Geyer:1993} as a
\textit{theory of differential urbanization} where it is assumed that
in general we observe the three regimes of urbanization, polarization
reversal and counter-urbanization, and that are characterized by a
gross migration which favors the larger, intermediate, and small-sized
cities, respectively.

Another approach to study urban changes is presented in the
\textit{stages of urban development} proposed
by~\cite{Hall:1971}. According to this model, the city has a life
cycle going from an early growing phase to an older phase of stability
or decline, and four main intermediate phases of development are
identified.  The first one called \textit{urbanization} consists of a
concentration of the population in the city core by migration of the
people from outer rings. The second phase of \textit{suburbanization}
is characterized by a population growth of the urban agglomeration as
a whole but with a population loss of the inner city and an increase
in urban rings. During the third phase of
(\textit{counterurbanization} or \textit{disurbanization}) the urban
population decreases both in the core and the ring. Finally, the last
phase of \textit{reurbanization} displays a re-increase of the urban
population.  Within this framework, we observe that for most
post-second war western countries \textit{urbanization} was dominating
in the $1950$s followed by a \textit{suburbanization} in the $1960$s
during which the population moved from the city core to the
suburbs. The standard theory of suburbanization suggests that it is
driven by a combination of technological progress (leading to
transport infrastructure development) and rising incomes
\cite{Anas:1998, Glaeser:2003, Antrop:2004}. In the $1970$s we observe
in many urbanized areas a regime of \textit{counter-urbanization}
where the population decreases. The significance of this regime and of
the re-urbanization period for the $1980$s and beyond, and more
generally the possibility of a cyclic development are controversial
topics (see for example~\cite{Champion:2001}).


Urban development and the spatial distribution of residences in urban
areas are obviously long-standing problems and were indeed discussed
in many fields such as geography, history and economics. Few of these
approaches tackled this problem from a quantitative point of view
(\cite{Helbert:1960,
  Mills:1967,Meuriot:1898,Taylor:1949,Wheaton:1982,Beckmann:1969,Harrison:1973,Tobler:1970,stanley:1995}). Among
the first empirical analysis on population density, Meuriot
\cite{Meuriot:1898} provided a large number of density maps of
European cities during the nineteenth century, and Clark
\cite{clark:1951} proposed the first quantitative analysis of
empirical data. Anas \cite{Anas:1978} presented an economic model for
the dynamics of urban residential growth where different zones of a
region exchange goods, capital, etc. according to some optimization
rule. In this same framework, the authors of \cite{Allen:1981}
proposed a dynamical central place model highlighting the importance
of both determinism and fluctuations in the evolution of urban
systems. For a review on different approaches, one can consult
\cite{Benenson:1999}, where studies are presented that use model
population dynamics in cities and in particular the ecological
approach, where ideas from mathematical ecology models are introduced
for modeling urban systems. An example is given
by~\cite{Dendrinos:1982} where phase portraits of differential
equations bring qualitative insights about urban systems
behavior. Other important theoretical approaches comprise the
classical Alonso-Muth-Mills model \cite{Anas:1998} developped in
urban economics, and also numerical simulations based on cellular
automata \cite{Sullivan:2001}. More recently, the fractal nature of
city structures (see the review \cite{Tannier:2005}) served as a guide
for the development of models
\cite{Batty:1994,stanley:1995,Makse:1998}. In particular, in
\cite{Makse:1998}, the authors proposed a variant of percolation
models for describing the evolution of the morphological structure of
urban areas. This coarse-grained approach however neglects all
economical ingredients and suggests that an intermediate way between
these purely morphological approaches and economical models should be
found.

For most of these quantitative studies however, numerical models
usually require a large number of parameters that makes it difficult
to test their validity and to identify the main mechanisms governing
the urbanization process. On the other hand, theoretical approaches
propose in general a large set of coupled equations that are difficult
to handle and amenable to quantitative predictions that can be tested
against data. In addition, even if a qualitative understanding is
brought by these theoretical models, empirical tests are often
lacking.

The recent availability of geolocalized, historical data (such as in
\cite{Perret:2015} for example) from world cities \cite{Angel:2012}
has the potential to change our quantitative understanding of urban
areas and allows us to revisit with a fresh eye long-standing
problems. Many cities created open-data websites \cite{Wired:2013} and
the city of New York (US) played an important role with the release of
the PLUTO dataset (short for Property Land Use Tax lot Output), where
tax lot records contain very useful information about the urbanization
process. For example, in addition to the location, property value,
square footage etc, this dataset gives access to the construction date
for each building. This type of geolocalized data at a very small
spatial scale allows to monitor the urbanization process in time and
at a very good spatial resolution.

These datasets allow in particular to produce `age maps' where the
construction date of buildings is displayed on a map (see Figure 1 for
the example of the Bronx borough in New York City). 
\begin{figure}[h!]
\includegraphics[width=0.9\columnwidth]{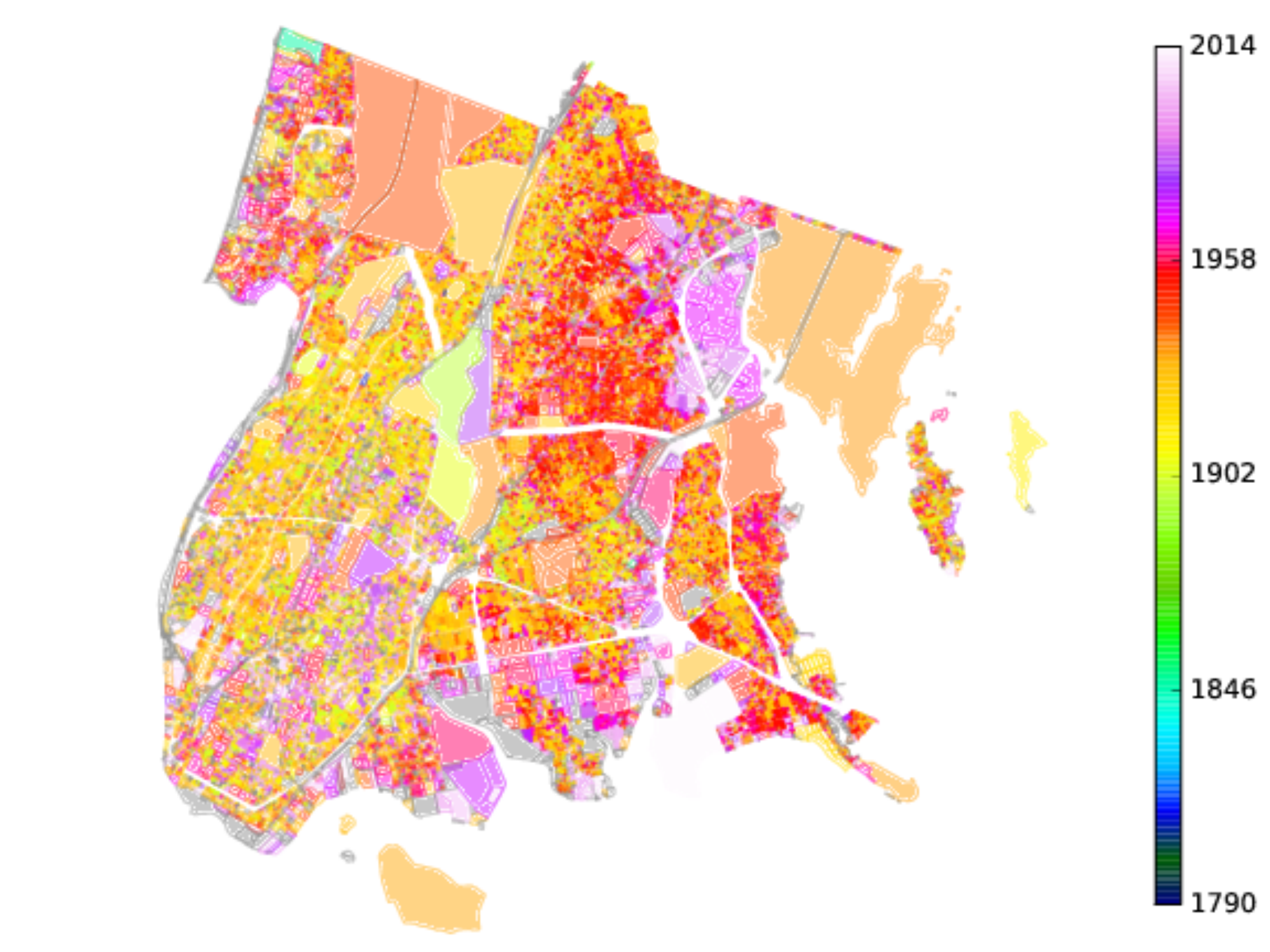}
\caption{Map of buildings construction date for the case of the Bronx
  (New York City, US). Most of the
  buildings were constructed during the beginning of the 20$^{th}$ century,
  followed by the construction in some localized areas of buildings in
  the second half of the 20$^{th}$ century. (See Materials and Methods for details on the dataset). 
}
\label{map}
\end{figure}
Many age building maps are now available:
Chicago~\cite{Chicago_map}. New York City US)~\cite{NY_map}, Ljubljana
(Slovenia)~\cite{ljubljana_map}, Reykjavik
(Iceland)~\cite{reykjavik_map}, etc.  In addition to be visually
attractive (see for example \cite{Portland_map,MN_map}), these maps
together with new mapping tools (such as the urban layers proposed in
\cite{MN_map}) provide qualitative insights into the history of
specific buildings and also into the evolution of entire
neighborhoods. \cite{Palmer:2013} studied the evolution of the city of
Portland (Oregon, US) from $1851$ and observed that only $942$
buildings are still left from the end of the $19^{th}$ century, while
$75,434$ buildings were built at the end of the $20^{th}$ century and
are still standing, followed by a steady decline of new buildings
construction since $2005$. Inspired by Palmer’s map, \cite{Plahuta}
constructed a map of building ages in his home town of Ljubljana,
Slovenia, and proposed a video showing the growth of this city from
$1500$ until now \cite{video_Ljub}. Plahuta observed that the number
of new buildings constructed each year displays huge spikes that
signalled important events: an important spike occurred when people
were able to rebuild a few years after a major earthquake hit the area
in $1899$, and other periods of rebuilding occurred after the two
world wars. In the case of Los Angeles (USA), the `Built:LA project'
shows the ages of almost every building in the city and allows to
reveal the city growth over time \cite{BUILT_LA}.

These different datasets allow thus to monitor at a very small spatial
resolution urban processes. In particular, we aim to focus on a
given district or zone, without considering for the moment their
position and their role in the whole urban agglomeration they belong
to. We ask quantitative questions about the evolution over time of
the population and of the number of buildings, and we aim to
understand if different districts of different cities can be
compared. Surprisingly enough, such a dual information is difficult to
find and -- up to our knowledge -- was not thoroughly studied at the
quantitative level (except at a morphological level with
fractal studies, \cite{Tannier:2005}). Here, we use data for different
cities (Chicago, $1930-2010$; London, $1900-2015$; New York City,
$1790-2013$; Paris, $1861-2011$) in order to answer questions about
these fundamental quantities. We want to remark that although these
cities are among the most urbanized ones, they are characterized by
quite different historical paths, with US cities being usually 'younger' compared to the
European ones. Chicago for example is a young city founded
at the beginning of the $19^{th}$ century, and Paris instead has an
history of about two thousands years.

More precisely, in this study we will show that the number of buildings versus the
population follows the same unique pattern for all the cities studied
here. Despite the small number of cities studied, the strong
similarities observed suggest the possibility of a universal behavior
that can be tested quantitatively. In order to go further in our
understanding of this unique pattern, we propose a theoretical model
and empirical evidences supporting it.

\section{Empirical results}

We investigate the urban growth of four different cities: Chicago
(US), London (UK), New York (US), and Paris (France). We discuss here
urbanization from the point of view of two dual aspects. First, we
consider the evolution of the population of urban areas and second,
the evolution of the number of buildings. These aspects thus concern
both an individual-related aspect (the population) and an important physical
aspect of cities, the buildings.

We do not study here age maps and in order to go beyond a simple visual
inspection of these objects, we study how the number of buildings
varies with the population. In most datasets, we essentially have
access to buildings that were built and survived until now. In this
respect we do not take into account the destruction, replacement or
modifications of buildings. Although replacement or modifications do
not alter our discussion, replacement with buildings of another
land-use certainly has an impact on the evolution of the population
and could potentially lead to a major impact on the evolution of
cities. As we will see in our model this can be in a way encoded in
the `conversion' process where a residential building is converted
into a non-residential one. The important point is to describe the
temporal evolution of buildings and their function, and we encode all
these aspects in the simpler quantity that is the number of
buildings. Further studies are however certainly needed in order to clarify
the impact of these points on our results.

The urbanization process can be described by many different aspects
and we will concentrate on two main indicators. Urbanization is about
concentration of individuals and the first natural parameter is the
population. Urbanization is also about built areas and in
order to describe the physical evolution of a city, the natural
parameter is the number of buildings (for a given area). Once both
these parameters are known (density of population and of buildings), we
already have an important piece of information. The following 
question is then how these two parameters relate to each other, and it
is then natural to plot the number of buildings versus the
population when the city evolves. This `fundamental' diagram contains
the core information about the urbanization process and will be the
focus of this study. 

\subsection{Choice of the areal unit}

An important discussion concerns the choice of the scale at which we
study the urbanization process. We have to analyse the processes of
urban change at a spatial scale that is large enough in order to
obtain statistical regularities, but not too large as different zones
may evolve differently. Indeed geographers observed that the
population density is not homogeneous and decreases in general with
the distance to the
center \cite{Bertaud:2003,Pumain_Guerois:2008}. Also, during the
evolution of most cities, they tend to spread out with the density
decreasing in central districts and increasing in the outer
ones \cite{clark:1951} and indeed in the literature the core of the
city is often analysed in relation to its suburbs.  In this study we
aim to simplify the analysis and we focus on a fixed area without
considering its role in the whole urban agglomeration; nevertheless we
would like this area to be mostly homogeneous and not mixing zones
behaving in different ways.

We choose to focus here on the evolution of administrative districts
of each city. At this level, data is available and we can hope to
exclude longer term processes. We will show in the following that even
if this choice appears as surprising, districts in the different
cities considered here display homogeneous growth.  More precisley, we
consider here the $5$ boroughs of New York, the $9$ sides of Chicago,
the $20$ arrondissements of Paris and the $33$ London districts. Also,
in this way we do not have to tackle the difficult problem of city
definition and its impact on various measures (see for example \cite{Arcaute:2015}) and
focus on the urban changes of a given zone with fixed surface area.
The datasets for these cities come from different sources (see
Materials and Methods) and cover different time periods. $1930-2010$
for Chicago, $1900 - 2015$ for London, $1790-2013$ for New York, and
$1861-2011$ for Paris. An important limitation that guided us for
choosing these cities is the simultaneous availability of building age
and historical data for district population.

\begin{figure}
\includegraphics[width=0.49\columnwidth]{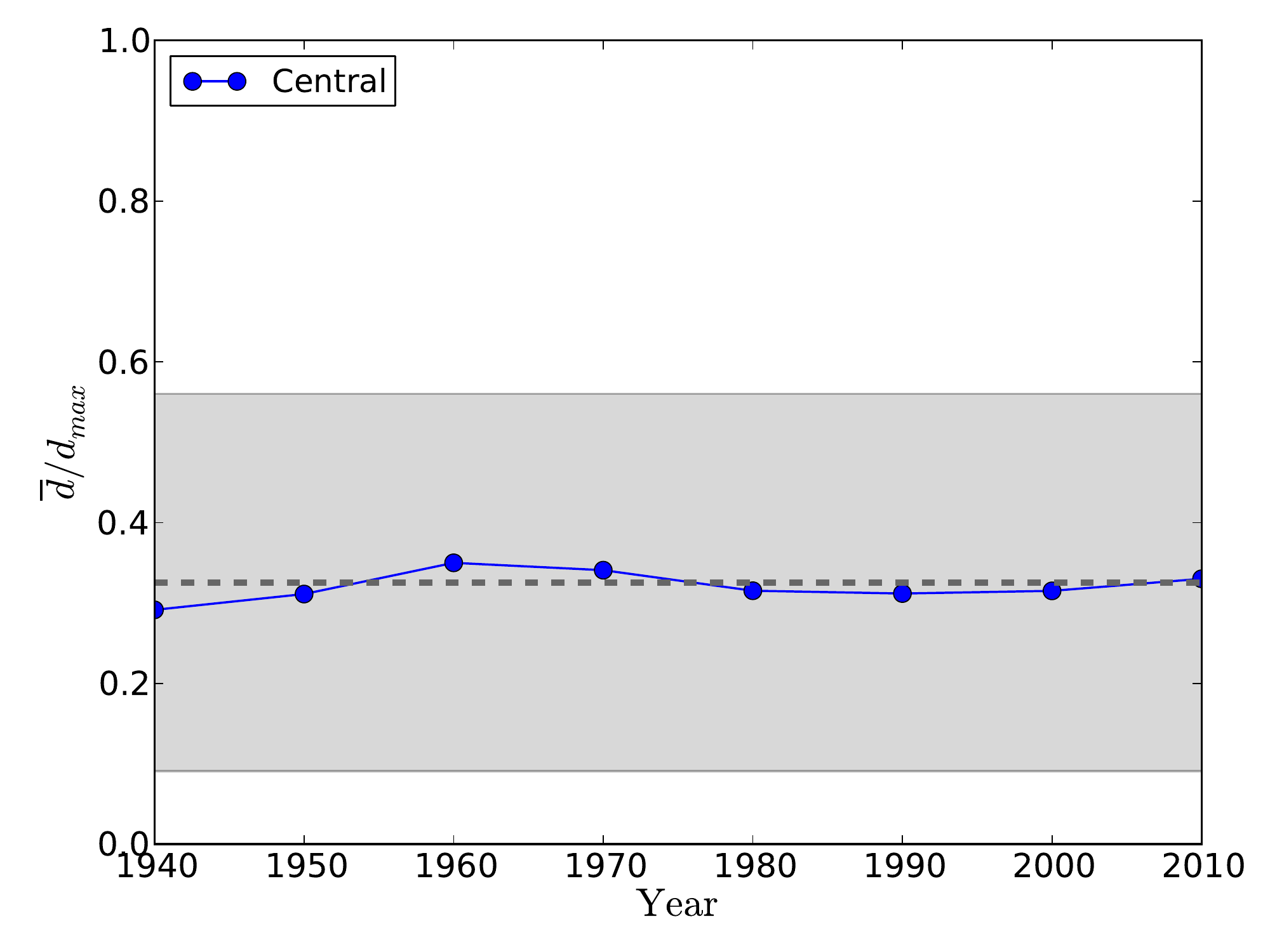}
\includegraphics[width=0.49\columnwidth]{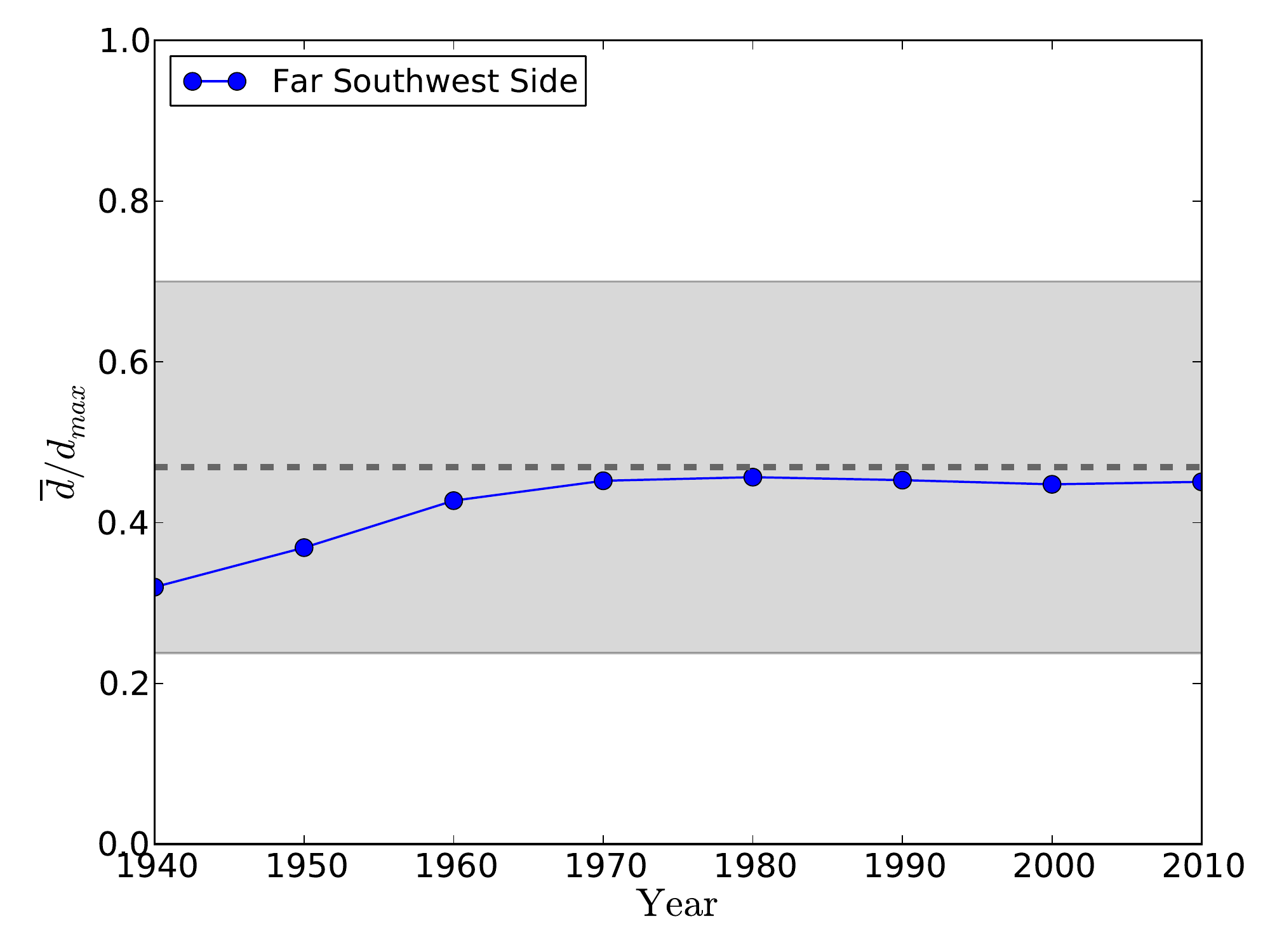}\\
\includegraphics[width=0.49\columnwidth]{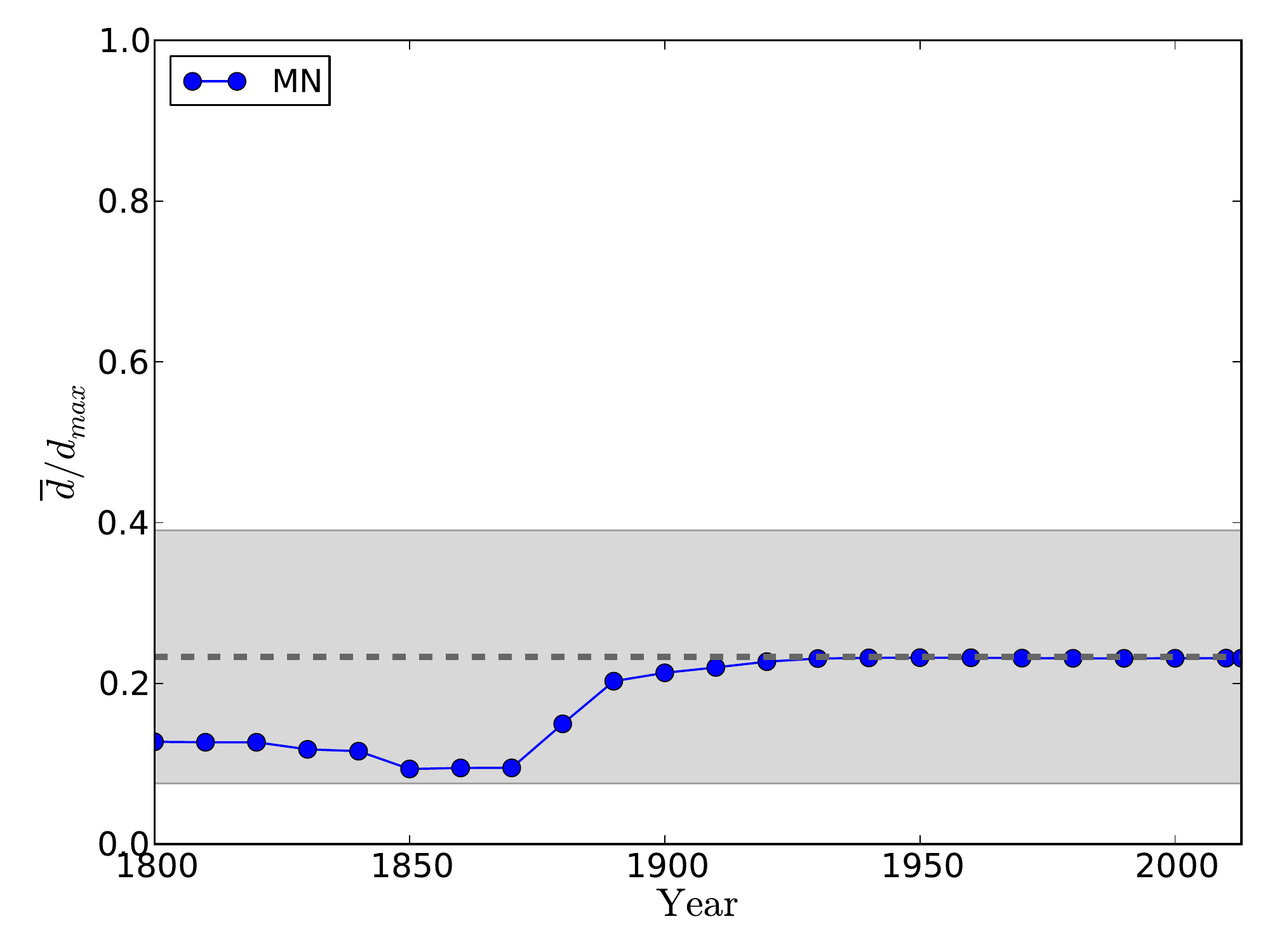}
\includegraphics[width=0.49\columnwidth]{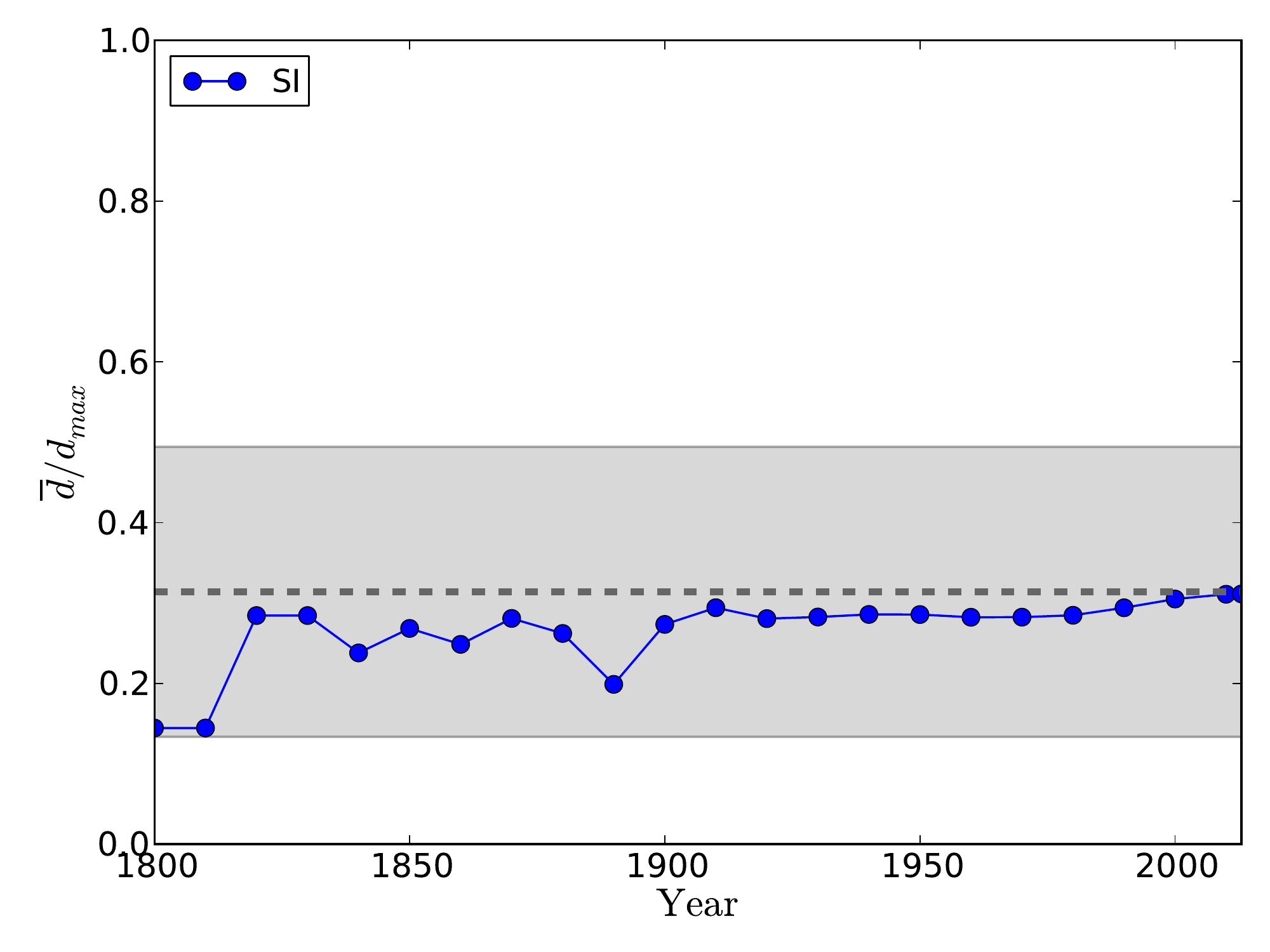}\\
\includegraphics[width=0.49\columnwidth]{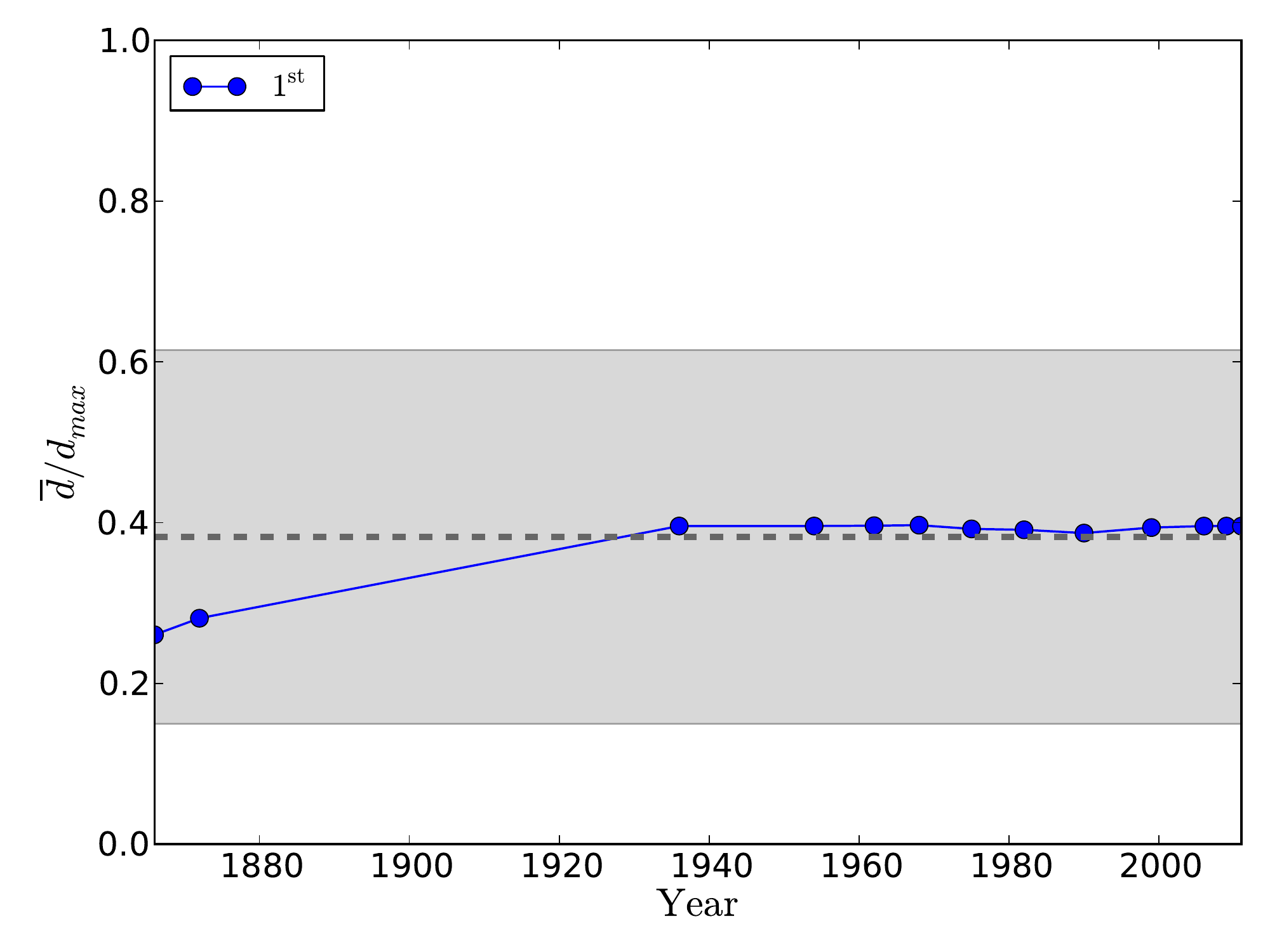}
\includegraphics[width=0.49\columnwidth]{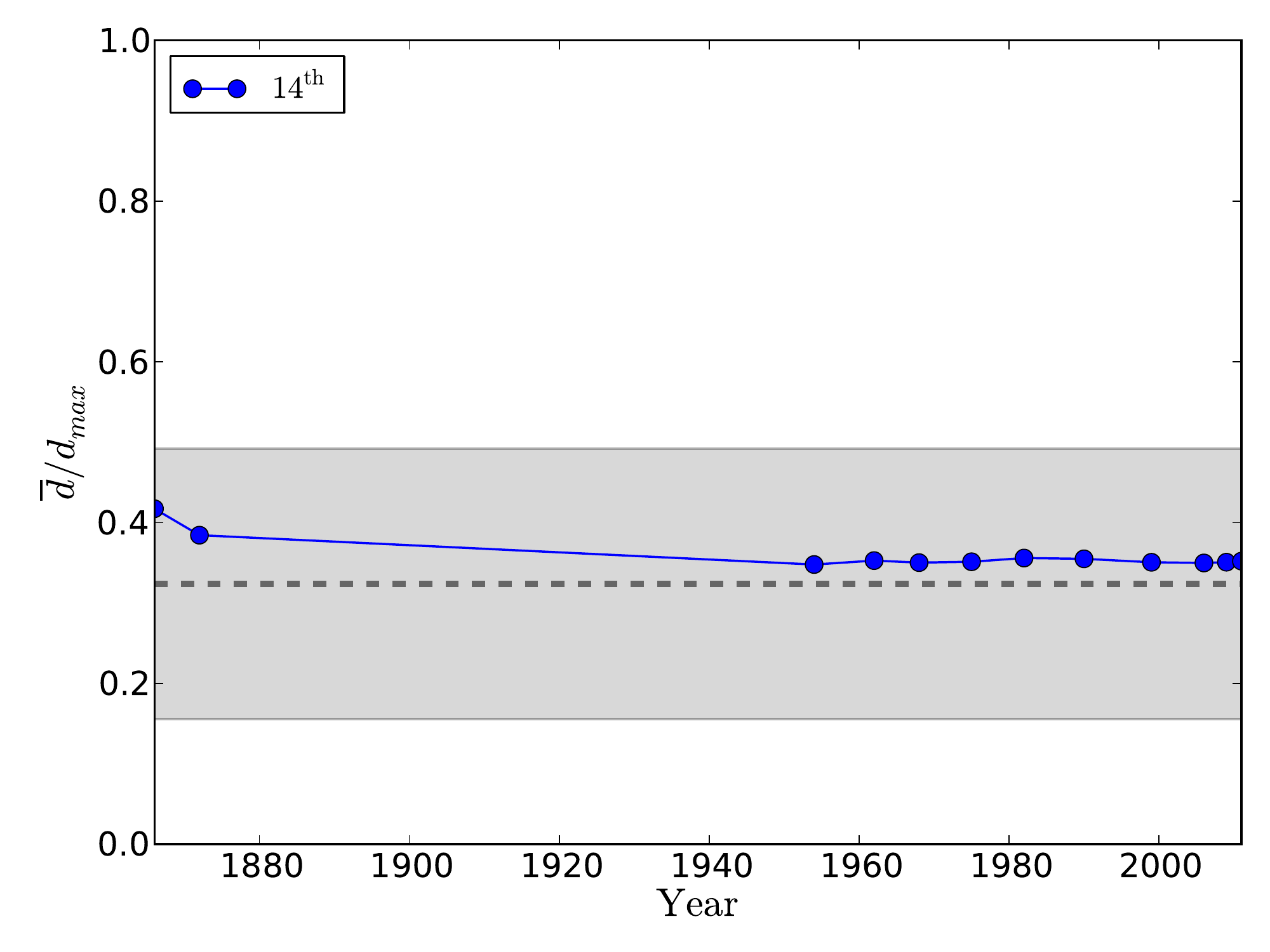}
\caption{\textbf{Homogeneity of growth in districts}. Average distance
  between buildings at a  given time (this distance is normalized by
  the maximum distance found for each district). Top: Chicago (centrral
  and far southwest sides). Middle: New York City (Manhattan and
  Staten Islands). Bottom: Paris (1st and 14th arrondissements). The dotted line represents the
  average value computed for a random uniform distribution and the
  grey zone the dispersion computed with this null model.}
\label{fig:null}
\end{figure}

The cities studied here display very different scales, ranging from
Paris with $20$ districts for $2-3$ millions inhabitants and an
average of $5km^2$ per district, to New York City with $5$ boroughs of very diverse
area (from $60km^2$ for Manhattan to $183km^2$ and $283km^2$ for
Brooklyn and Queens, respectively). The most important assumption that
we will use here is that the development in each of these districts is relatively
homogeneous. We test this assumption on Chicago, New York City and
Paris for which we have the exact localization of new buildings (which
we don't have for London). For each district and at each point in time
we compute the average distance $\overline{d}$ (normalized by the
maximum distance  in the district $d_{max}$) between new buildings in this
district. We also compute the same quantity for a `null' model for
which the new buildings are distributed uniformly. The results are
shown (Fig~\ref{fig:null}) for a selection of districts in these different cities (the full
set of results is shown in the Supplementary material).
We see on the Fig.~\ref{fig:null} that despite the very diverse sizes
of these districts, in all cities studied here, the development of new
buildings is consistent with a uniform distribution, within these
districts. This is a rather unexpected result as for example in Paris we have
relatively homogeneous districts while in New York City, the boroughs 
are much larger and aggregate together a variety of urban
spaces. These results therefore show that despite the variety of
cases, this choice of aerial unit provides a reasonable partition of
space where the growth is homogeneous. In particular, it implies that
a smaller area is not necessarily a good choice for studying the
evolution of the number of buildings as it would suffer from strong
sampling effects. 

\subsection{Population density growth}

\begin{figure}[ht!]
\includegraphics[width=0.8\columnwidth]{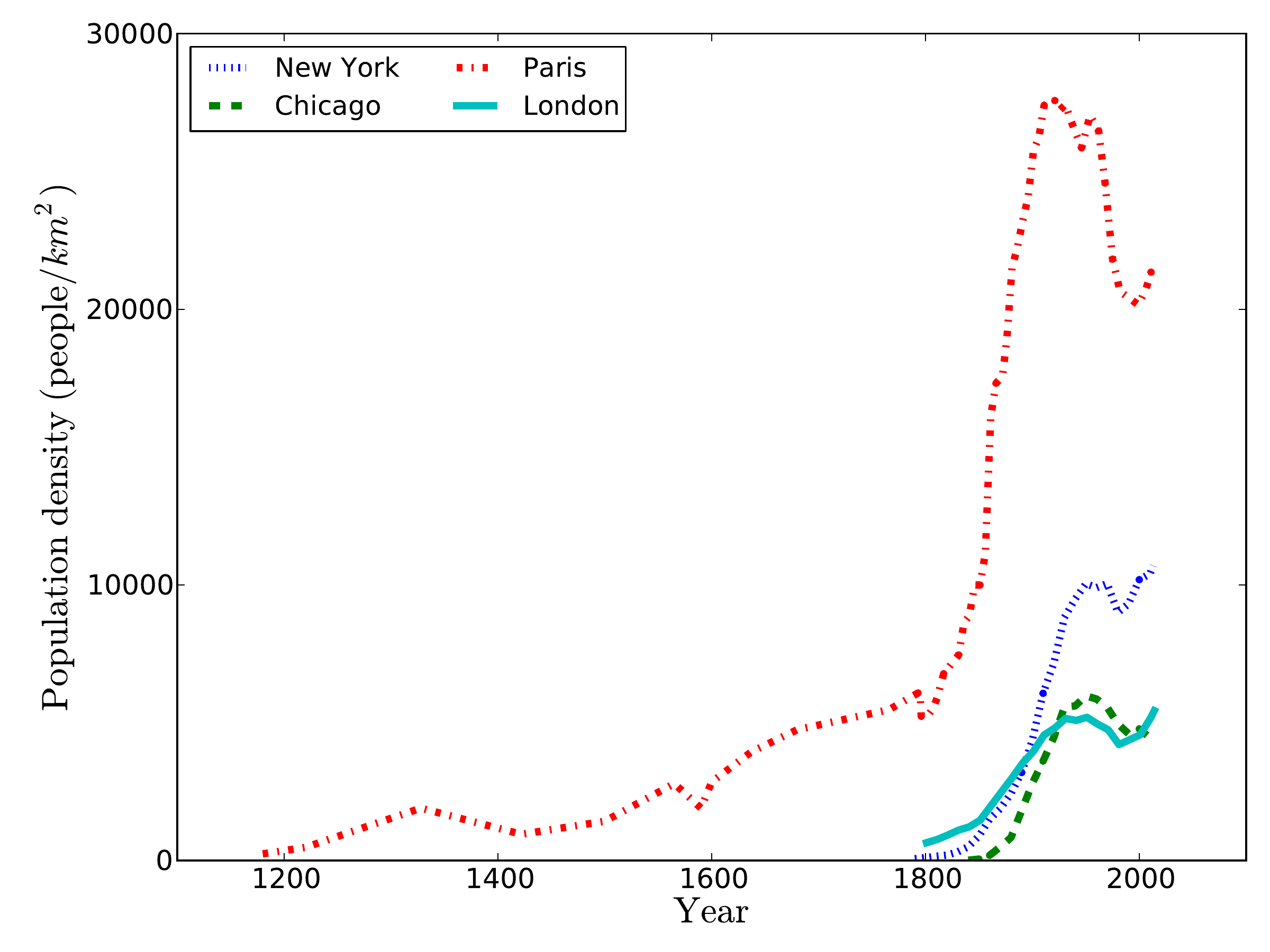}
\caption{\textbf{Population density versus time}. The average
  population density versus time for the four cities studied in the
  paper. All these cities display a density peak in the first
  half of the 20$^{th}$ century (see Material and Methods for details on datasets).}
\label{fig:popdensity}
\end{figure}

In order to provide an historical context, we first measure the
evolution of the population density and then analyse the evolution of
the number of buildings in a given district and its population. In
Fig.~\ref{fig:popdensity} we show the average population density for
the four cities studied here.
This plot reveals that these different cities follow similar dynamics,
at least at a coarse-grained level. After a positive growth and a
population increase that accelerates around $1900$, we observe a
density peak. After this peak, the density decreases (even sharply in
the case of NYC) or stays roughly constant. This decreasing regime is
associated to the post World War years, defined by geographers
as the suburbanization/counter-urbanization period.  In the last
years, New York City, Paris and London display a re-densification
period. The possibility of this latter period has been proposed in
some cyclic model as the \textit{stages of urban development}
one~\cite{Hall:1971}. Nevertheless, evidences or interpretations about
this phase are still an highly discussed topic.  At least, this first figure
highlights the existence of a seemingly `universal' pattern governing
the urban change process, probably driven by technological changes. 

However, at the smaller scale of districts, these large cities display
different behavior shown in Fig.~\ref{fig:3} where we plot the time
evolution of some district densities (all results are presented in the
Supplementary material). 
\begin{figure}[ht!]
\includegraphics[width=1\columnwidth]{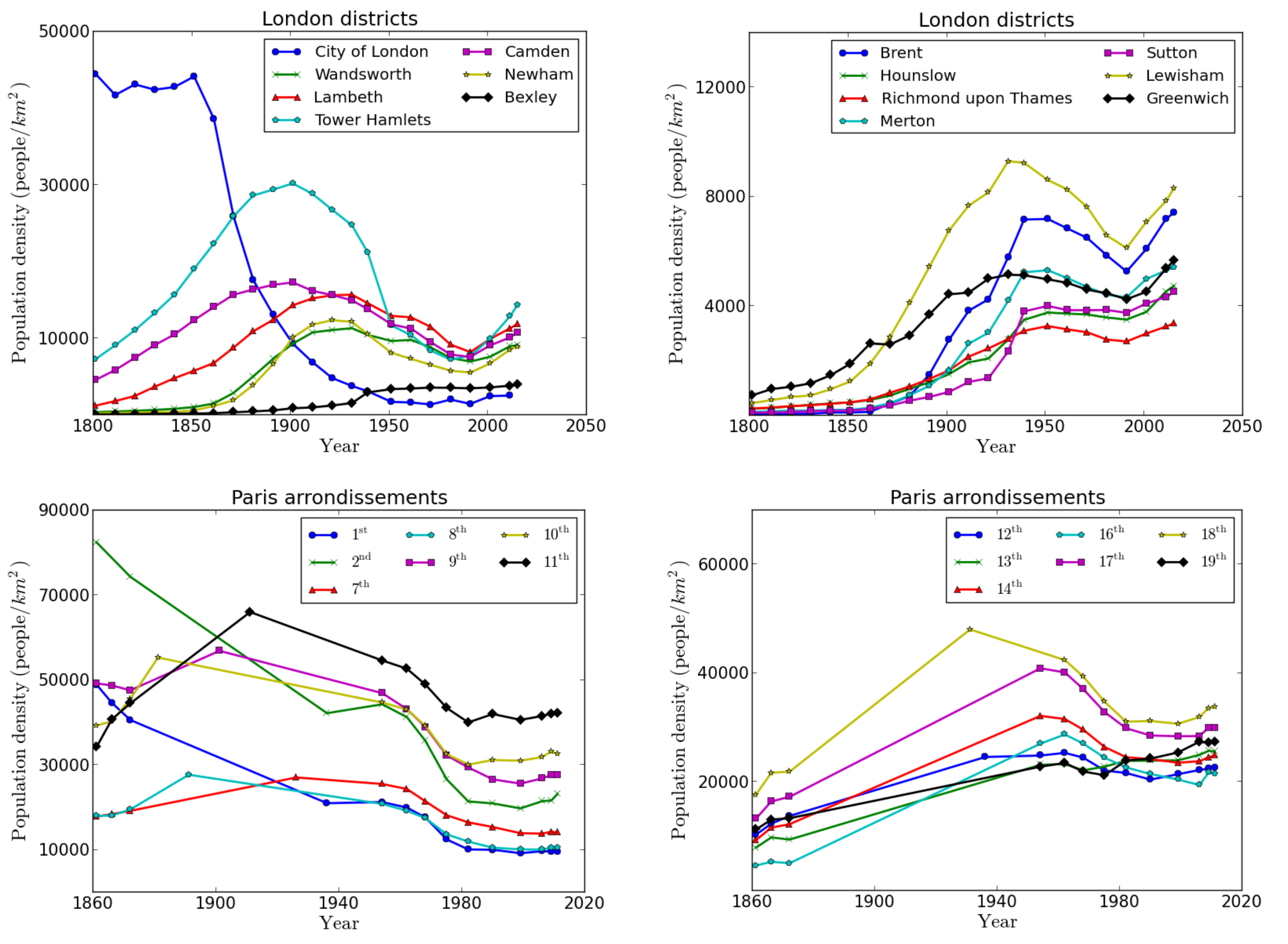}

\caption{\textbf{Population density versus time}. Local
  population densities for a selection of London districts (top), and a
  selection of Paris arrondissements (bottom). For the sake of clarity we did not plot all the
districts studied and additional results can be found in the Supplementary material.
}
\label{fig:3}
\end{figure}
In the case of London (Fig.~\ref{fig:3}, top panels), we note that the
district City of London reached a density peak before $1800$ while
other districts (for example Lewisham, Brent and Newham) display all
the different phases of urbanization described above. For Chicago (see
the Supplementary material) and Paris (Fig.~\ref{fig:3}, bottom panels), the
different districts are not all synchronized and display
simultaneously different urbanization phases. The central districts of
Paris (the $1^{st}$ and the $4^{th}$ for example) typically reached
their density peak before $1860$, while less central districts
(11$^{th}$ to 20$^{th}$) reached their density peak in the first half
of the 20$^{th}$ century, consistently with the idea of a centrifugal
urbanization process.

For the five boroughs of New York (see the Supplementary material), we observe that Manhattan
(MN), the Bronx (BX) and Brooklyn (BK) already passed through
the different phases of urbanisation, and are now in a
re-densification period. In contrast, Staten Island (SI) and Queens
(QN) are still in the urbanization period characterized by a positive
population growth rate and didn't reach yet a density peak. 

These preliminary results highlight the importance of spatial
delimitations when studying a city. The dynamics of different
districts might be the same as also suggested by qualitative models
presented in the introduction, but are not necessary simultaneous
mainly because of the difference between districts belonging to the
core of the city and districts belonging to the ring, and further the
distance from the core of the city, later the district will reach the
second phase. For this reason, we will not consider in the following
cities as a whole, but rather follow the evolution of various
quantities for each district which display a better level of
homogeneity.

We note here that a large number of empirical studies have already been performed where
the densification and the disurbanization phase were
observed~\cite{Fielding:1982, Frey:1990, Nucci:1995, Champion:1989,
  Beale:1975, Berry:1976}. In most of these studies, the analysis was
performed focusing in the dependence between the behavior of the core
and of the ring districts or on the size of the urban agglomeration.

\subsection{Number of building vs. population}

We now turn to the main result of this paper which is the
characterization of the urbanization from the point of view of both
the physical aspect via the number of buildings, and the individual
aspect described here by the population.

\begin{figure*}[ht!]
\includegraphics[width=2\columnwidth]{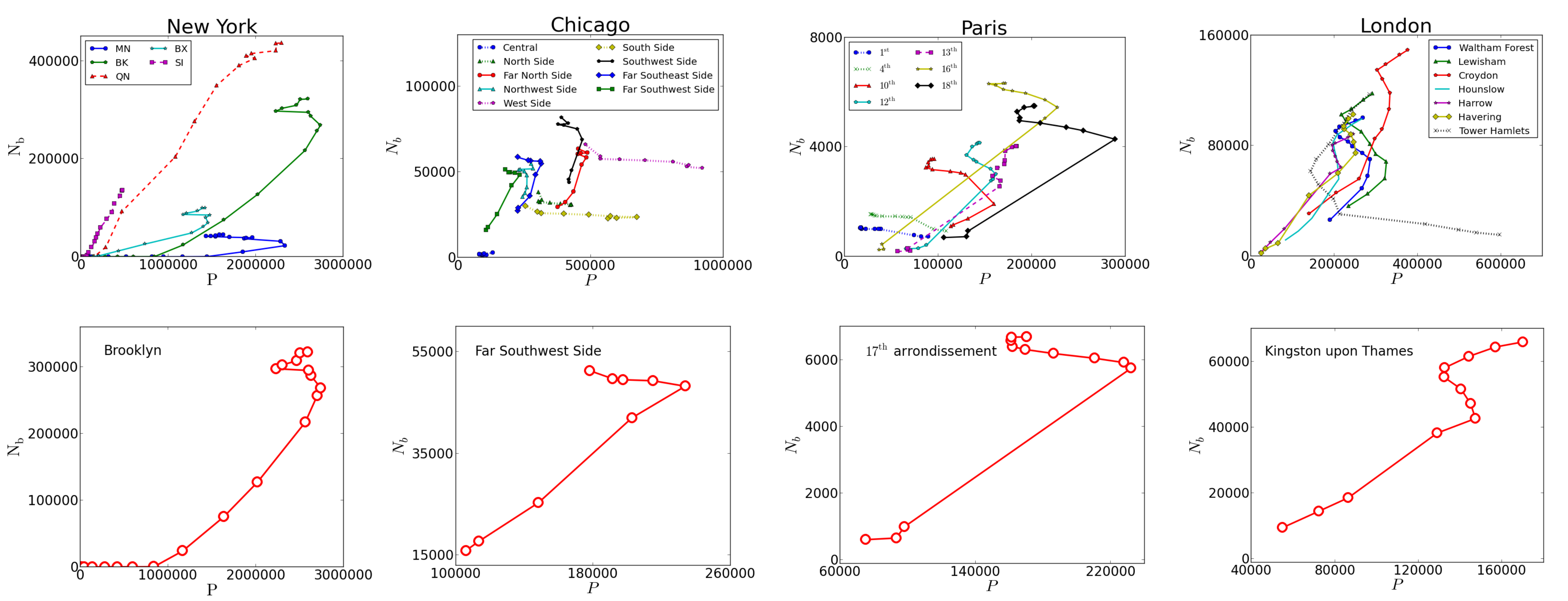}
\caption{\textbf{Number of buildings versus population}. We represent
  with continuous lines the districts that have reached their density
  peak, with dashed lines for districts that are still in the
  growing phase. We use dotted line for the districts that reached the
  density peak before the first year available in the dataset. (Top
  panels) Results for districts in the cities studied here. (Bottom)
  We show examples illustrating the `universal' diagram for districts
  in different cities that display all the regimes described in the text.}
\label{fig:nb}
\end{figure*}

For each district, we then study the relation between the number of
buildings $N_b$ and the population $P$ of different districts
(Fig.~\ref{fig:nb}), and plot $N_b$ versus $P$. We thus connect an
element of the infrastructure - the building - to the population
which allows us to get rid of exogenous effects that governs the time
evolution of population for example. This plot encodes these two basic
fundamental aspects of the urbanization process and we refer to this
representation as `the fundamental diagram'.
In Fig.~\ref{fig:nb}, we observe an apparent diversity of behaviors
but, as we will see in the following, they can all be interpreted and
compared in the framework of a simple quantitative model. In
Fig.~\ref{fig:nb} top-left we show the result for the five boroughs of
New York City. We observe that Staten Island and Queens (dashed lines)
are in a growing phase characterized by a positive value of $dN_b/dP$,
while Manhattan, Brooklyn and Bronx (plotted in continuous line)
reached other dynamical regimes. In Fig.~\ref{fig:nb} top-center-left
we plot the nine sides of Chicago, and we observe a clear growth phase
followed by a `saturation' (corresponding to the density peak) for the
Far North, Northwest, Southwest, Far Southeast and Far Southwest sides
(plotted in continuous line). In contrast, the other sides (Central,
North, West and South), in dotted line, seem to have reached a
saturation before $1930$. Indeed, the dotted lines (that have to be
read chronologically from the right to the left) do not display the
growth regime, suggesting that it stopped before $1930$, year of the
earliest available data. In Fig.~\ref{fig:nb} top-centre-right, we
represent the evolution for some Paris arrondissements. We observe the
growth regime followed by a saturation for the $10^{\mathrm {th}}$,
$12^{\mathrm {th}}$, $16^{\mathrm {th}}$ and $18^{\mathrm {th}}$
arrondissement (in continuous line), the $13^{th}$ seem not having
reach a saturation yet, while the others have reached saturation
before $1861$. In the top-right plot of Fig.~\ref{fig:nb} for London
districts, we observe that all districts displayed here reached a
saturation, but that the district Tower Hamlets (dotted line) reached
it before $1900$, year of the first available data.

\begin{figure}[ht!]
\includegraphics[width=1.0\columnwidth]{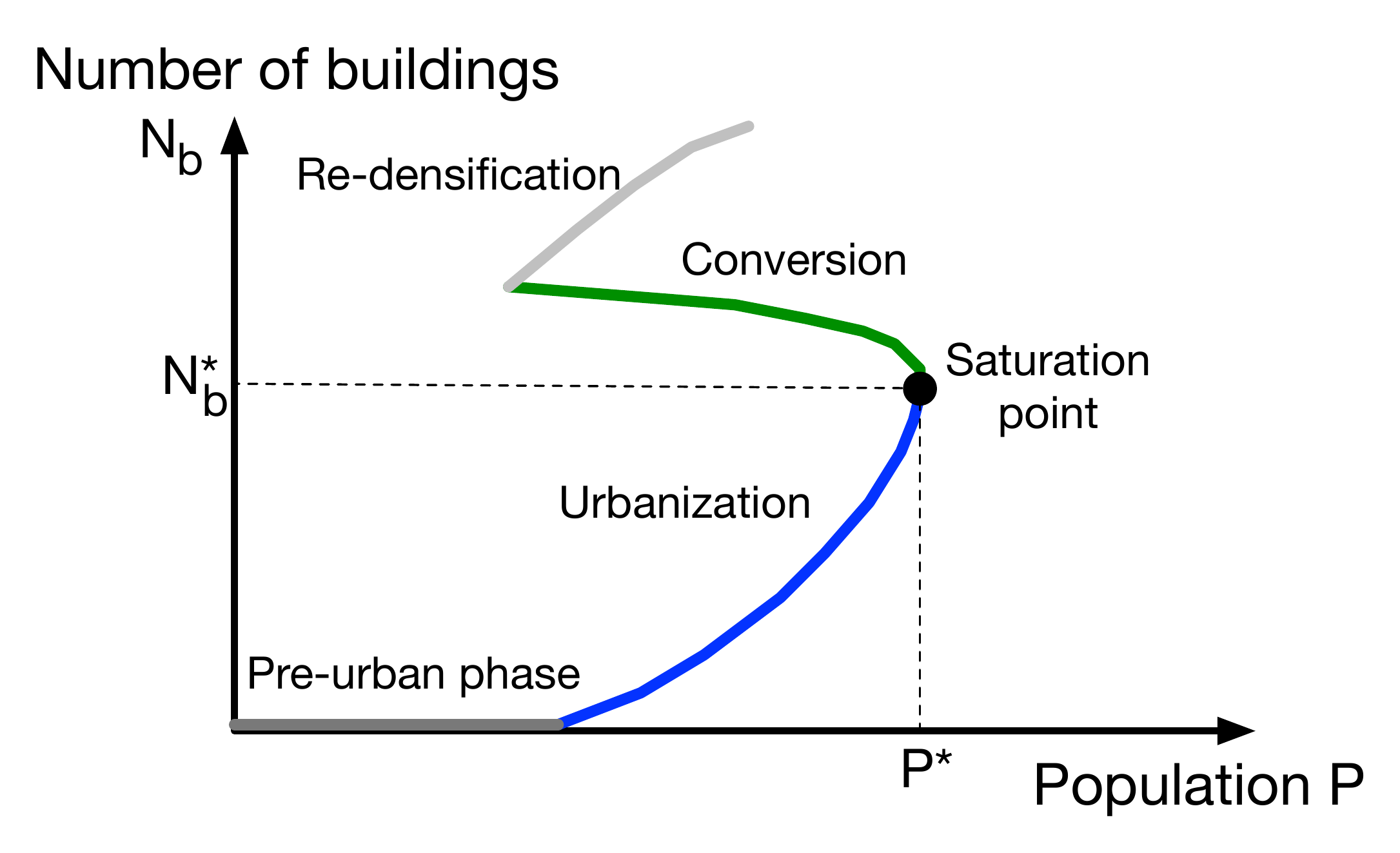}
\caption{\textbf{Schematic representation of the fundamental
    curve}. We represent here the typical district growth curve
  characterized by three main phases: after a pre-urbanization period,
  there is first an urbanization phase with a positive growth rate
  $dN_b/dP$ that stops at the `saturation point' $(P^*,N_b^*)$. A
  second `conversion' phase follows, during which the population
  decreases. Finally, we observe a last re-densification phase where
  both the population and the number of buildings increase.}
\label{fig:diag}
\end{figure}

These various plots show that for different districts we have
essentially the same trajectory in the plane $(P,N_b)$ at different
stage of their evolution. We show illustrative examples for various
cities in Fig.~\ref{fig:nb}(bottom) that reached the second regime
after a saturation point (while other districts are still in the first
regime). The evolution of these `mature' districts of these different
cities can thus be represented by a typical path shown in
Fig.~\ref{fig:diag}. This path is characterized by a first phase of
rapid growth of the number of buildings versus population. In a second
regime, the population decreases while the number of buildings stays
roughly constant. In a last -- and more recent -- phase, the number of
buildings and population both grow again. The behavior of the urban
changes emerging by studying the relation between population and
number of buildings in a fixed area is thus analogous to the one
described in the \textit{stages of urban development}
model of \cite{Hall:1971}, in which a qualitative understanding of the
first two phases is widely recognized, while the last one remain
widely discussed.  We remark that the year at which the second or the
third phase begins is not necessary the same for all districts and
depends mainly on the role and function of the district in the whole
urban agglomeration.


\section{Theoretical model}

The data studied in the previous section display a pattern that seems
to encompass specific features of the different cities and we propose
a theoretical model based on the following interpretation for these
different regimes. The first regime corresponds to the urbanization
where buildings are constructed on empty lots until the `saturation
point' $(P^*,N_b^*)$, which signals the beginning of the second regime
(we note that not all districts reached this saturation point and can
still be in the first growing phase). In this second regime, land-use
is modified (for example from residential to stores or offices) and
the population naturally decreases while the number of buildings stays
approximately constant. We emphasize here that this `conversion' is
meant here as a generic term that describes the process in which a
part or the whole of a building changes from a residential use to a
non-residential one. In the last regime, both the number of
buildings and the population grow again, corresponding to the
`re-densification' of cities. This last phase which occurs mostly at
the same period for the different cities seems to be triggered by
external factors such as governance. This is why we will focus on the
first two regimes and to understand the reasons and control parameters
of the saturation point. In order to provide quantitative evidences
for these first two phases, we propose a simple model based on the
simple interpretation described above and -- very importantly -- that allows us to make
predictions that we can test against data. 

We model the evolution of a given zone of surface area $A$ by a
two-dimensional square grid where each cell of surface $a_\ell$
represents an empty, constructible lot. The maximum number of lots is
then given by $N_{max}=A/a_\ell$. Each cell can be empty or occupied
(a building has already been built) and each building on a lot $i$ is
characterized by its number of residential floors $h_r(i)$, commercial
floors $h_c(i)$ (the total number of floors is
$h(i)=h_r(i)+h_c(i)$). At each time step $t \rightarrow t + \Delta t$
(in the following we count the time $t$ in units of $\Delta t$), we
pick at random a cell $i$ and if it is empty we update it with
\begin{align}
\begin{cases}
P \rightarrow P  + \Delta P ~, \\
N_b \rightarrow N_b + 1 ~,\\
h(i) = h_{r}(i) = 1 ~,\\
h_c(i)=0~,
\end{cases}
\end{align}
where $\Delta P$ is the number of people per residential floor (we
assume here that the number of person per floor does not change too
much in time which is certainly true in terms of order of magnitude). If a
building is already present on the chosen cell, we add an extra
residential floor with probability $p_h$ or convert a residential floor into a
non-residential one (such as offices or stores) with probability
$p_c$:
\begin{align}
\begin{cases}
h_r(i) \rightarrow h_r(i) + 1 \\
h_c(i) \rightarrow h_c(i)  \\
P \rightarrow P  + \Delta P
\end{cases}
\mathrm{with\;prob.}\; p_h
\\
\begin{cases}
h_r(i) \rightarrow h_r(i) - 1 \\
h_c(i) \rightarrow h_c(i) + 1 \\
P \rightarrow P  - \Delta P 
\end{cases}
\mathrm{with\;prob.}\; p_c
\end{align}
Finally, nothing happens with probability $1 - p_h - p_c$. Each
district is thus characterized by the parameters $\Delta P$, $p_c$ and
$p_h$. The mean-field equations describing the evolution of $H_r=\sum_ih_r(i)$ (the
total number of residential floors in the district), the total number of buildings
$N_b$ and the total population $P$ in the district are
\begin{align}
&\frac{dH_r}{dt} = \frac{N_b}{N_{max}} (p_h - p_c) + \left(1 - \frac{N_b}{N_{max}}\right) ~,\label{eq:M_H}\\ 
&\frac{dN_b}{dt} = 1 - \frac{N_b}{N_{max}} ~, \label{eq:M_Nb}\\ 
&\frac{dP}{dt} = \Delta P \frac{dH_r}{dt} ~. \label{eq:M_P} 
\end{align}
Solving Eq.~\eqref{eq:M_Nb} and Eq.~\eqref{eq:M_P} leads to
\begin{align}
& N_b(t) = N_{max} \left( 1 - e^{-t/N_{max}} \right)~, \label{eq:Nb_t}
  \\
\nonumber
& P(t) = \Delta P [(p_h - p_c) t \\
& + N_{max}(1 + p_c - p_h)(1 - e^{-t/N_{max}})]~. 
\label{eq:P_t1}
\end{align}
Eqs.~\eqref{eq:M_H},\eqref{eq:M_Nb},\eqref{eq:M_P} imply that the population is an
increasing function of the number of building up to a saturation value
$N_b^*$ corresponding to the population $P^*$, after which the
population decreases (ie. above which $dP/dt$ becomes negative). After
simple calculations (see Supplementary material for details), we
obtain
\begin{align}
&N_b^* = \frac{N_{max}}{1 + p_c - p_h}~, \label{eq:N_star} \\
&\frac{P^*}{\Delta P N_{max}}= (p_h - p_c) \log \left(\frac{1 +
      p_c - p_h}{p_c - p_h} \right) + 1 ~. \label{eq:P_star}
\end{align}
The saturation happens only if $N_b^* < N_{max}$ and thus if
$p_c > p_h$ which expresses the fact that the conversion rate should
be large enough in order to observe a saturation point (if the
conversion rate is too small, the first phase of growth will continue
indefinitely).  Defining the normalized variables $N_b^{*'} = N_b^* /N_{max}$ and
$P^{*'} = P^* /N_{max}$, we can rewrite the above equation as
\begin{equation}
\label{eq:P_star_1}
P^{*'} = \Delta P \left[ 1 + (1/N_b^{*'} - 1) \log(1 - N_b^{*'}) \right]~.
\end{equation}
This relation allows us to determine the average number of people per
building floor $\Delta P$ for each district (see the Supplementary
material for a discussion about this parameter). Also, the theoretical
results given by Eq.~\eqref{eq:Nb_t} and Eq.~\eqref{eq:P_t1} imply a
scaling that can be checked empirically. Indeed, if we make the
following change of variables
\begin{align}
\nonumber
&X(t) = \frac{N_b(t)}{ N_{max} }~, \\
&Z(t) = \frac{ \frac{P(t)}{\Delta P N_{max}} - \frac{N_b(t)}{N_b^*} }{
  \frac{1}{N_b^{*'}} - 1}~,
\label{eq:rescaled}
\end{align}
then the curves for the different districts at different times should
all collapse on the same curve given by
\begin{equation}
Z = \log{\left(1 - X\right)}~.
\label{eq:Z}
\end{equation} 

\begin{figure}[!]
\centerline{\includegraphics[width=1\linewidth]{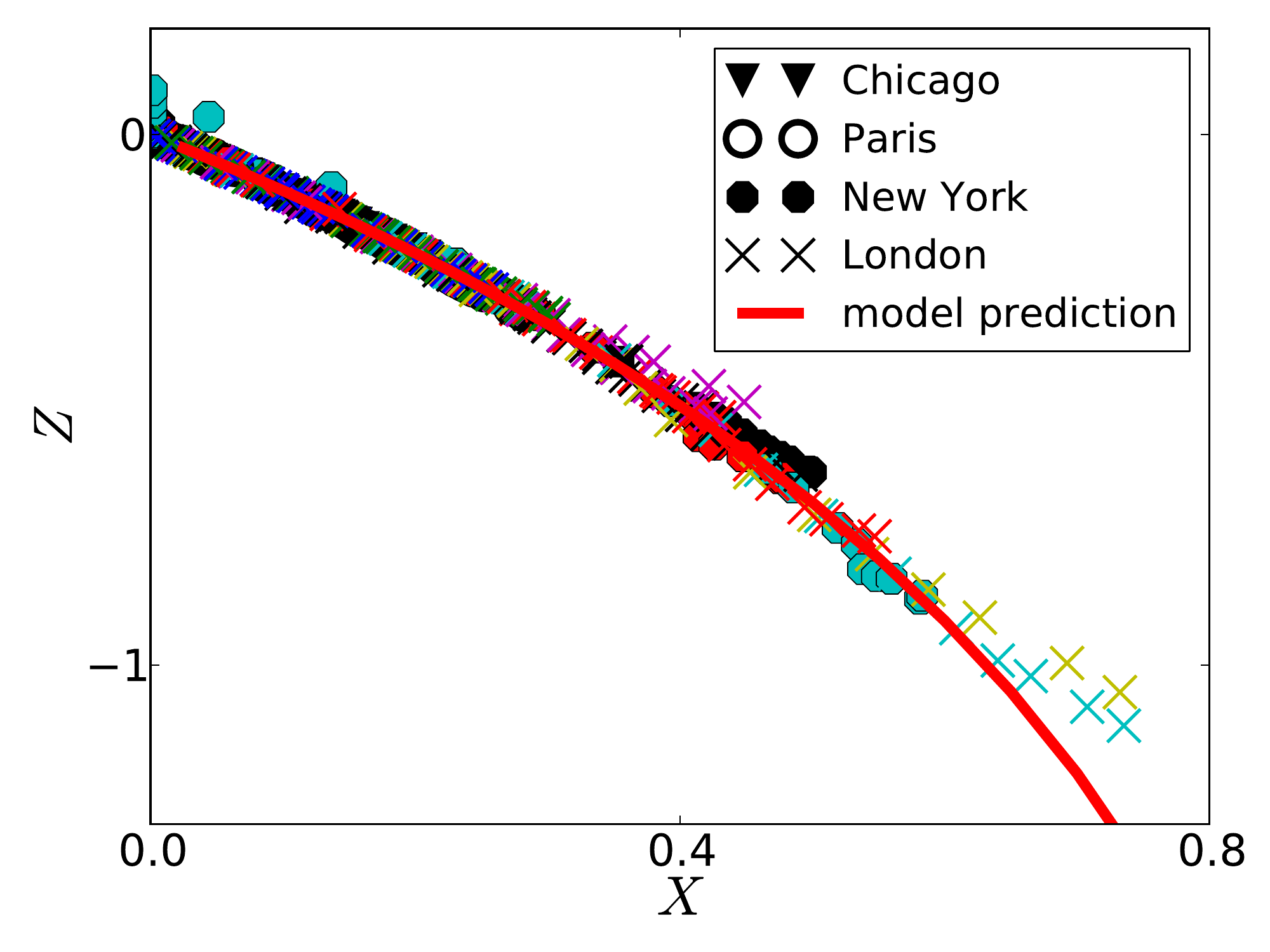}}
\caption{\textbf{Collapse for the rescaled variable $Z$ and $X$.} We
  plot the rescaled variables $Z$ versus $X$
  (Eqs. \eqref{eq:rescaled}) for all the $47$
  saturated districts of all cities. Each city is characterized by
  a different symbol and each district by a different color. The
  continuous red line is the theoretical prediction given by
  Eq.~\eqref{eq:Z}. All the cities considered in this study are present
  and we kept the districts that have saturated and for which we can
  compute $(P^*,N_b^*)$. (We give the list of the $47$ districts shown here in
  the Supplementary material).}
\label{fig:collapse}
\end{figure}

In order to test this model, we focus on all districts that have already
reached saturations (the others are still in the first growth
phase). From the data we know the area $A$ of each district and the
average building footprint surface $a_l$ of each district. This allows
us to compute the maximum number of buildings $N_{max} = A/a_l$ of the
district.  Moreover, the empirical curves allow us to determine the
saturation values $(P^*, N_b^*)$, corresponding to the value of the
population and the number of buildings after which the density growth
rate becomes negative (and we can then compute $(P^{*'}, N_b^{*'})$).
At this point, we thus have estimated from empirical data all the
parameters that characterize a district, without performing any fit.
We can now test the scaling Eq.~\eqref{eq:Z} predicted by the
model. As explained above, the curves obtained for different districts
should all collapse on the theoretical one. In Fig.~\ref{fig:collapse}
we plot the theoretical prediction (red line) and the values for the
different districts (represented by different symbols and different
colors for the different districts). An excellent collapse is
observed, supporting the validity of the model.

This collapse is a validation of the model: it shows that the
non-trivial relation between variables (Eq. \ref{eq:Z}) predicted by
the model is in agreement with the data. We observe deviations for
larger values of $X$ for districts in London that might be explained
by the uncertainty in determining the area of buildings in this city.

\section{Discussion and perspectives}

Theoretical urban models can be roughly divided in two categories. On
one hand there are economics models characterized by complex
mathematical equations rarely amenable to quantitative
predictions that can be tested against data. On the other hand, there
are computer simulations (such as agent-based models or cellular automata) that are
characterized by a large number of parameters, preventing to
understand the hierarchy of processes governing the phenomenon. In the
approach presented here, we build a simple model with the smallest
number of parameters and able to describe quantitatively the evolution
of various quantities such as the number of buildings and the
population for a given district.  

The agreement with data is tested with a data collapse which does
not rely on a parameter fit. The excellent agreement observed shows
that the model is able to explain empirical data. However, this
agreement is not a definitive proof that the model described here is
the fundamental one. Ideally one should compare with other existing
models but in this case our proposal seems to be the first attempt to
describe quantitatively the evolution of fundamental quantities with
the help of simple fundamental mechanisms. Interestingly enough this
random model relies on a set of simple reasonable assumptions such as
growth and conversion and also on non-correlated growth of buildings
inside districts, an assumption that seems to be both supported by
empirical measures on districts and the theoretical model.

Further quantitative studies are however needed and are of two types. First other
datasets for other cities are needed in order to test for the validity
of the quantitative behavior observed here. Also, the comparison with
other competing theoretical models could be very fruitful and we can
only encourage the construction of such models.

Our empirical analysis confirms that there are essentially three
different phases of the urbanization process: a growth phase where we
observe an increase of both the number of buildings and the
population; a second regime where the population decreases while the
number of buildings stays roughly constant, and a last phase where
both population and the number of buildings are increasing.  The first
two phases are well described by the simple model proposed here and
which integrate the crucial ingredient of converting residential space
into commercial activities. We observe empirically the existence of a
`re-densification' phase where both population and the number of
buildings increase after the conversion phase. This phase seems to
happen simultaneously for the different districts in a city which
suggests that it is an effect due to planning decisions and not
resulting from self-organization. Modelling the appearance of this
regime is thus at this point a challenge for future studies.

Beside showing that a minimal modeling for describing urbanization is
possible despite the large variety of cities, we believe that this
approach could constitute the basis for more elaborated models. These
models could then be thoroughly tested against data, could describe
the impact of various parameters and also help to understand some
features of the possible future evolution of cities.

\section{Materials and Methods}

\subsection{Data description}

\subsubsection{New York data}

We used data from the Primary Land Use Tax Lot Output (PLUTO) data
file, developed by the New York City Department of City Planning’s
Information Technology Division (ITD)/Database and Application
Development Section~\cite{dataNY}. It contains extensive land use and
geographic data at the tax lot level. PLUTO data files contain three
basic types of data: tax lot characteristics, building characteristics
and geographic/political/administrative districts. In particular for
each building of the city we focused on the building's borough, the
building age and the surface of the lot. For each borough we compute
the average building surface $a_l$ (assumed to be given by the average
building lot surface) over all the buildings in the borough and with
known age (for New York city we have this information for $94 \%$ of
buildings).  New York data cover the period from $1790$ to $2013$. For
the historical population data, we used different sources \cite{data_P_NY_1, data_P_NY_2, data_P_NY_3,
  data_P_NY_4} 

\subsubsection{Chicago data}

We used the Building Footprints dataset (deprecated August 2015)
provided by the Data portal of the City of Chicago~\cite{data_Chicago_1}. For each building
we have the information on the geometrical shape from which we compute
the building surface, the year built and the position. By using the
shapefiles of the $77$ Chicago communities~\cite{data_Chicago_2}, we can deduce the community
(and thus the side) where the building is located in. For each side we
compute the average building surface $a_l$ and average this quantity
over all the buildings with known year built, situated in the side.
For Chicago the percentage of buildings with known built year is $54 \%$.
Population data from each community area comes from ~\cite{data_P_Chicago} and they cover the period from
$1930$ to $2010$.

\subsubsection{Paris data}

We used the dataset `Emprise Batie Paris' provided by the open data
initiative of the `Atelier Parisien d'urbanisme (APUR)'~\cite{data_Paris}. For each
building we have the information on the geometrical shape, from which
we compute the building surface, the year built and the arrondissement
the building is situated in. For each arrondissement we compute the
average building surface $a_l$ averaging this quantity over all the
buildings with known year built (i.e. the $ 57 \% $ of the buildings),
situated in the arrondissement. Population data comes
from~\cite{data_P_Paris} and since the actual
arrondissements where defined in $1859$, population data at the level
of the arrondissements covers the period from $1861$ to $2011$.

\subsubsection{London data}

We used the dataset `Dwelling Age Group Counts
(LSOA)'~\cite{dataLondon1}, which contain the residential dwelling
ages, grouped into approximately 10-year age bins from pre-$1900$ to
$2015$ (the bin $1940-1944$ is missing). The number of properties is given for each LSOA area and each
age bin. From these data we deduced the number of buildings for each
London district as function of the year. Data for the historical
population of the London boroughs were obtained from `A Vision of
Britain through time'~\cite{data_P_London}. Finally we used
OSOpenMapLocal~\cite{dataLondon3} containing the geometrical shape of the buildings in
London for computing the average footprint surface for each
district. We note that in this last dataset some buildings are
aggregated and rendered as homogeneized zones. For this reason we
computed the average building surface of each district by averaging
over all the buildings belonging to the district having a footprint
surface smaller than $700 m^2$. In order to locate the district to
which a building belongs to, we used the shapefile of London districts
boundaries~\cite{london_boundary}. 

\section{Acknowledgements}
GC thanks the Complex Systems Institute in Paris (ISC-PIF) for hosting her during part
of this work and Riccardo Gallotti for interesting discussions. MB thanks the program Paris 2030 for financial support.

\bibliographystyle{prsty}

\begin{thebibliography}{99}


\bibitem[Champion(2001)]{Champion:2001}
T. Champion, {\em Urbanization, suburbanisation, counterurbanisation
and reurbanisation.} In: Paddison, R. (Ed.), Handbook of
Urban Studies. Sage, London, pp. 143–161. (2001).

\bibitem[Antrop(1994)]{Antrop:2004}
M Antrop, {\em Landscape change and the urbanization process in
Europe}, Landscape and urban planning 67.1 (2004): 9-26.

\bibitem[Fielding(1982)]{Fielding:1982}
A.J. Fielding, {\em Counterurbanisation in Western
Europe}, Progress in Planning, 17 (1): 1–52. (1982).

\bibitem[Geyer and Kontuly(1993)]{Geyer:1993}
H.S. Geyer and T.M. Kontuly, {\em A theoretical
foundation for the concept of differential urbanisation},
International Regional Science Review, 15 (12):
157–77. (1993).

\bibitem[Hall(1971)]{Hall:1971}
P. Hall, {\em Spatial structure of metropolitan
England and Wales}, in M. Chisholm and G.
Manners (eds), Spatial Policy Problems of the British
Economy. Cambridge: Cambridge University Press.
pp. 96–125. (1971).

\bibitem[Anas et al.(1998)]{Anas:1998}
A. Anas, R. Arnott., K.A. Small, {\em Urban spatial
structure}, Journal of economic literature 36.3 (1998): 1426-1464.

\bibitem[Glaeser and Kahn(2003)]{Glaeser:2003}
E.L.~Glaeser and M.E.~Kahn, {\em Sprawl and Urban Growth}, the National Bureau of Economic Research, Working Paper 9733 http://www.nber.org/papers/w9733 (2003).

\bibitem[Herbert and Stevens(1960)]{Helbert:1960}
J.D.~Herbert and B.H.~Stevens, {\em A model for the distribution of residential activity in urban areas}. Journal of Regional Science, 2-2 (1960).

\bibitem[Mills(1967)]{Mills:1967}
E.S.~Mills, {\em An Aggregative Model of Resource Allocation in a Metropolitan Area}. The American Economic Review, 57-2 (1967), pp.~197-210.

\bibitem[Meuriot(1898)]{Meuriot:1898}
Meuriot,~M.P. {\em Des Agglomérations Urbaines Dans L'europe Contemporaine: Essai Sur Les Causes, Les Conditions, Les Conséquences De Leur Développement}. Paris: Belin, (1898).  

\bibitem[Taylor(1949)]{Taylor:1949}
Taylor,~T.G. {\em Urban geography}. London: Methuen, (1949). 

\bibitem[Wheaton(1982)]{Wheaton:1982} W.C.~Wheaton, {\em Urban
    Residential Growth under Perfect Foresight}Journal of Urban
  Economics, 12 (1982), pp.~1--21.

\bibitem[Beckmann(1969)]{Beckmann:1969} M.J.~Beckmann, {\em On the
    distribution of urban rent and residential density }Journal of
  Economic Theory, 1-1 (1969), pp.~60--67.

\bibitem[Harrison and Kain(1973)]{Harrison:1973}
D.~Harrison and J.F.~ Kain, {\em Cumulative Urban Growth and Urban Density Functions}. Journal of Urban Economics, 1 (1974), pp.~61--98.

\bibitem[Tobler(1970)]{Tobler:1970}
Tobler, Waldo R. {\em A computer movie simulating urban growth in the Detroit region.} Economic geography 46.sup1 (1970): 234-240.

\bibitem[Makse et al.(1995)]{stanley:1995}
Makse H.A., Havlin S., Stanley H.E., {\em Modelling urban growth.} Nature, 337:1912 (1995), 779-782.

\bibitem[Clark(1951)]{clark:1951}
C.~Clark, {\em Urban population densities }. Journal of the Royal Statistical Society, Series A (General), 114-4 (1951), pp.~490--496.

\bibitem[Anas(1978)]{Anas:1978}
A.~Anas, {\em Dynamics of Urban Residential Growth}, Journal of Urban Economics, 5 (1978), pp.~66--87.

\bibitem[Allen and Sanglier(1981)]{Allen:1981}
P.M.~Allen and M.~Sanglier, {\em Urban evolution, self-organization,
  and decision making}, Environment and Planning A,
  13 (1981), pp.~167--183.

\bibitem[Benenson(1999)]{Benenson:1999}
I.~Benenson, {\em Modeling population dynamics in the city: from a regional to a multi-agent approach}, Discrete dynamics in nature and society, 3 (1999), pp.~149--170.

\bibitem[Dendrinos(1982)]{Dendrinos:1982} D.~Dendrinos and
  H.~Mullally, {\em Evolutionary patterns of urban populations},
  Geographical Analysis, 13 (1982), pp.~328--344.

\bibitem[Sullivan and Torrens(2001)]{Sullivan:2001} D.~O'Sullivan and
  P.M. Torrens, {\em Theory and Practical Issues on Cellular Automata}
  chapter {\em Cellular Models of Urban Systems}, Springer London
  (2001). url="http://dx.doi.org/10.1007/978-1-4471-0709-5\_13"

\bibitem[Tannier and Pumain (2005)]{Tannier:2005}
C. Tannier, D. Pumain. Fractals in urban geography: a
theoretical outline and an empirical example. Cybergeo: European
Journal of Geography (2005).

\bibitem[Batty and Longley(1994)]{Batty:1994}
M. Batty, P.A.  Longley. Fractal cities: a geometry of
form and function. Academic Press, 1994.

\bibitem[Makse et al.(1998)]{Makse:1998}
HA. Makse, J.S. Andrade, M. Batty, S. Havlin, H.E. Stanley. Modeling urban growth patterns with correlated
percolation. Physical Review E, 58(6), 7054 (1998).


\bibitem[Perret et al.(2015)]{Perret:2015}
J. Perret, M. Gribaudi, M. Barthelemy. {\em Roads and cities of 18th
  century France} Scientific data 2 (2015).

\bibitem[Angel et al.(2012)]{Angel:2012}
Angel, S., J. Parent, D.L. Civco and A.M. Blei. {\em The Atlas of Urban Expansion}. Lincoln Institute of Land Policy (2012).

\bibitem[Mason(2013)]{Wired:2013}
Mason B. (2013) {\em Brilliant Maps Reveal Age of the World’s Buildings.} Available at http://www.wired.com/2013/10/building-ages-map-gallery/. Accessed 31/08/2016.

\bibitem[Jacobsen(2013)]{Chicago_map} Jacobsen S. (2013) {\em Chicago Building Age Map.} Available at http://transitized.com/chibld/index.html?utm\_source=-//Transitized\&utm\_medium=\\post\&utm\_campaign=New\%20CBAM\%
20Post (Accessed 31/08/2016).

\bibitem[Brandom(2016)]{NY_map} Brandon L. {\em Building Age NYC.} Available at http://pureinformation.net/building-age-nyc/\#12/40.7392/-73.9651. Accessed 31/08/2016.

\bibitem[ljubljana(2016)]{ljubljana_map} Available at
  http://www.virostatiq.com/data/ljubljana-building-ages/. Accessed
  31/08/2016.

\bibitem[Riggott(2016)]{reykjavik_map} Riggott M. {\em The Age of
    Greater Reykjav\'ik.} Available at
  http://tiles.flother.is/2013/reykjavik-age/ . Accessed 31/08/2016.

\bibitem[Palmer(2014)]{Portland_map} Palmer J. (2014) {\em Portland
    Oregon: The Age of a City.} Available at
  http://labratrevenge.com/pdx/\#12/45.4483/-122.7139. Accessed
  31/08/2016.

\bibitem[Morphocode(2016)]{MN_map} Morphocode {\em Urban
    Layers}. Available at
  http://io.morphocode.com/urban-layers/. Accessed 29/08/2016.

\bibitem[Palmer(2013)]{Palmer:2013} Palmer J. (2013) {\em Portland
    Oregon: The Age of a City.} Available at
  http://dealloc.me/2013/06/30/the-making-of-pdx/. Accessed
  31/08/2016.

\bibitem[Plahuta(2013)]{Plahuta} Plahuta M. (2013) {\em Building ages
    in Ljubljana, Slovenia.} Available at
  http://virostatiq.com/structure-ages-in-ljubljana-slovenia/. Accessed
  29/08/2016.

\bibitem[O'Hara(2016)]{video_Ljub} O'Hara M. {\em City of Ljubljana -
    growth between years 1500 - 2013.} Available at
  https://vimeo.com/72249300. Accessed 29/08/2016.

\bibitem[Ureta(2016)]{BUILT_LA} Ureta O. {\em built: LA Building Age
    // 1890-2008.}  Available at
  http://cityhubla.github.io/LA\_Building\_Age/\#12/34.0267/-118.2621.
  Accessed 29/08/2016.

\bibitem[Bertaud and Malpezzi(2003)]{Bertaud:2003} Bertaud A. and Malpezzi S. (2003). {\em The spatial distribution of
population in 48 world cities:  Implications for economies in transition.} Report, World Bank.

\bibitem[Gu\'erois and Pumain(2008)]{Pumain_Guerois:2008} Gu\'erois
  M., Pumain D. (2008). {\em Built-up encroachment and the urban
    field: a comparison of forty european cities.}  Environment and
  Planning A 40: 2186-2203.

\bibitem[Arcaute et al.(2015)]{Arcaute:2015} Arcaute, E., Hatna, E., Ferguson, P., Youn, H.,
  Johansson, A., Batty, M. (2015). Constructing cities,
  deconstructing scaling laws. Journal of The Royal Society Interface,
  12(102), 20140745.
  
\bibitem[Frey(1990)]{Frey:1990}
W.H. Frey, {\em Metropolitan America: beyond the
transition}, Population Bulletin 45 (2). Washington,
DC: Population Reference Bureau. (1990).

\bibitem[Nucci and Long(1995)]{Nucci:1995}
A. Nucci and L. Long, {\em Spatial and demographic
dynamics of metropolitan and nonmetropolitan
territory in the United States}, International Journal
of Population Geography, 1 (2): 165–81. (1995).

\bibitem[Champion(1989)]{Champion:1989}
A.G. Champion, {\em Counterurbanisation: The
Changing Pace and Nature of Population Deconcen-
tration.} London: Edward Arnold. (1989).

\bibitem[Beale(1975)]{Beale:1975}
C.L. Beale, {\em The Revival of Population Growth in
Non-metropolitan America.} Washington, DC:
Economic Research Service, US Department of
Agriculture. (1975).

\bibitem[Berry(1976)]{Berry:1976}
B.J.L. Berry, {\em Urbanization and Counterur-
banization.} Beverly Hills, CA: Sage. (1976). 
  
  

\bibitem[PLUTO dataset (2016)]{dataNY}
  {\em Primary Land Use Tax Lot Output (PLUTO) dataset}, developed by the New York City Department of City Planning’s Information Technology Division (ITD)/Database and Application Development Section. Retrieved from http://www1.nyc.gov/site/planning/data-maps/open-data.page. Accessed 2016.

\bibitem[data NY (2016)]{data_P_NY_1}
{\em Table PL-P1 NYC: Total Population New York City and Boroughs, $2000$ and $2010$} (PDF). nyc.gov. Retrieved $16$ May $2016$.

\bibitem[data NY (2010)]{data_P_NY_2}
U.S. Census Bureau, Population Division, Table 5. Annual Estimates of the Resident Population for Minor Civil Divisions in New York, Listed Alphabetically Within County: April 1, 2000 to July 1, 2009 (SUB-EST2009-05-36) and Table 1. Annual Estimates of the Resident Population for Incorporated Places Over 100,000, Ranked by July 1, 2009 Population: April 1, 2000 to July 1, 2009 (SUB-EST2009-01), Release Date: June 2010, retrieved on July 31, 2010.

\bibitem[data NY(2011)]{data_P_NY_3}
Forstall, Richard L., Population of States and Counties of the United States: 1790 to 1990, U.S. Bureau of the Census, Washington, DC, 1996 ISBN 0-934213-48-8, (Part III, Kentucky to Oklahoma) retrieved April 3, 2011.
\bibitem[data NY(1995)]{data_P_NY_4}
"Population", article by Jane Allen with tables by Nathan Kantrowitz in The Encyclopedia of New York City, edited by Kenneth T. Jackson, New-York Historical Society and Yale University Press, 1995, pages 910-914, ISBN 0-300-05536-6.

\bibitem[data1 Chicago(2016)]{data_Chicago_1}
City of Chicago Data Portal. {\em Building Footprints (deprecated August 2015)}. Available at https://data.cityofchicago.org/Buildings/Building-Footprints-deprecated-August-2015-/qv97-3bvb/data.
Accessed 31/08/2016

\bibitem[data2 Chicago(2016)]{data_Chicago_2}
City of Chicago Data Portal. {\em Boundaries - Community Areas (current).} Available at https://data.cityofchicago.org/Facilities-Geographic-Boundaries/Boundaries-Community-Areas-current-/cauq-8yn6/data. Accessed 31/08/2016.

\bibitem[data3 Chicago(2016)]{data_P_Chicago} http://www.robparal.com/downloads/ACS0509/Histori-\\calData/Chicago. Accessed 2016.

\bibitem[data1 Paris(2016)]{data_Paris} Platforme open data de l'atelier parisien d'urbanisme. {\em Emprise batie Paris.} Available at http://cassini.apur.opendata.arcgis.com/datasets/002f14-\\c0cf28435296a341d9921adf99\_0 . Accessed 31/08/2016

\bibitem[data2 Paris(2016)]{data_P_Paris} Data from Wikipedia. 
Available at {https://fr.wikipedia.org/wiki/1er\_arrondissement\_de\_Paris}. Accessed 31/08/2016.
\bibitem[data London1 (2016)]{dataLondon1}
    Consumer Data Research Centre, {\em Dwelling Age Group Counts (LSOA)}, retrieved from https://data.cdrc.ac.uk/dataset/house-ages-and-prices/resource/4f1956b2-3128-4297-ba97-059e1fbc1fcc . Accessed 31/08/2016.

\bibitem[data London2(2016)]{data_P_London} A vision of Britain through time. Available at http://www.visionofbritain.org.uk/. Accessed 31/08/2016.

\bibitem[data London3(2016)]{dataLondon3}
    Ordnance Survey, {\em OS Open Map - Local}, retrieved from https://www.ordnancesurvey.co.uk/business-and-government/products/os-open-map-local.html. Accessed 31/08/2016.

\bibitem[data London4(2016)]{london_boundary} London DataStore. http://data.london.gov.uk/dataset/-\\statistical-gis-boundary-files-london.  Accessed 2016.
\end{thebibliography}


\clearpage
\onecolumngrid
\section{Supplementary material}
\subsection{The model: calculations}

We analyze here Eq.~[6]. In particular, we show that there is in this model a critical
value of $N_b$ above which the population decreases (ie. above which
$dP/dt$ becomes negative).
We thus solve:
\begin{eqnarray*}
&\frac{dP}{dt}  \geq  0 \\
& \Delta P \frac{N_b}{N_{max}} (p_h - p_c) + \Delta P (1 - \frac{N_b}{N_{max}}) \geq& 0 ~, \\
\end{eqnarray*}
and obtain the following condition
\begin{equation}
N_b \leq \frac{N_{max}}{1 + p_c - p_h}
\end{equation}
which then implies that
\begin{equation}
N_b^* = \frac{N_{max}}{1 + p_c - p_h}~.
\end{equation}
We thus observe a saturation effect only if $N_b^* < N_{max}$ and thus if $p_c > p_h$.\\
We can then compute the time $t^*$ for which saturation happens: knowing that
\begin{equation}
N_b^* = N_{max}\left( 1 - e^{-t^*/N_{max}}  \right) 
\end{equation}
we get
\begin{equation}
t^* = N_{max} \log \left( \frac{1 + p_c - p_h}{p_c - p_h} \right)~.
\end{equation}
Using Eq. [8], we then obtain
\begin{equation}
P^* = \Delta P (p_h - p_c) N_{max} \log{\left(\frac{1 + p_c - p_h}{p_c - p_h} \right) + \Delta P N_{max} }~.
\end{equation}

\subsection{Additional measures}
Analysing the relation between the number of buildings and the population during a city district growth, we observed the emergence of a 'universal' pattern characterized by three regimes: urbanization, conversion and densification.

\subsubsection{Chicago sides}

Concerning Chicago, $5$ sides saturated, three of them in $1970$ and two of them in $1960$. Just one of them began a re-densification process in $1980$. In the converting period we have a variation of the population equals to $-16992$ corresponding to a decrease of $6.5 \%$ of the population.

\subsubsection{Paris arrondissements}
The first four arrondissements seem to have saturated before the first data available. The $13^{th}$ seems not yet saturated and the others with the exception of the $6^{th}$ present all three phases even if often the re-densification one is quite recent.

The average value of the saturation year is
\[
Year_s = 1932 \pm 28~,
\]
the average value of the re-densification date is 
\[
Year_d = 1993 \pm 10~,
\]
the average period of conversation in years is
\[
\Delta t_{c} = 60 \pm 30 ~, 
\]
the average period of conversation in population is
\[
\Delta P_c = -58940 \pm 25434~,
\]
that corresponds to an average decrease of
\[
\frac{\Delta P_c}{P^*} = -0.33 \pm 0.16 \%~.
\]
The behaviors seem thus quite various.

\subsubsection{London districts}
Eight of the $33$ London districts seem saturated before the first data available. The others show all the three regimes.

We have

\[
Year_s = 1949 \pm 15~,
\]

\[
Year_d = 1992 \pm 2.43~,
\]

\[
\Delta t_c = 43 \pm 15~,
\]

\[
\Delta P_c = -63054 \pm 62908~,
\]

\[
\frac{\Delta P_c}{P^*} = -0.2 \pm 0.15 \%~.
\]
In particular one remarks that $23$ of the $25$ districts have $Year_d = 1992$

\subsubsection{New York boroughs}
In New York city only three of the five boroughs reached saturation and all of these present a densification regime.
We have
\[
Year_s(MN) = 1910 \qquad Year_s(BK) = 1950 \qquad Year_s(BX) = 1970~,
\]
all the boroughs began de re-densification phase in $1980$.

\subsection{List of districts in the data collapse and measure of $\Delta P$}
The parameter $\Delta P$ introduced in the model, defined as the number of people per residential floor has been estimated as explained in the main text from the equation
\begin{equation}
\label{eq:1}
\frac{P^*}{N_{max}} = \Delta P \left[ 1 + ( N_{max}/N_b^* - 1 )\log{( 1 - N_b^*/N_{max} )}  \right] = \Delta P / f(N_b^*, N_{max}).
\end{equation}
In the tables below we reported for each district the estimation we obtained. These latter allowed us to test the validity of the model through a data collapse in which we used data for all the districts that already reached a saturation point. These correspond to the districts for which we estimated the $\Delta P$ value in the tables below.
 
\subsubsection{New York}
\begin{center}
    \begin{tabular}{| l | l |}
    \hline
   	borough & $\Delta P$ \\ \hline
	MN &  $184.84$ \\ \hline
	BK &  $16.73$ \\ \hline
	BX & $29.36$ \\ \hline
    \end{tabular}
\end{center}

\subsubsection{Paris}
\begin{center}
    \begin{tabular}{| l | l |}
    \hline
   	arrondissement &  $\Delta P$ \\ \hline
	
	7  & $104$ \\ \hline
	8 & $146$ \\ \hline
	9  & $131$ \\ \hline
	10  & $159$ \\ \hline
	11  & $149$ \\ \hline
	12 & $103$ \\ \hline
	13 & $116$ \\ \hline
	14 & $85$ \\ \hline
	15 & $98$ \\ \hline
	16 & $76$ \\ \hline
	17 & $75$ \\ \hline
	18 & $128$ \\ \hline
	19 & $118$ \\ \hline
	20 & $127$ \\ \hline
    \end{tabular}
\end{center}

\subsubsection{Chicago}
\begin{center}
    \begin{tabular}{| l | l |}
    \hline
   	side  & $\Delta P $ \\ \hline
	Far North Side & $16.1$ \\ \hline
	Far Southeast Side  & $11.2 $ \\ \hline
	Northwest Side  & $10.2$ \\ \hline
	Far Southwest Side  & $9.2$ \\ \hline
	Southwest Side & $13 $ \\ \hline

    \end{tabular}
\end{center}

\subsubsection{London}
\begin{center}
    \begin{tabular}{| l | l |}
    \hline
    district & $\Delta P $ \\ \hline
	
	Lambeth & $10.7$ \\ \hline
	Greenwich & $15$ \\ \hline
	Sutton & $6.6 $ \\ \hline
	Lewisham & $8$ \\ \hline
	Barnet & $6.4 $ \\ \hline
	Hammersmith and Fulham & $6.6$ \\ \hline
	Barking and Dagenham & $6.6$ \\ \hline
	Enfield & $6.7$ \\ \hline
	Croydon & $5.2 $ \\ \hline
	Merton & $ 6.4$ \\ \hline
	Haringey & $ 6.9$ \\ \hline
	Harrow & $6.3$ \\ \hline
	Hounslow & $7.1$ \\ \hline
	Kingston upon Thames & $6.5$ \\ \hline
	Havering & $6.5$ \\ \hline
	Waltham Forest & $ 7$ \\ \hline
	Hillingdon & $5.8$ \\ \hline
	Bexley  & $5.4$ \\ \hline
	Ealing & $6.1 $ \\ \hline
	Bromley & $ 5.6$ \\ \hline
	Redbridge & $6.4 $ \\ \hline
	Newham & $14.9 $ \\ \hline
	Wandsworth & $8.8 $ \\ \hline
	Richmond upon Thames & $6.3 $ \\ \hline
	Brent & $6.9 $ \\ \hline
    \end{tabular}
\end{center}

\subsection{Additional empirical results}

For the sake of completness, we present here additional results for
the cities studies here.

\begin{figure}[b!]
\begin{center}
\begin{tabular}{c}
\includegraphics[angle=0, width=0.5\textwidth]{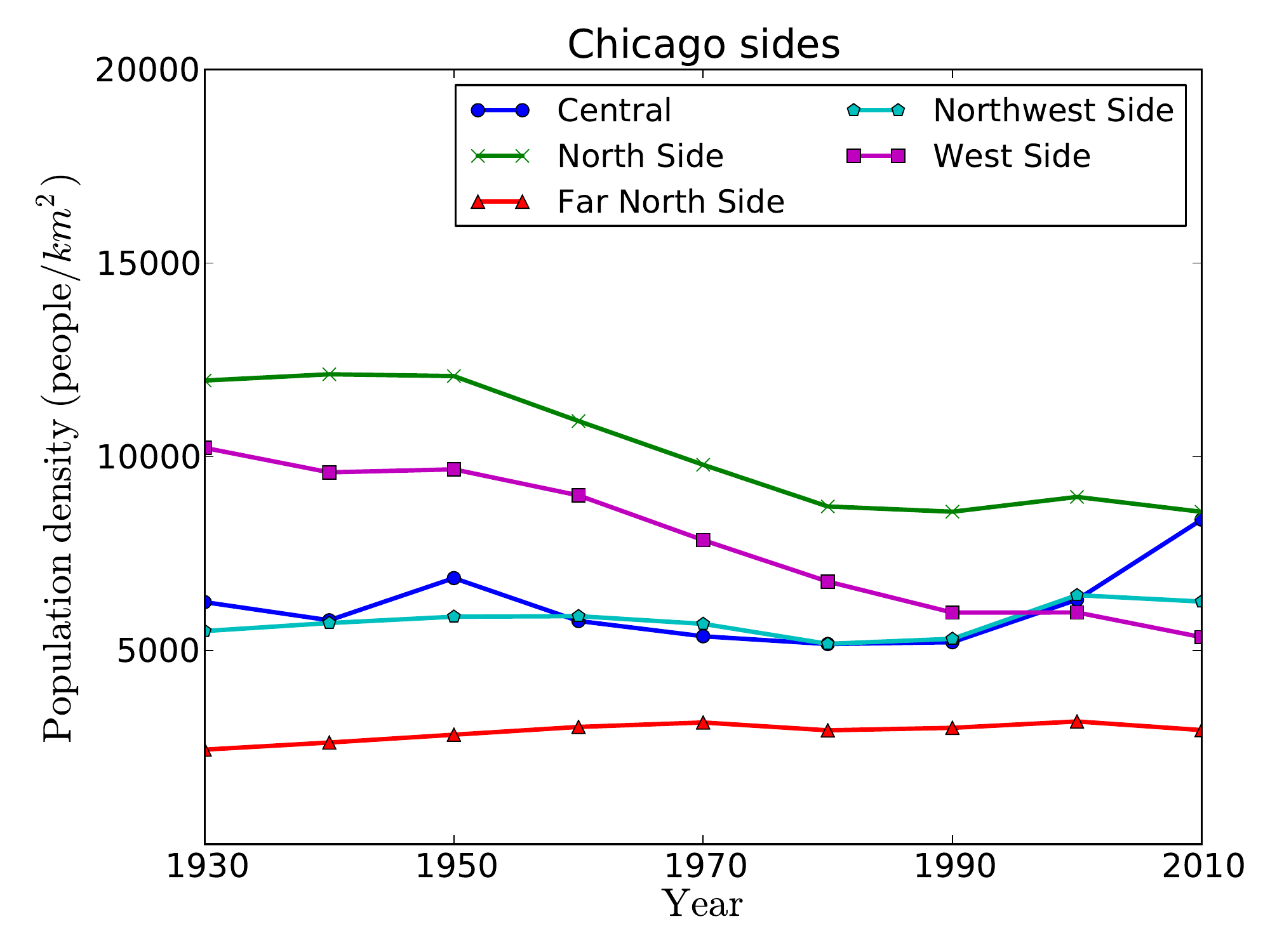} \\
\includegraphics[angle=0, width=0.5\textwidth]{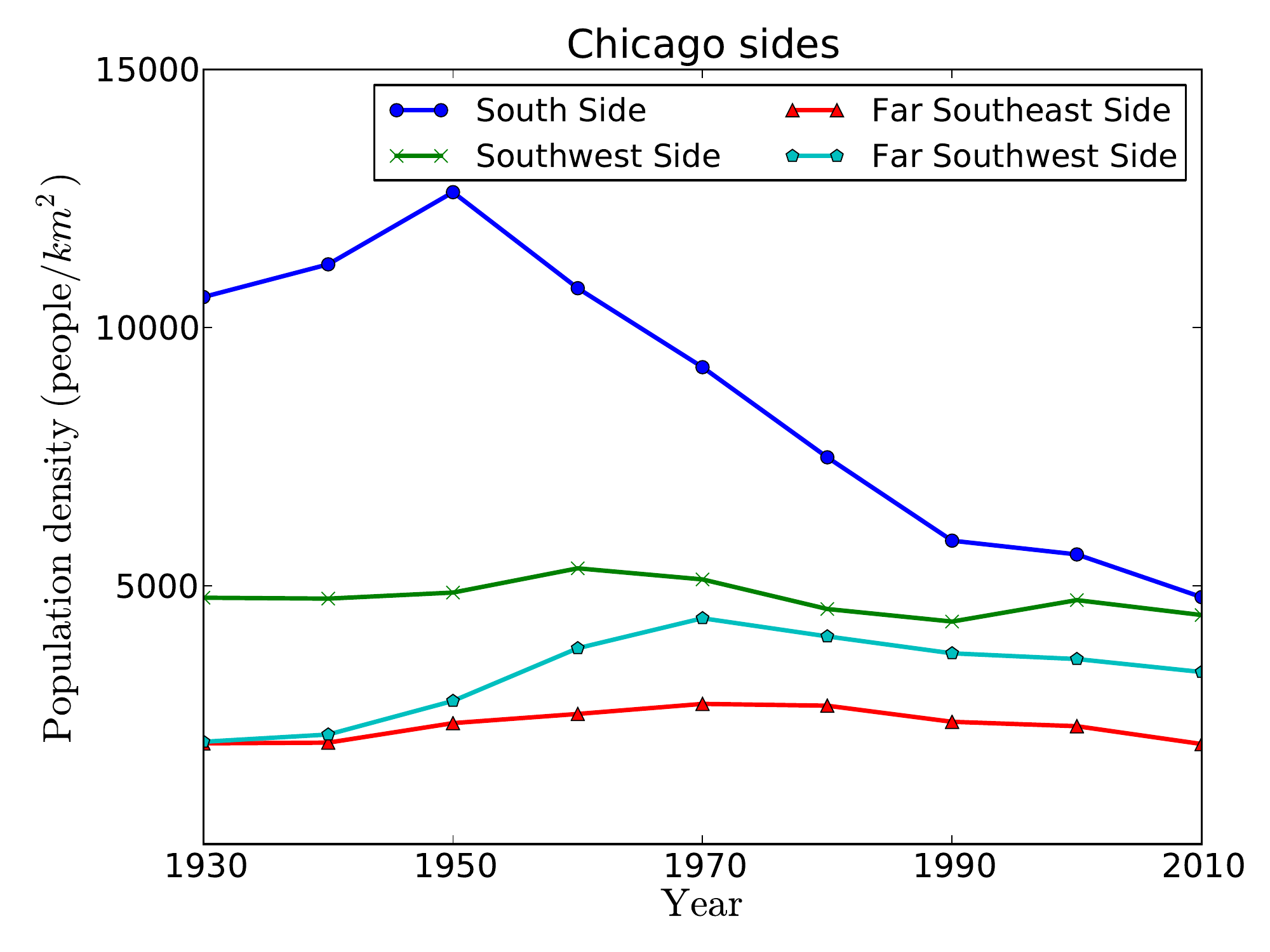} 
\end{tabular}%
\end{center}
\caption{\textbf{Chicago districts: population density VS year}}
\label{figS1}
\end{figure}

\begin{figure}[!]
\begin{center}
\begin{tabular}{c}
\includegraphics[angle=0, width=0.5\textwidth]{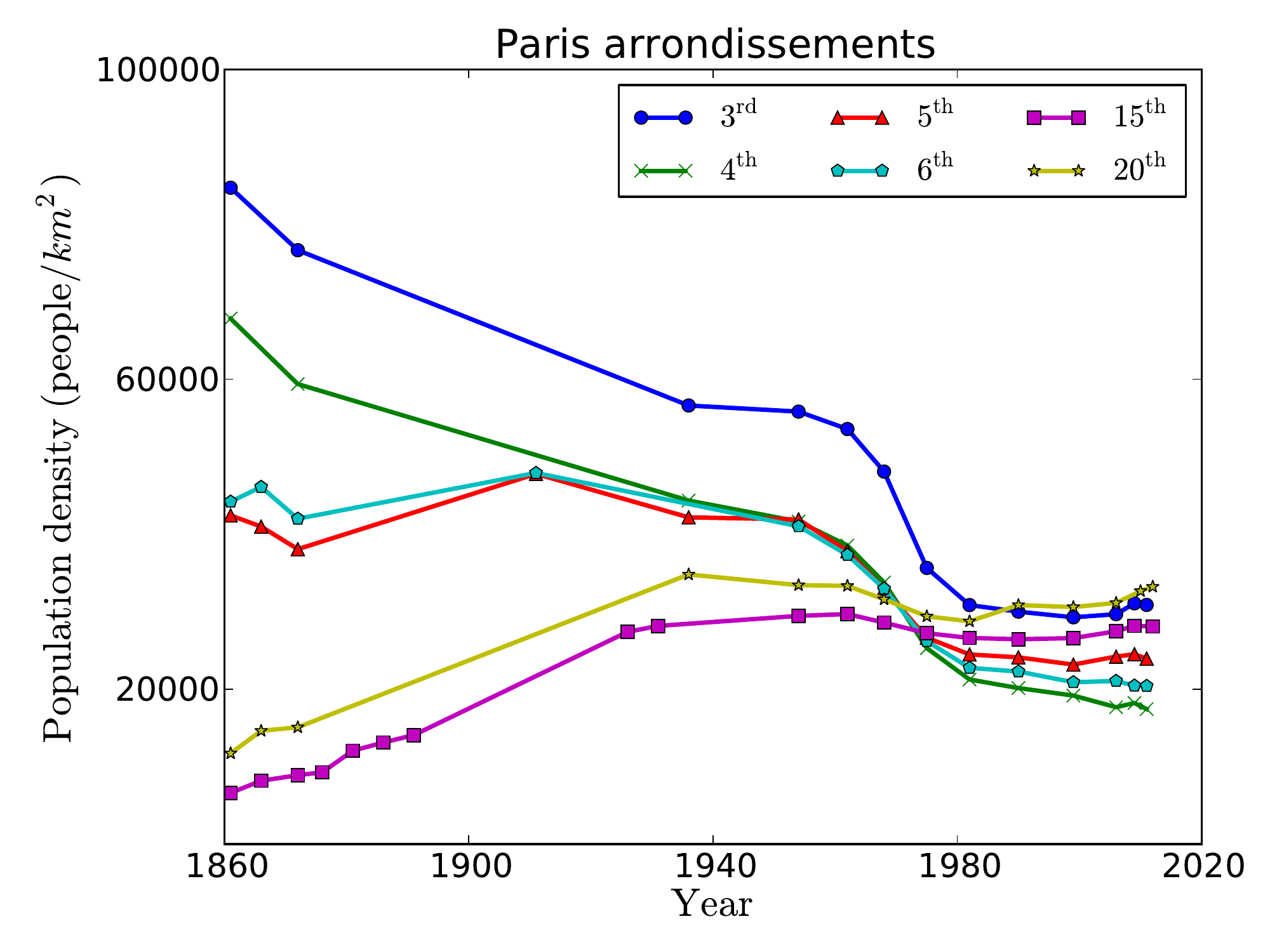} 
\end{tabular}%
\end{center}
\caption{\textbf{Paris arrondissements: population density VS year}}
\label{figS2}
\end{figure}

\begin{figure}[!]
\begin{center}
\begin{tabular}{c}
\includegraphics[angle=0, width=0.5\textwidth]{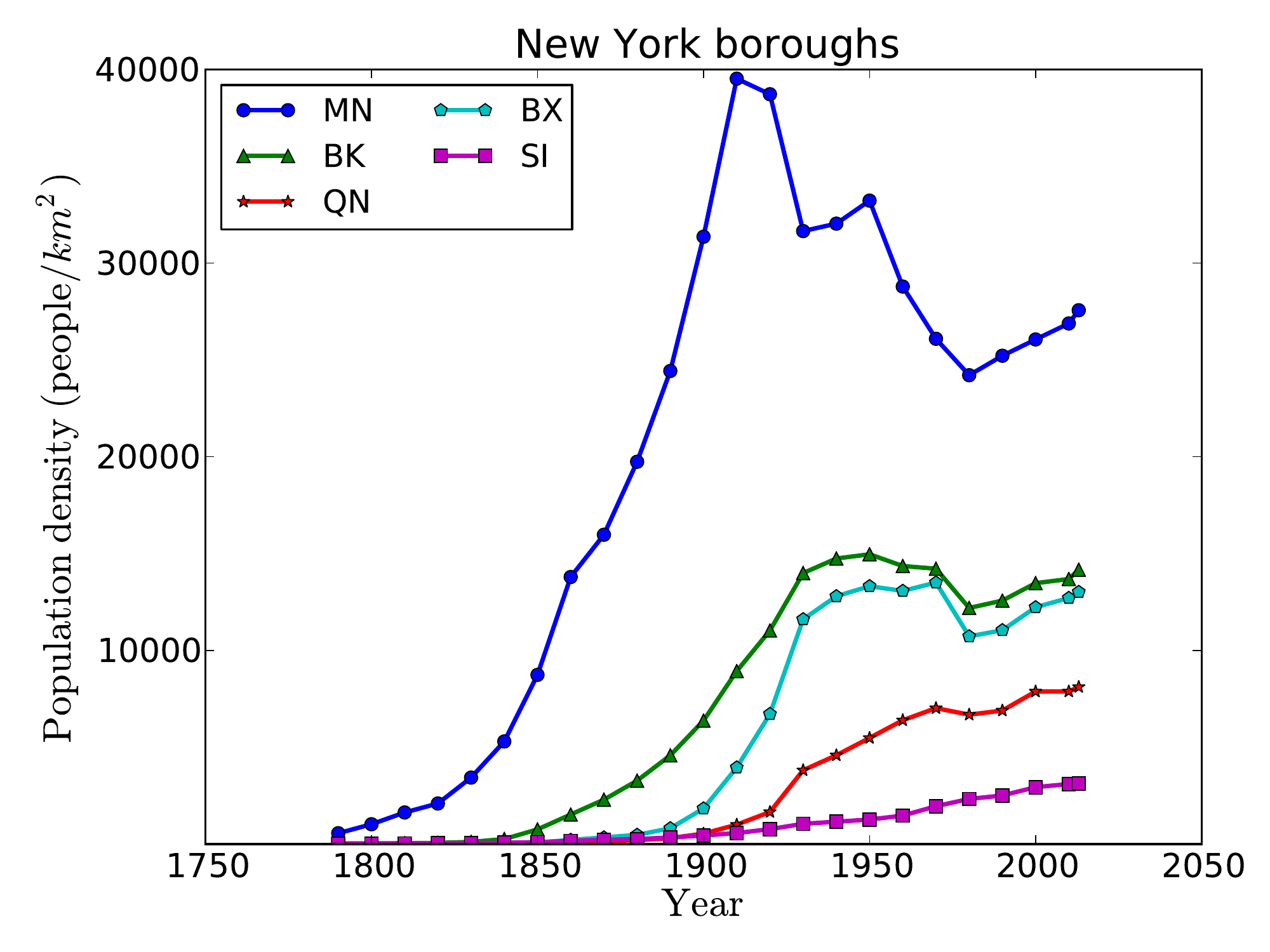} 
\end{tabular}%
\end{center}
\caption{\textbf{New York boroughs: population density VS year}}
\label{figS3}
\end{figure}

\begin{figure}[!]
\begin{center}
\begin{tabular}{c}
\includegraphics[angle=0, width=0.5\textwidth]{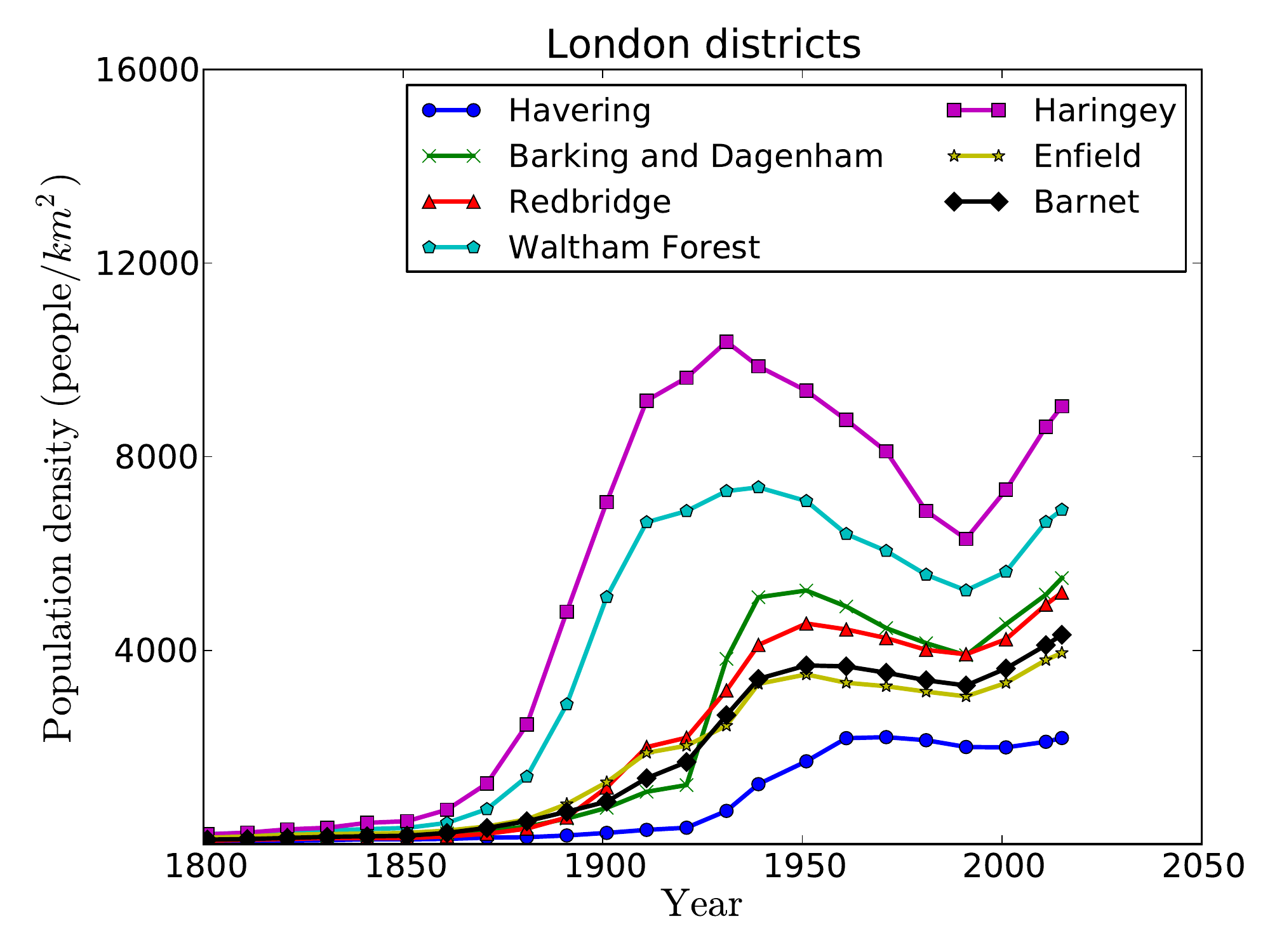} \\
\includegraphics[angle=0, width=0.5\textwidth]{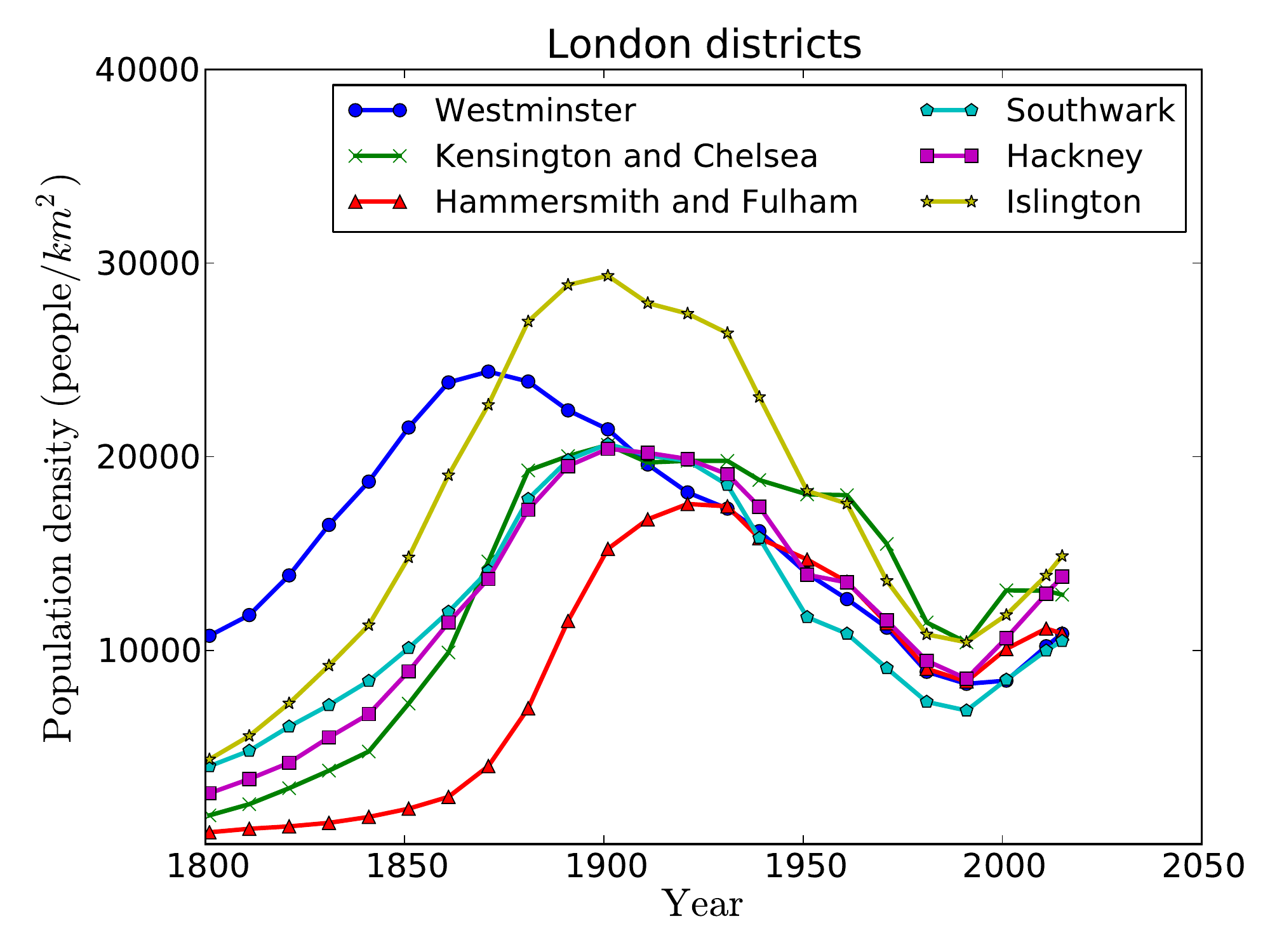} \\
\includegraphics[angle=0, width=0.5\textwidth]{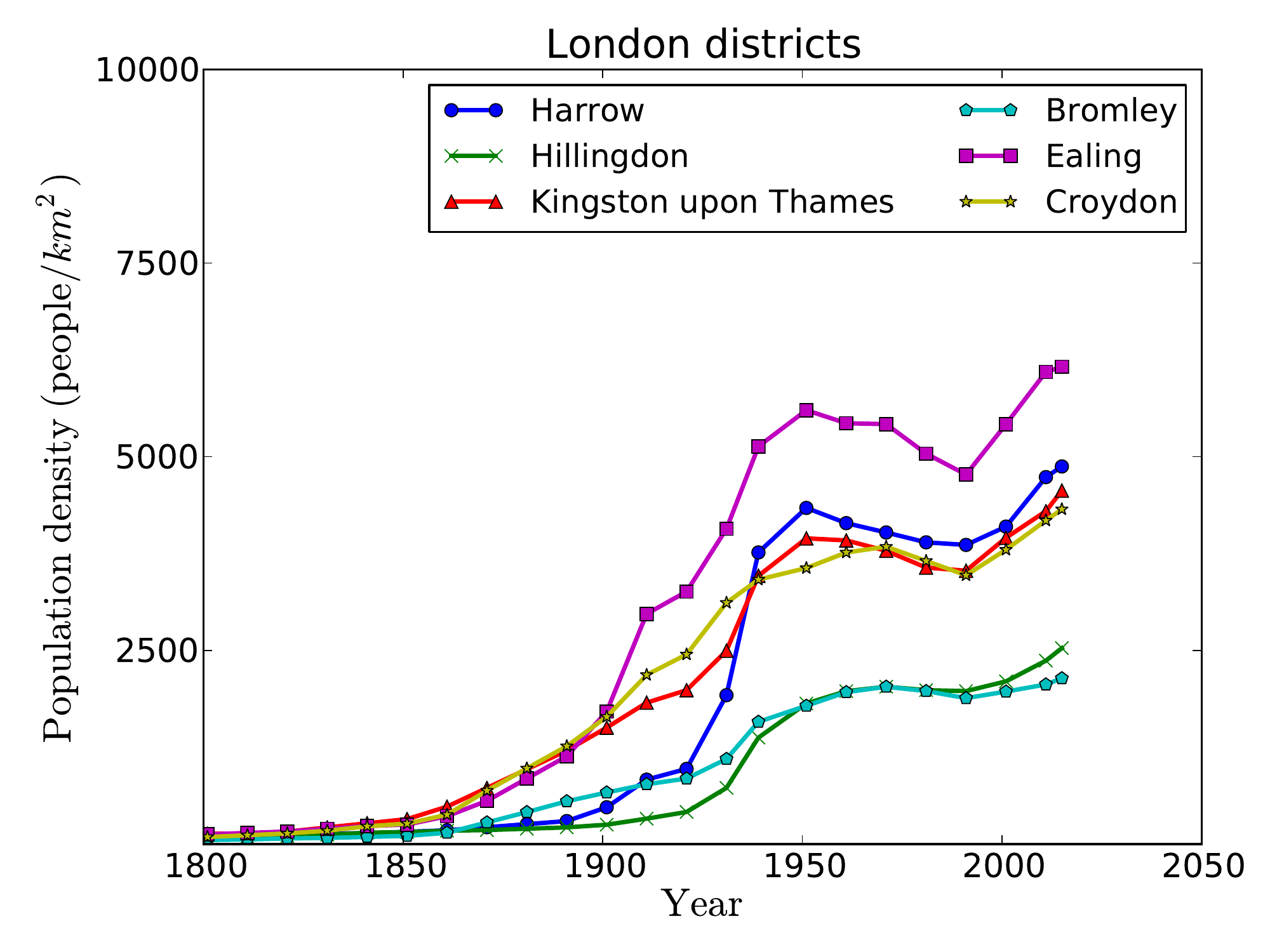} 
\end{tabular}%
\end{center}
\caption{\textbf{London districts: population density VS year}}
\label{figS4}
\end{figure}

\begin{figure}[b!]
\begin{center}
\begin{tabular}{cc}
\includegraphics[angle=0, width=0.5\textwidth]{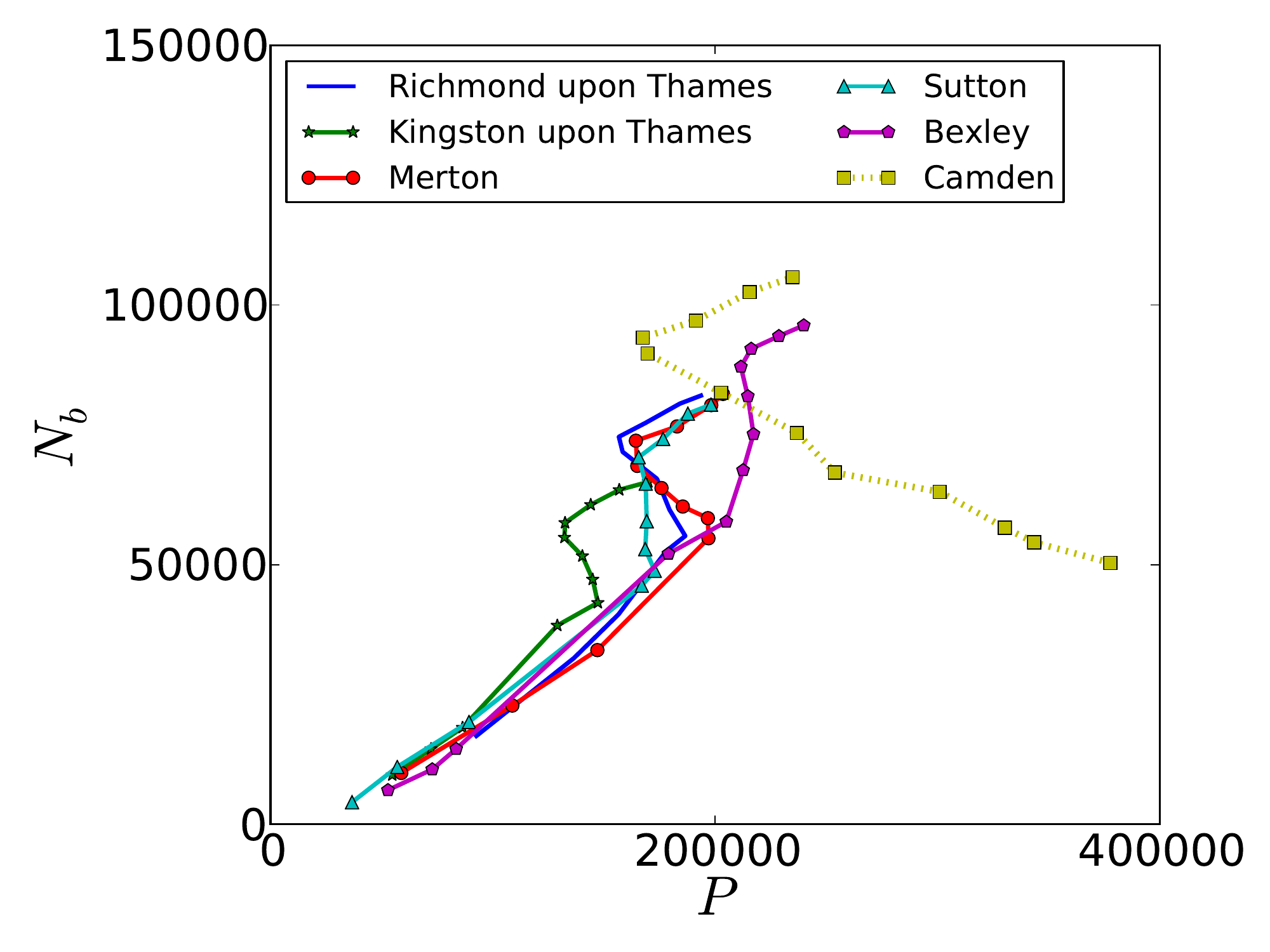} &
\includegraphics[angle=0, width=0.5\textwidth]{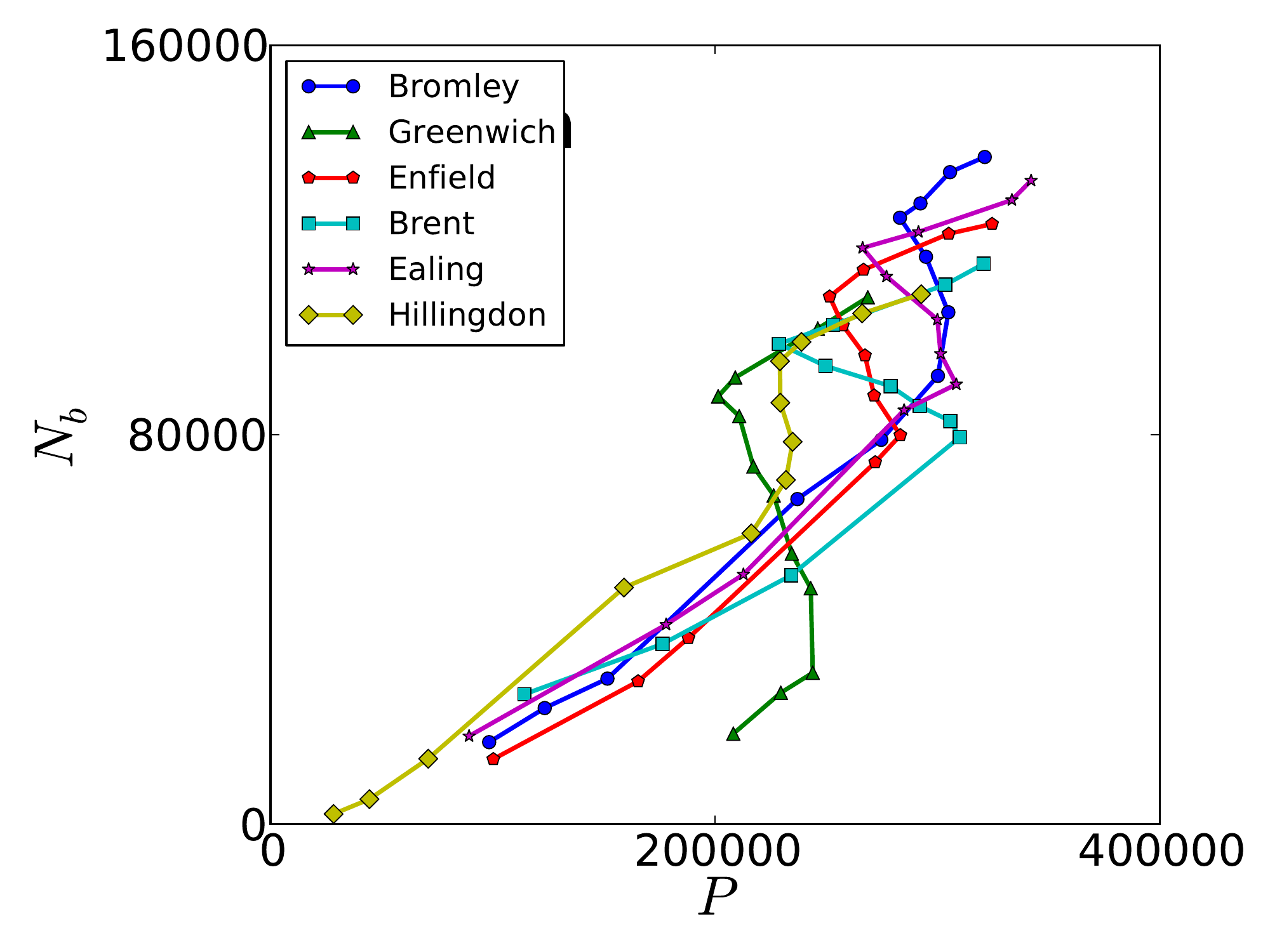} \\
\includegraphics[angle=0, width=0.5\textwidth]{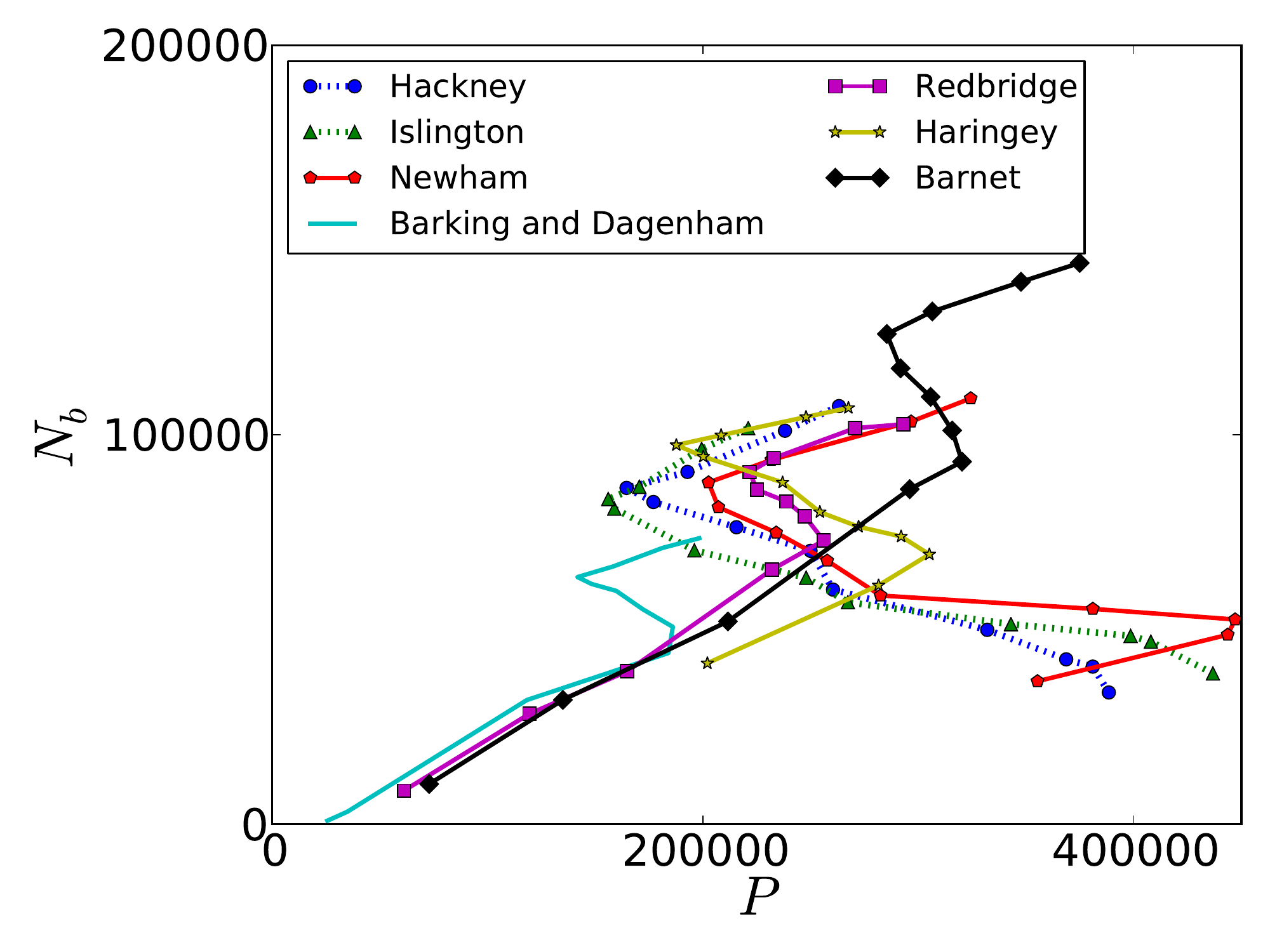}  &
\includegraphics[angle=0, width=0.5\textwidth]{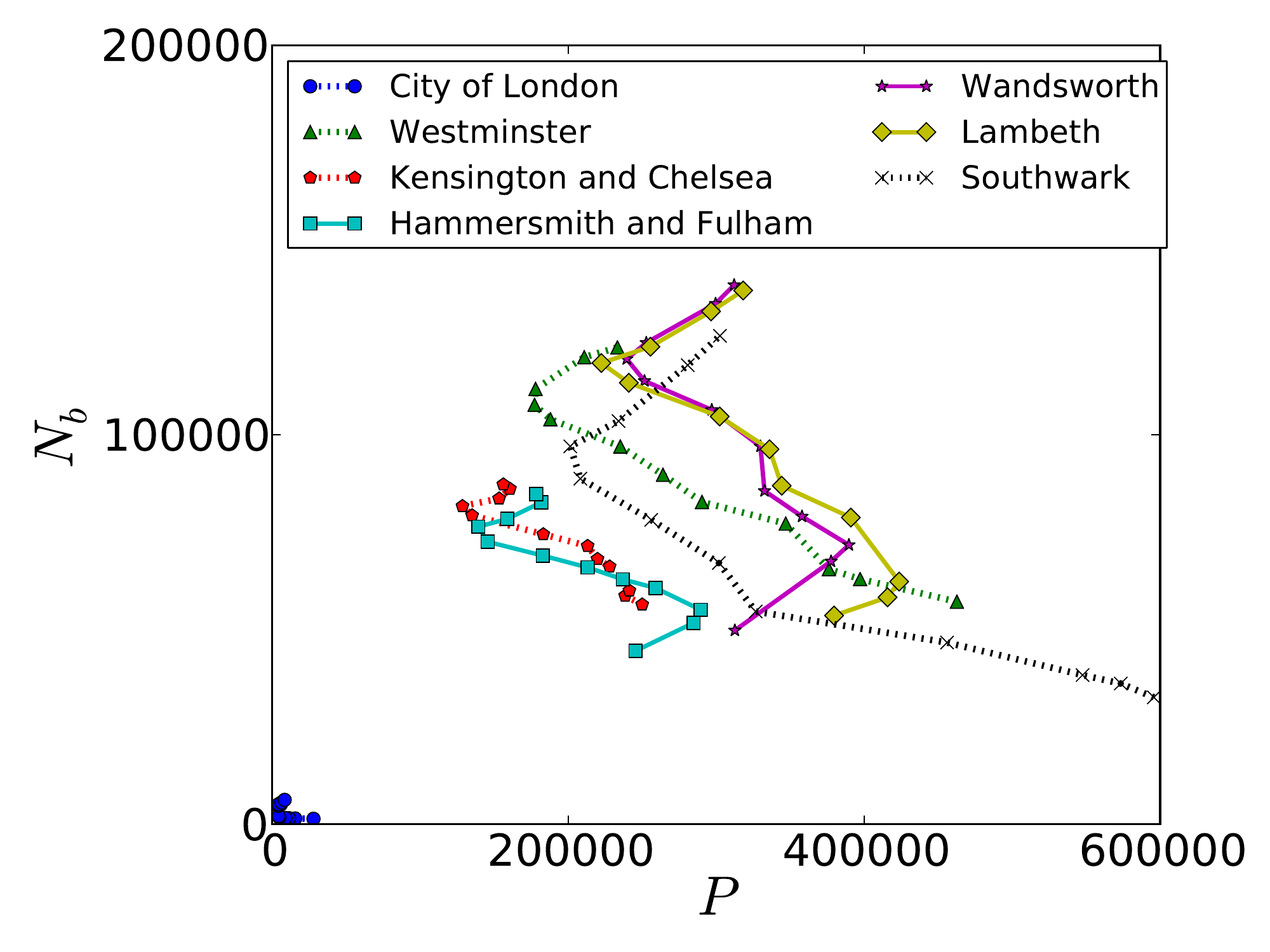}
\end{tabular}%
\end{center}
\caption{\textbf{London districts: number of buildings VS population.} In continuous line we have districts that reached the saturation point. In dashed line we have districts that are still in the growing phase and in dotted line the ones that reached the saturation points before the year of the first available data. }
\label{figS5}
\end{figure}

\begin{figure}[t!]
\begin{center}
\begin{tabular}{cc}
\includegraphics[angle=0, width=0.5\textwidth]{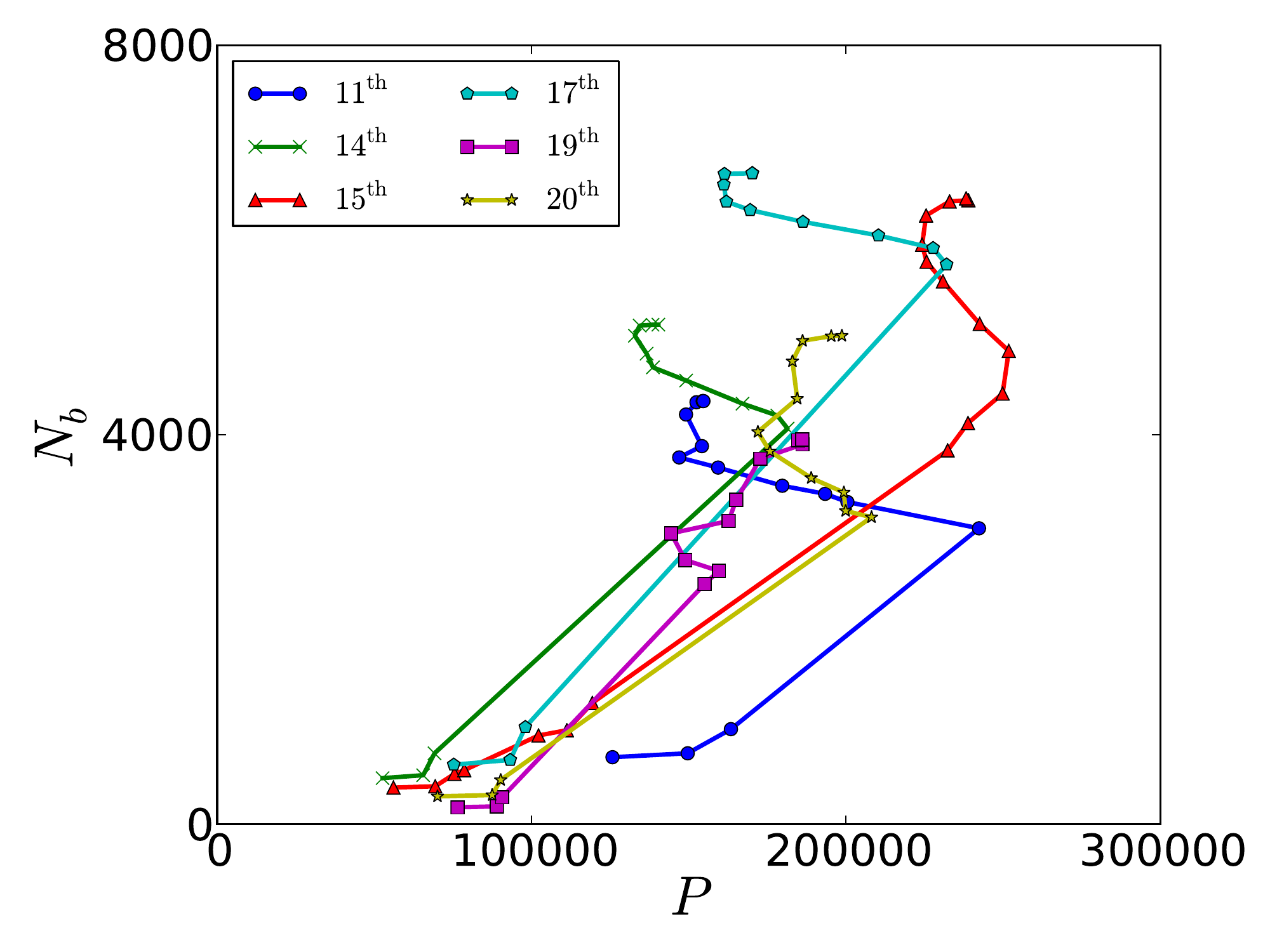} &
\includegraphics[angle=0, width=0.5\textwidth]{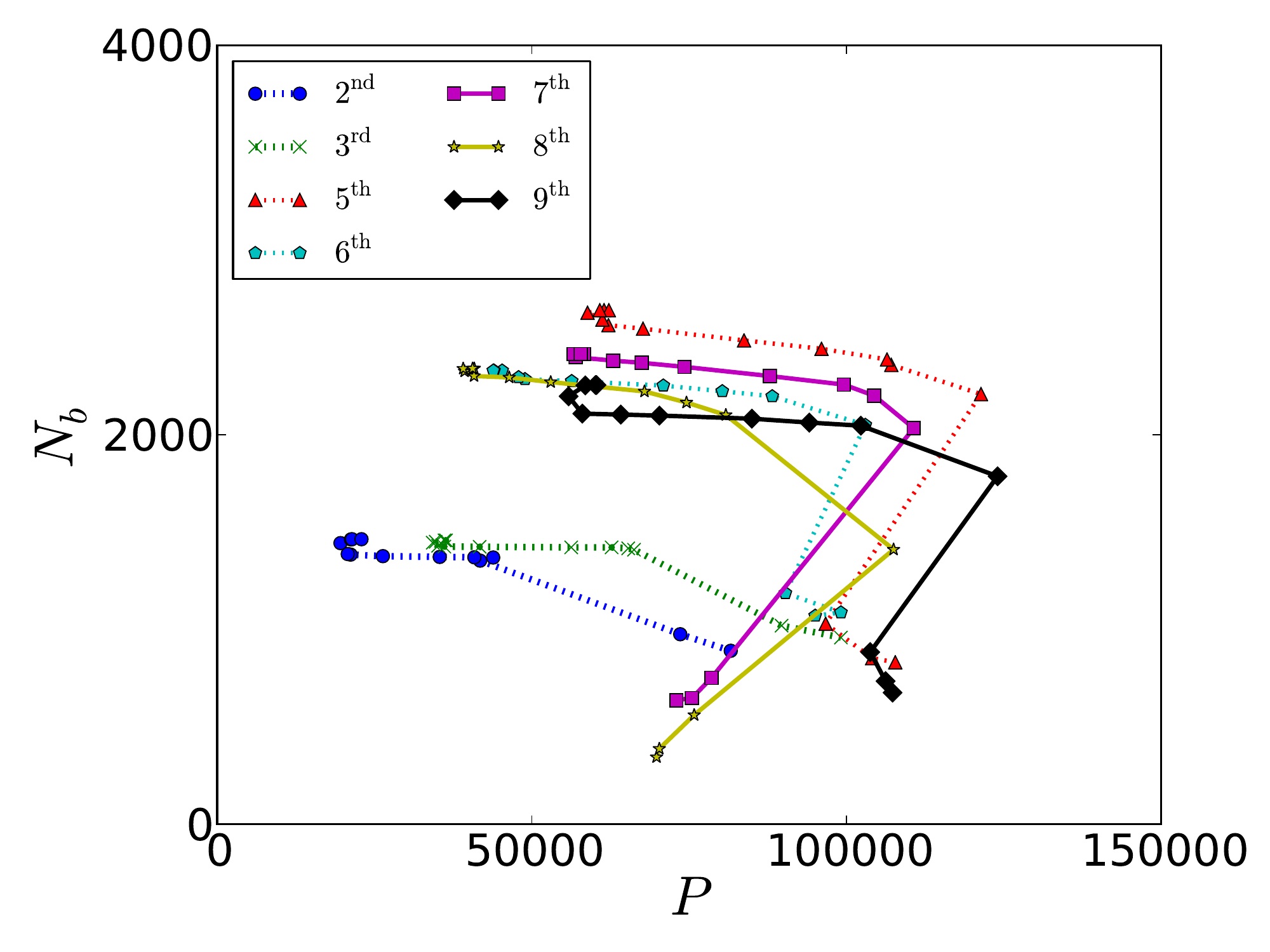} 
\end{tabular}%
\end{center}
\caption{\textbf{Paris arrondissements: number of buildings VS population.} In continuous line we have districts that reached the saturation point. In dashed line we have districts that are still in the growing phase and in dotted line the ones that reached the saturation points before the year of the first available data.}
\label{figS6}
\end{figure}

\begin{figure}[!]
\begin{center}
\begin{tabular}{ccc}
\includegraphics[angle=0, width=0.3\textwidth]{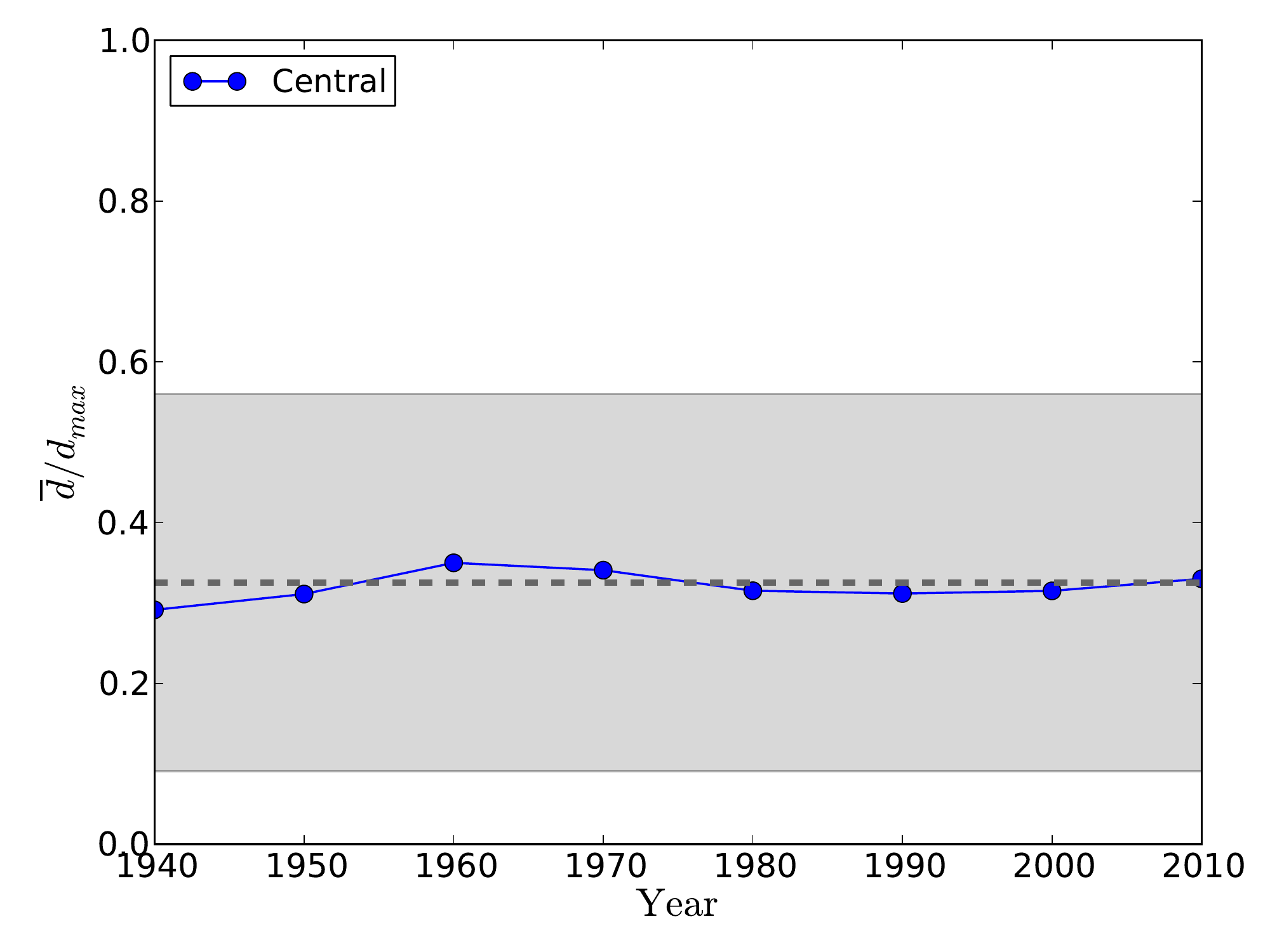} &
\includegraphics[angle=0, width=0.3\textwidth]{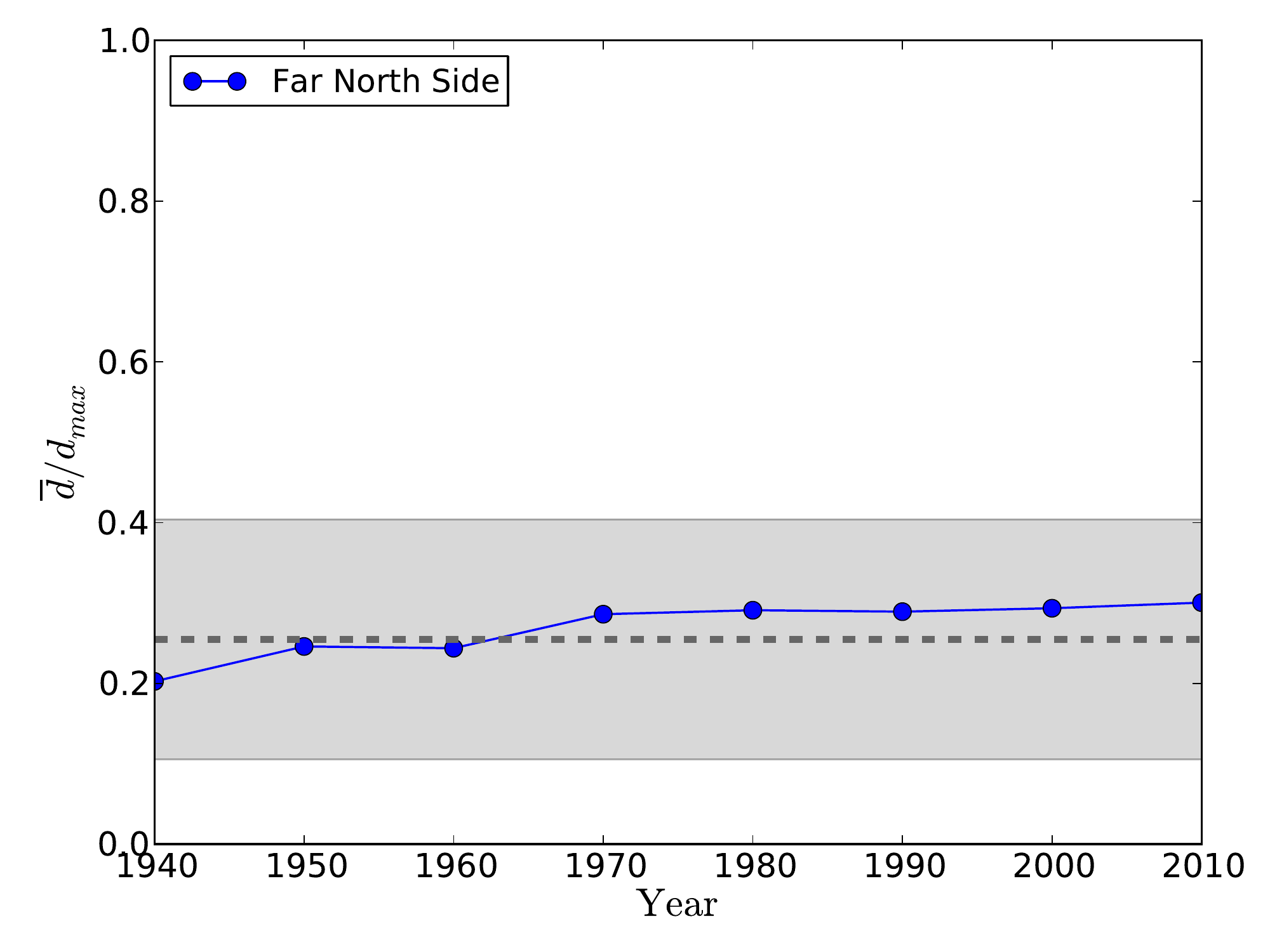} &
\includegraphics[angle=0, width=0.3\textwidth]{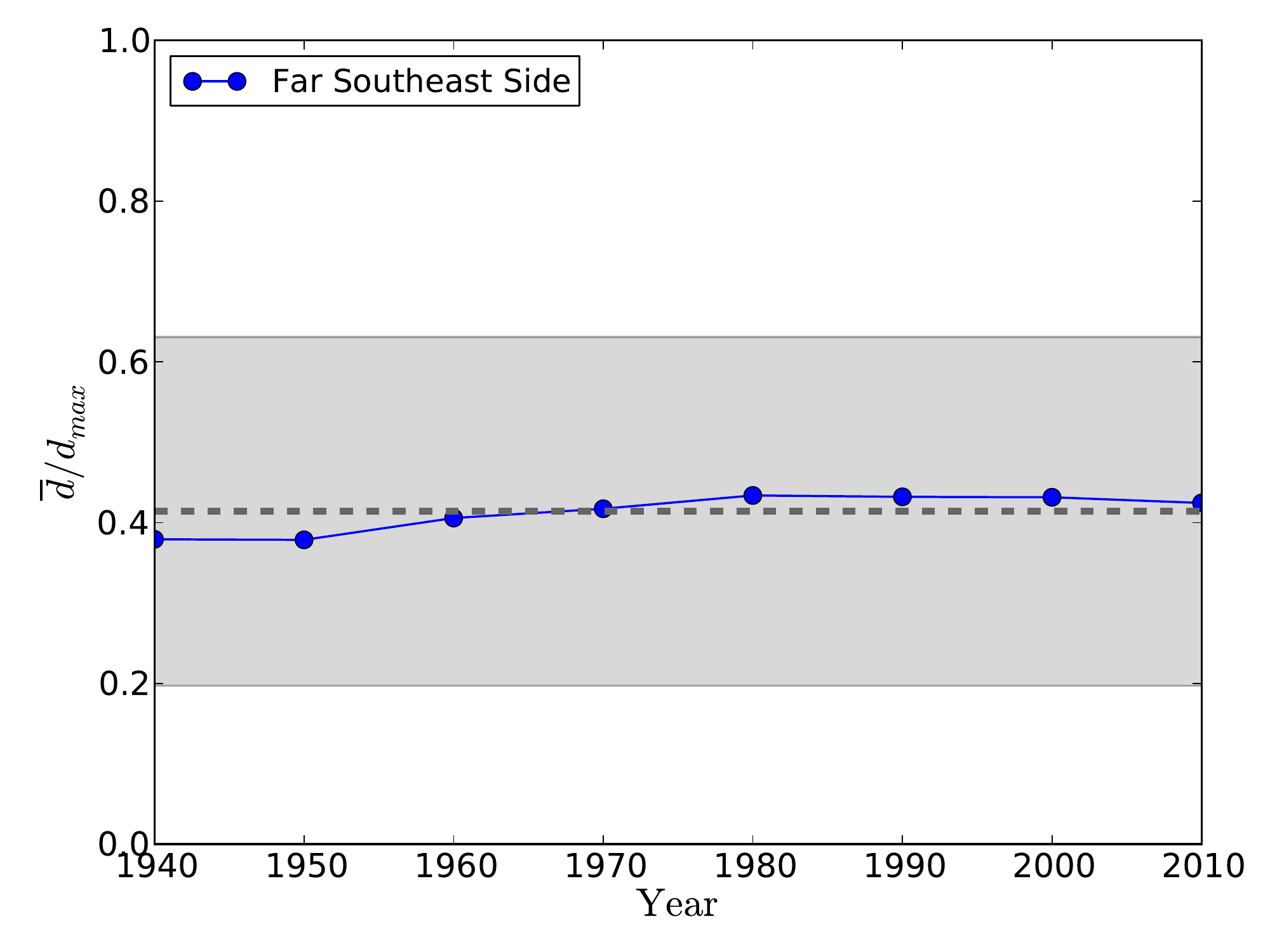}\\
\includegraphics[angle=0, width=0.3\textwidth]{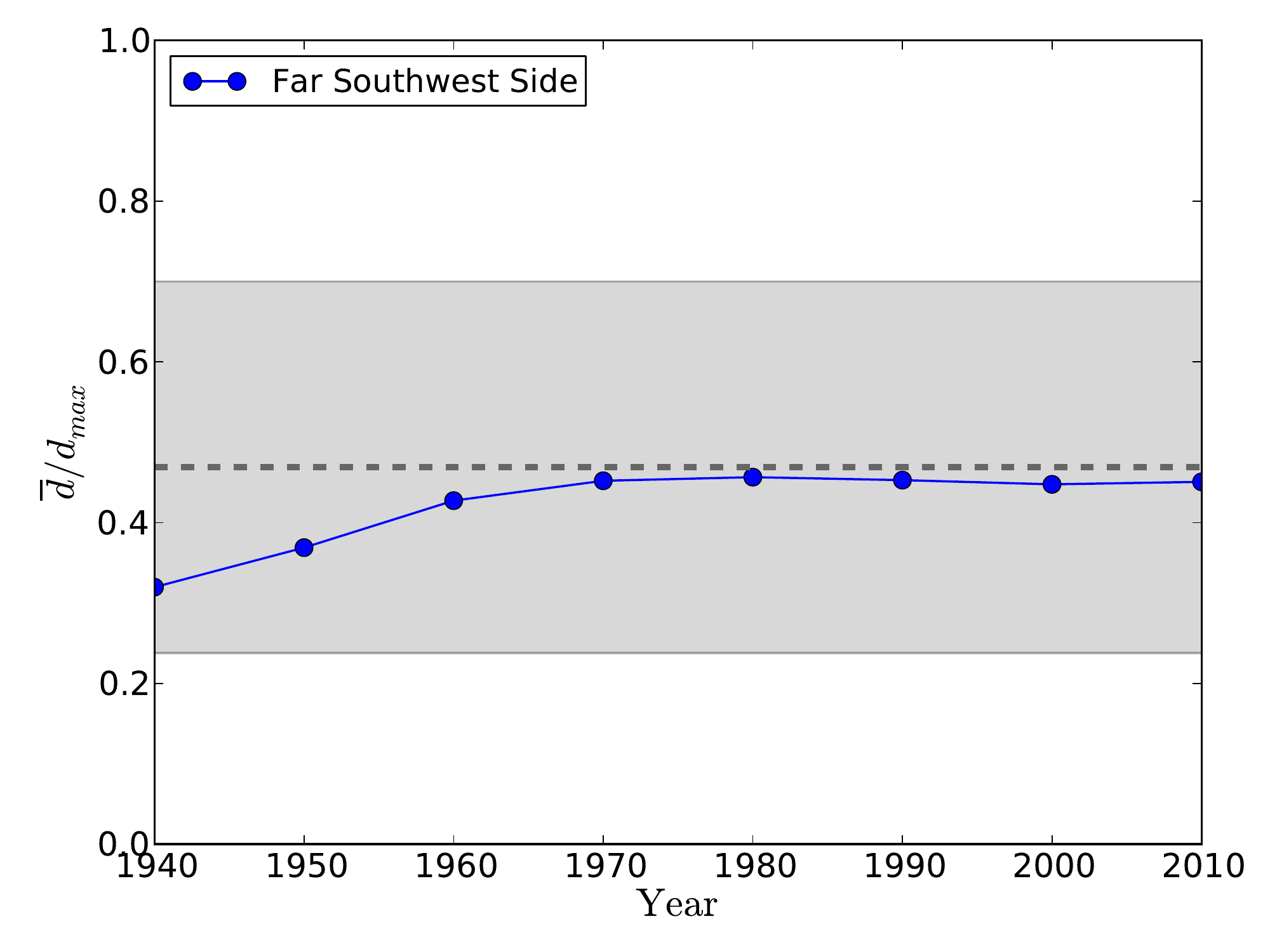} &
\includegraphics[angle=0, width=0.3\textwidth]{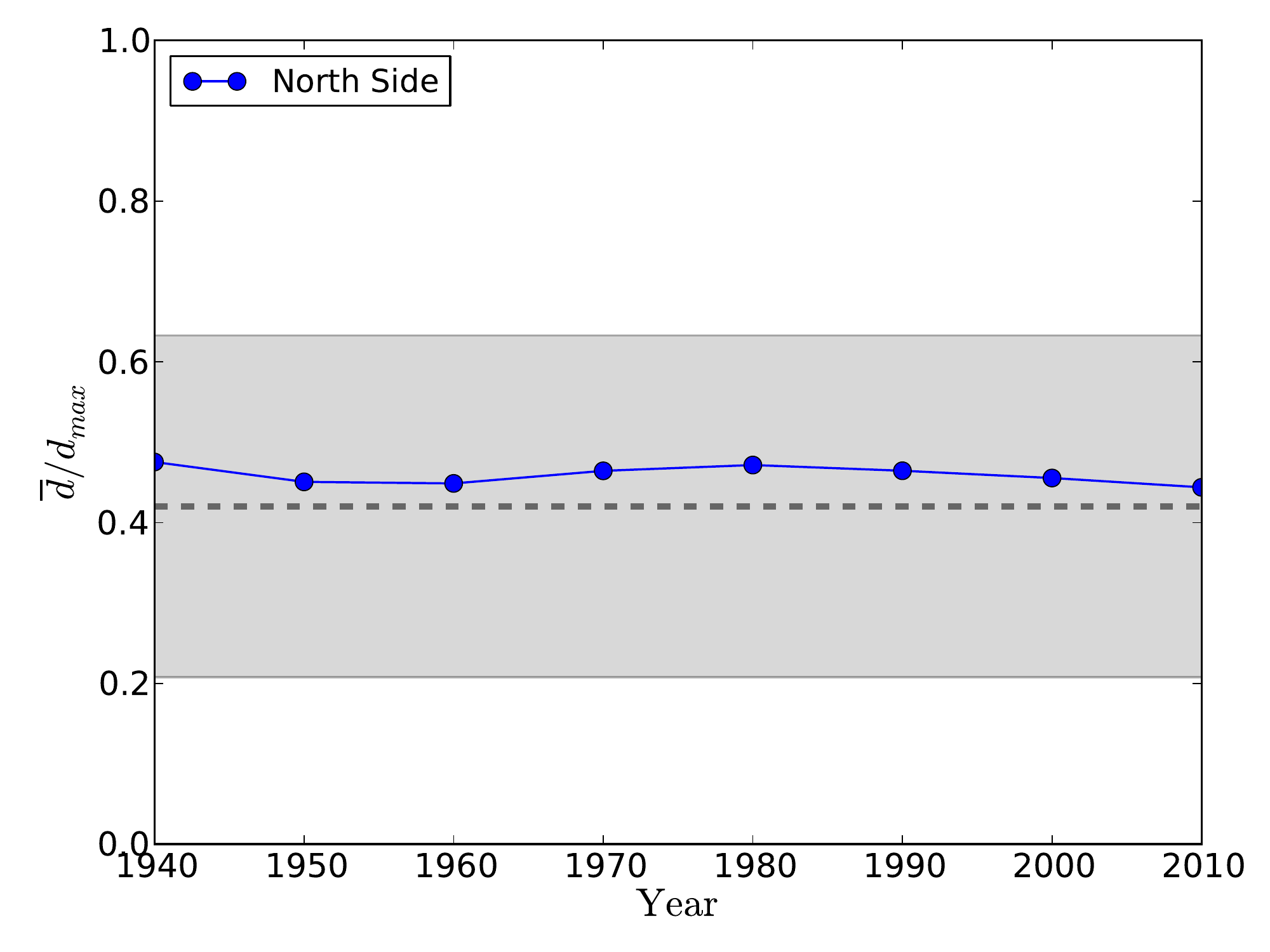} &
\includegraphics[angle=0, width=0.3\textwidth]{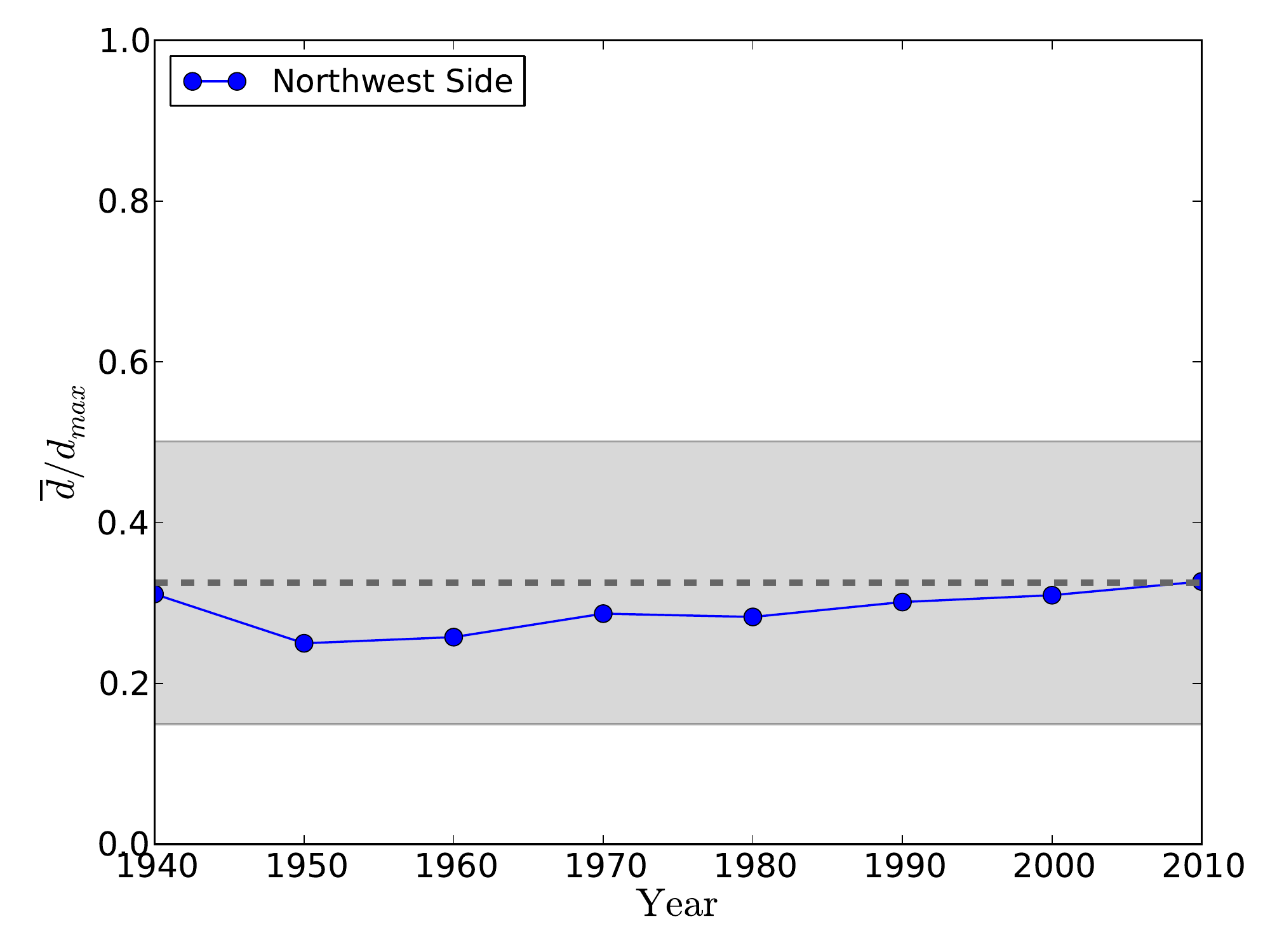} \\
\includegraphics[angle=0, width=0.3\textwidth]{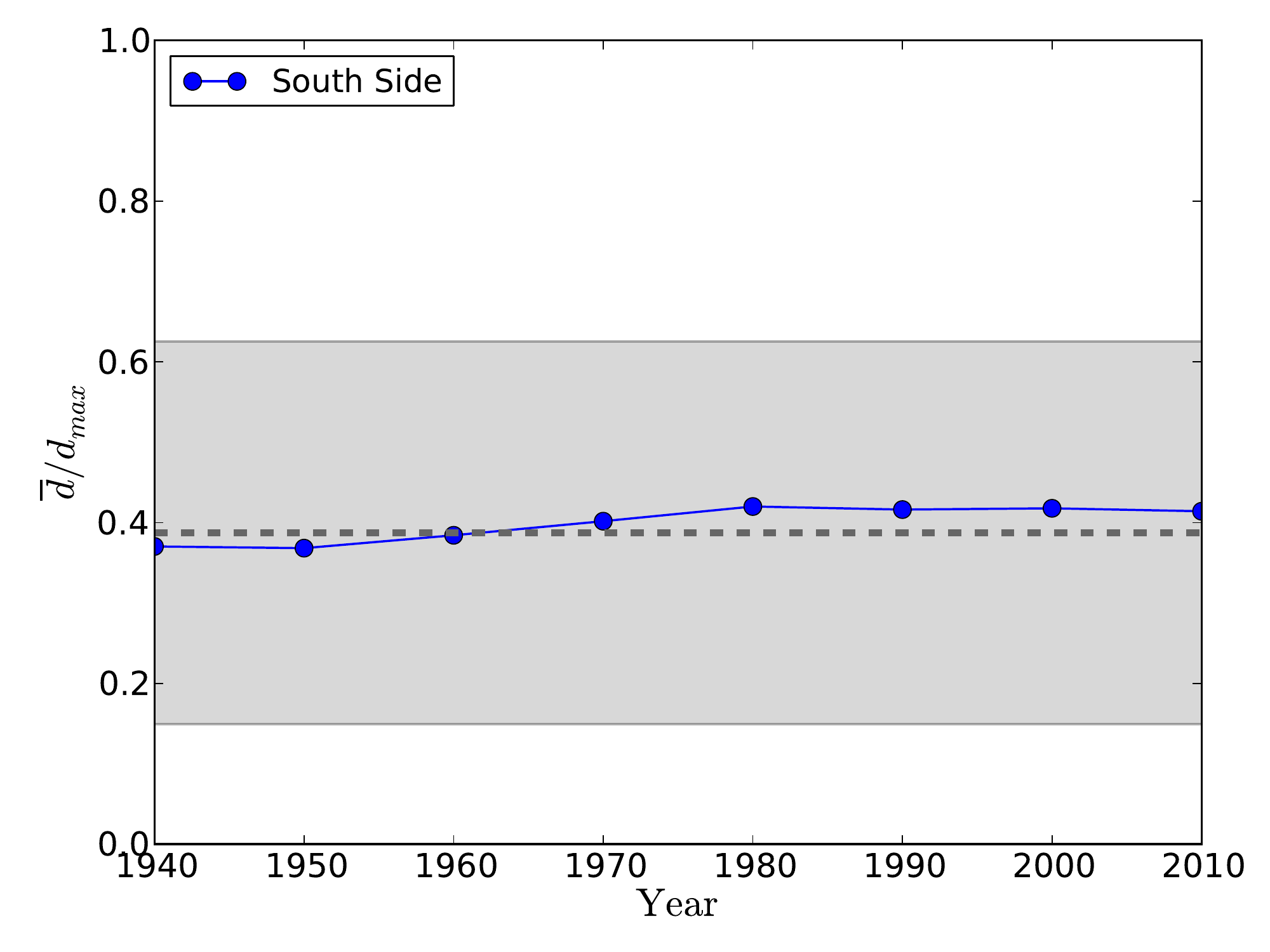} &
\includegraphics[angle=0, width=0.3\textwidth]{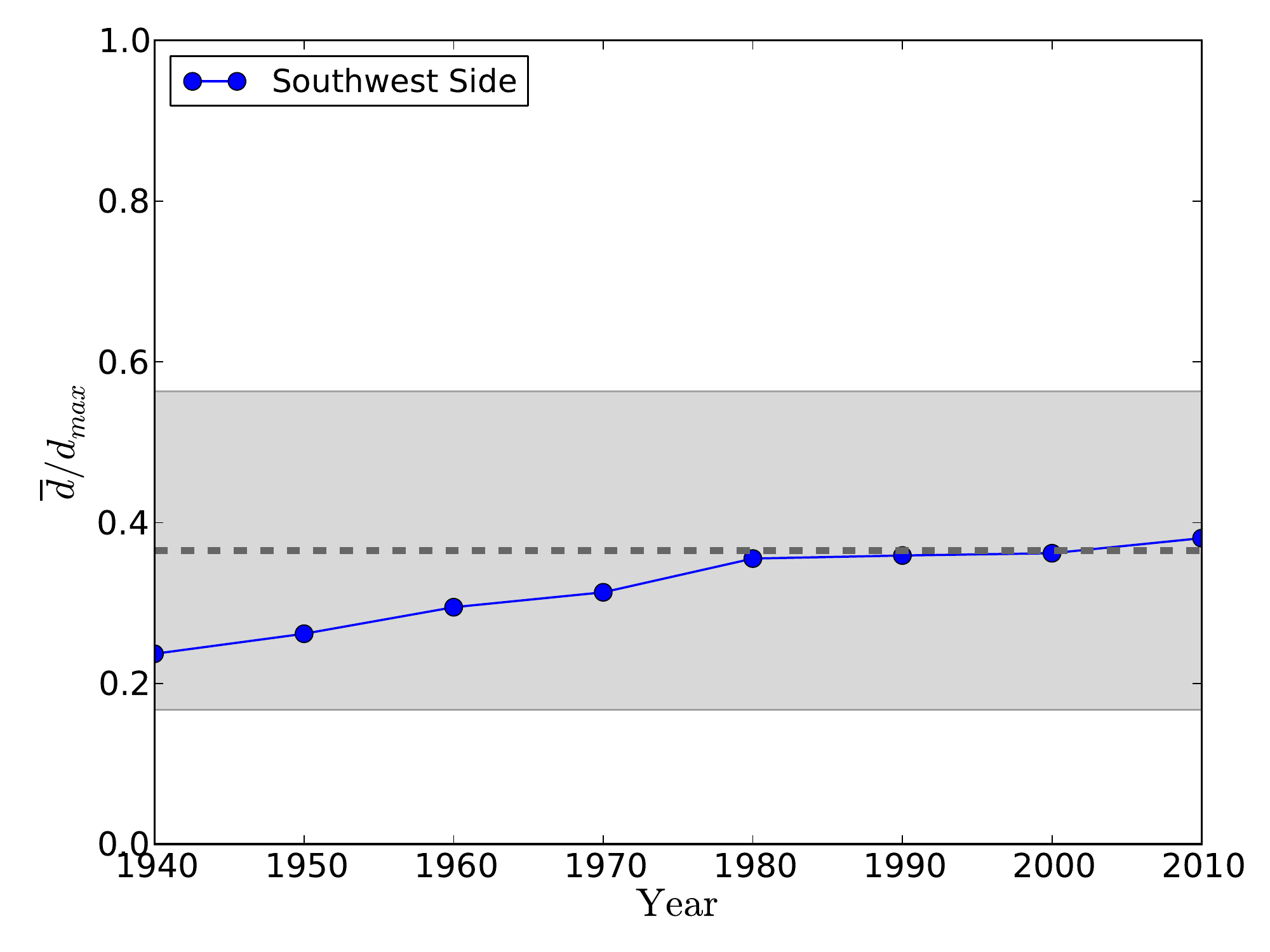} &
\includegraphics[angle=0, width=0.3\textwidth]{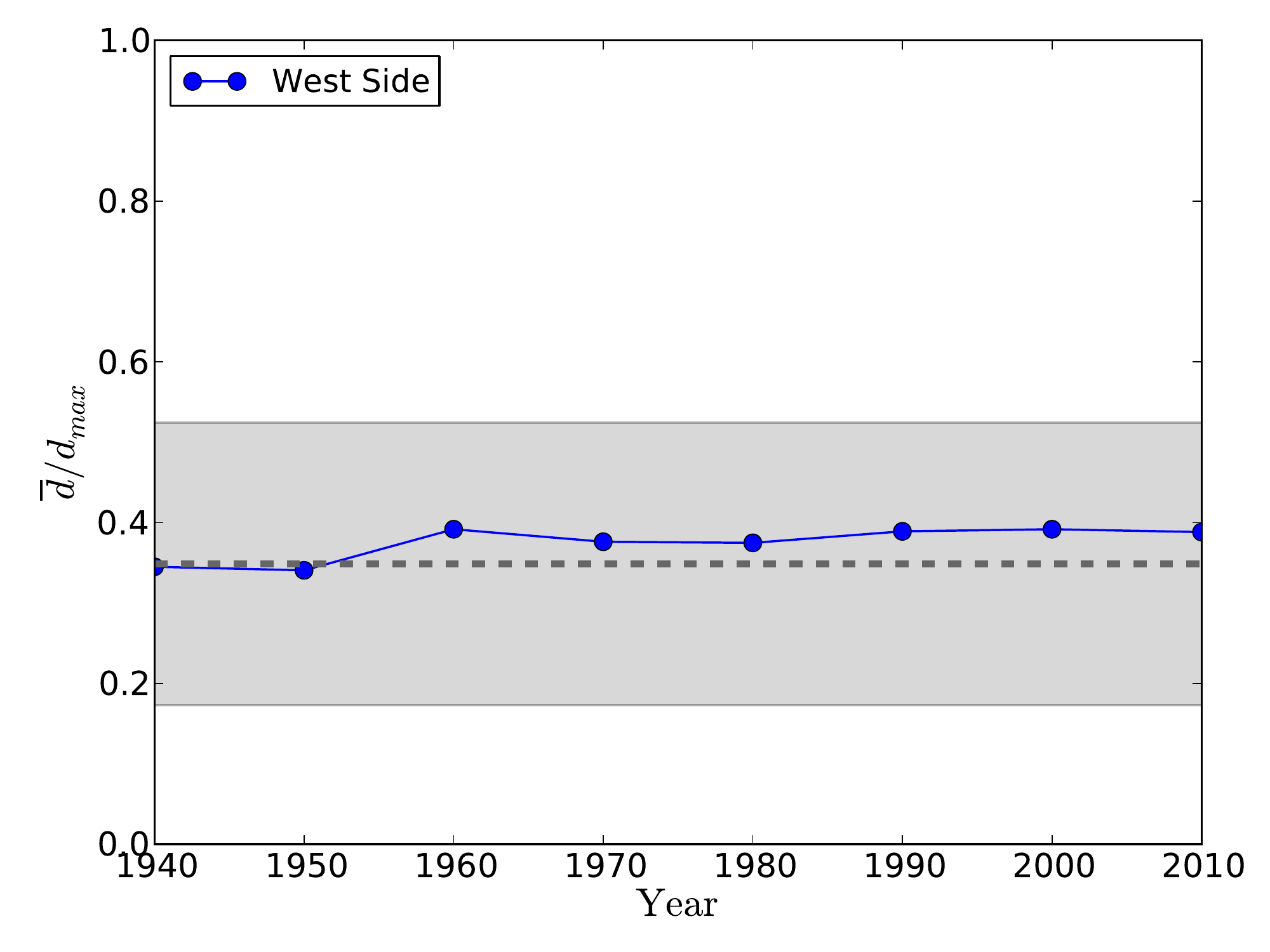} 
\end{tabular}%
\end{center}
\caption{\textbf{Chicago sides: homogeneity of growth in districts.} Average
distance between buildings at a given time (this distance is
normalized by the maximum distance found each district). The dotted line represents the average
value computed for a random uniform distribution and the grey
zone the dispersion computed with this null model.}
\label{figS7}
\end{figure}

\begin{figure}[!]
\begin{center}
\begin{tabular}{cccc}
\includegraphics[angle=0, width=0.25\textwidth]{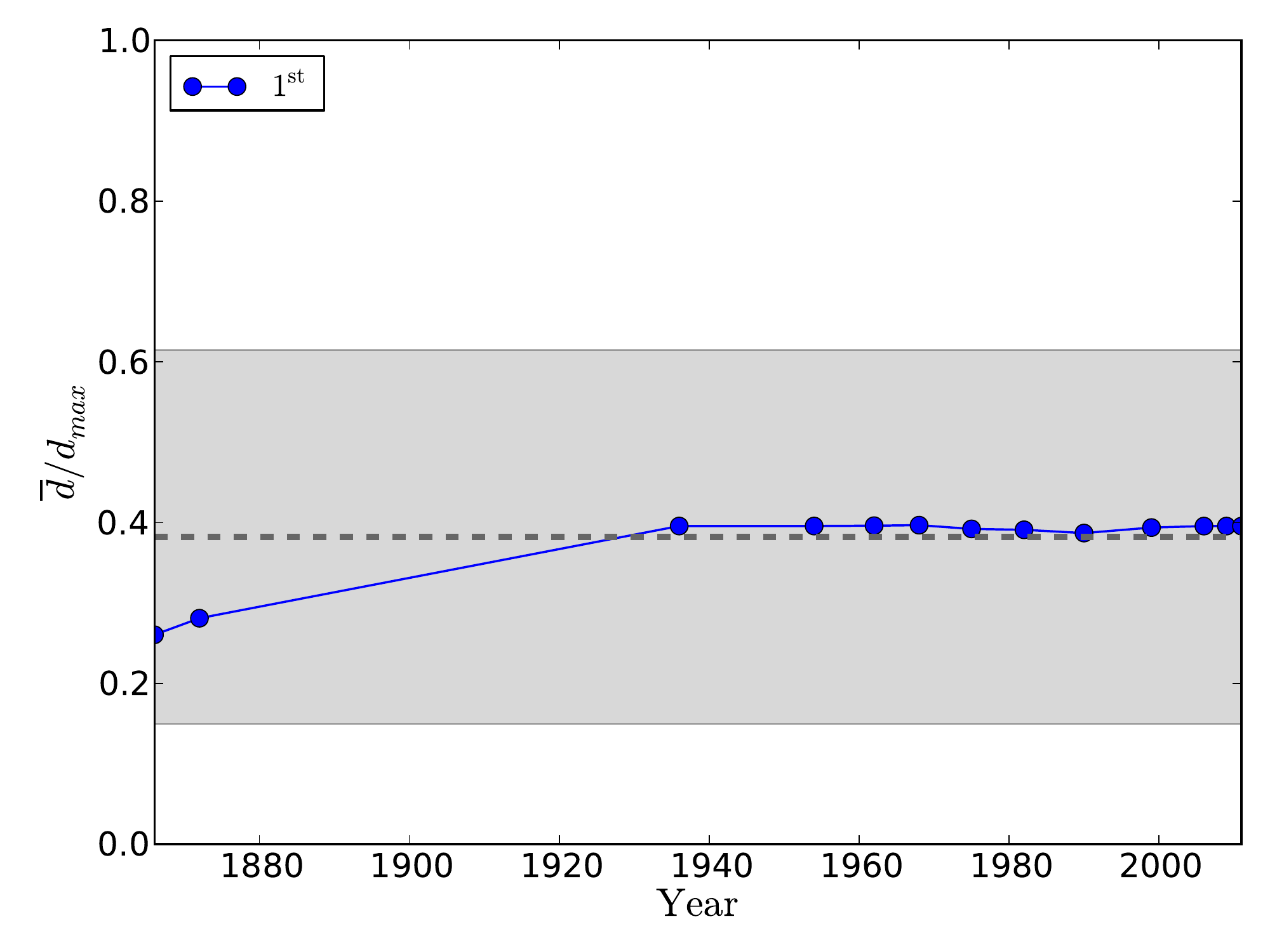} &
\includegraphics[angle=0, width=0.25\textwidth]{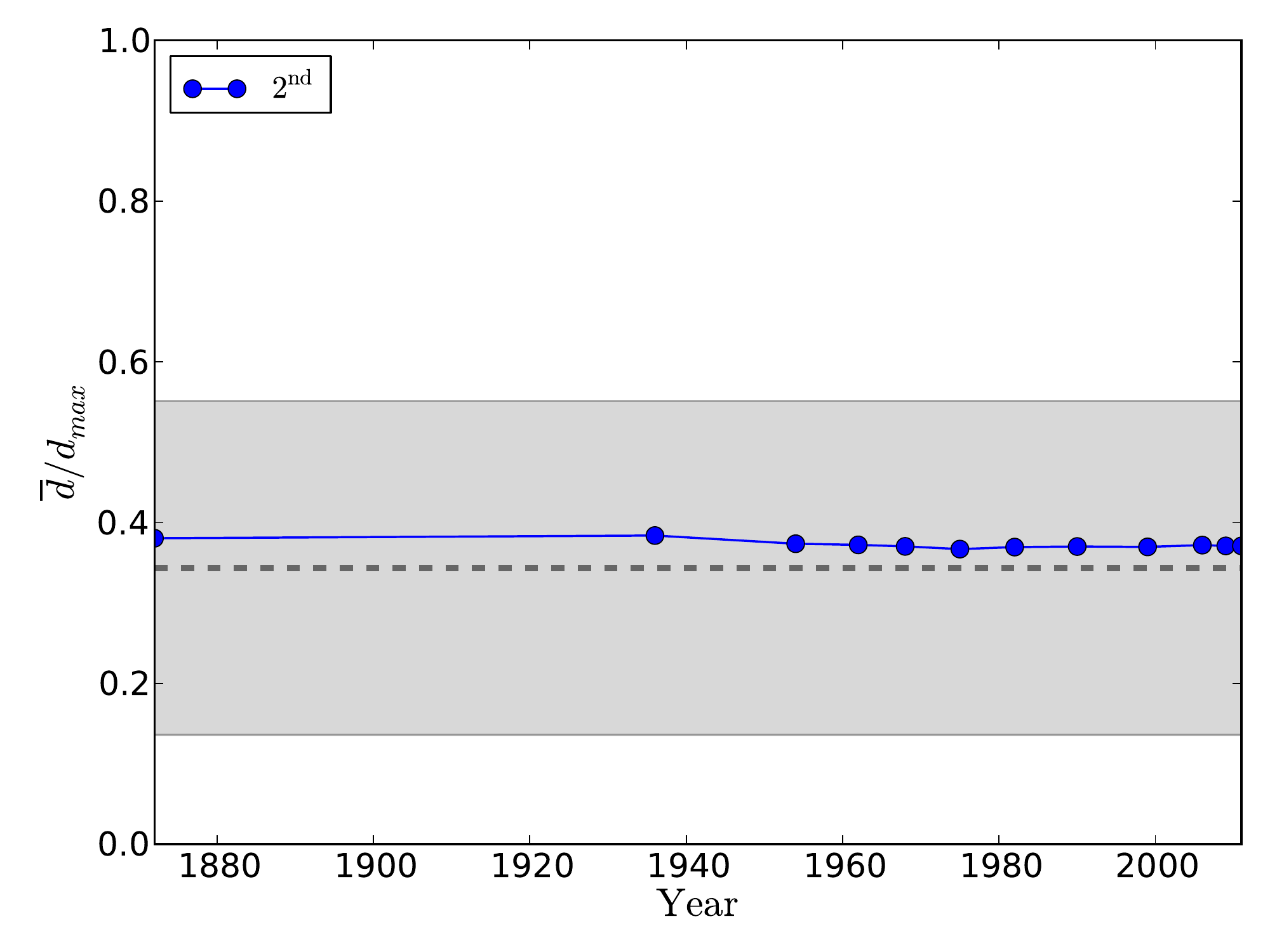} &
\includegraphics[angle=0, width=0.25\textwidth]{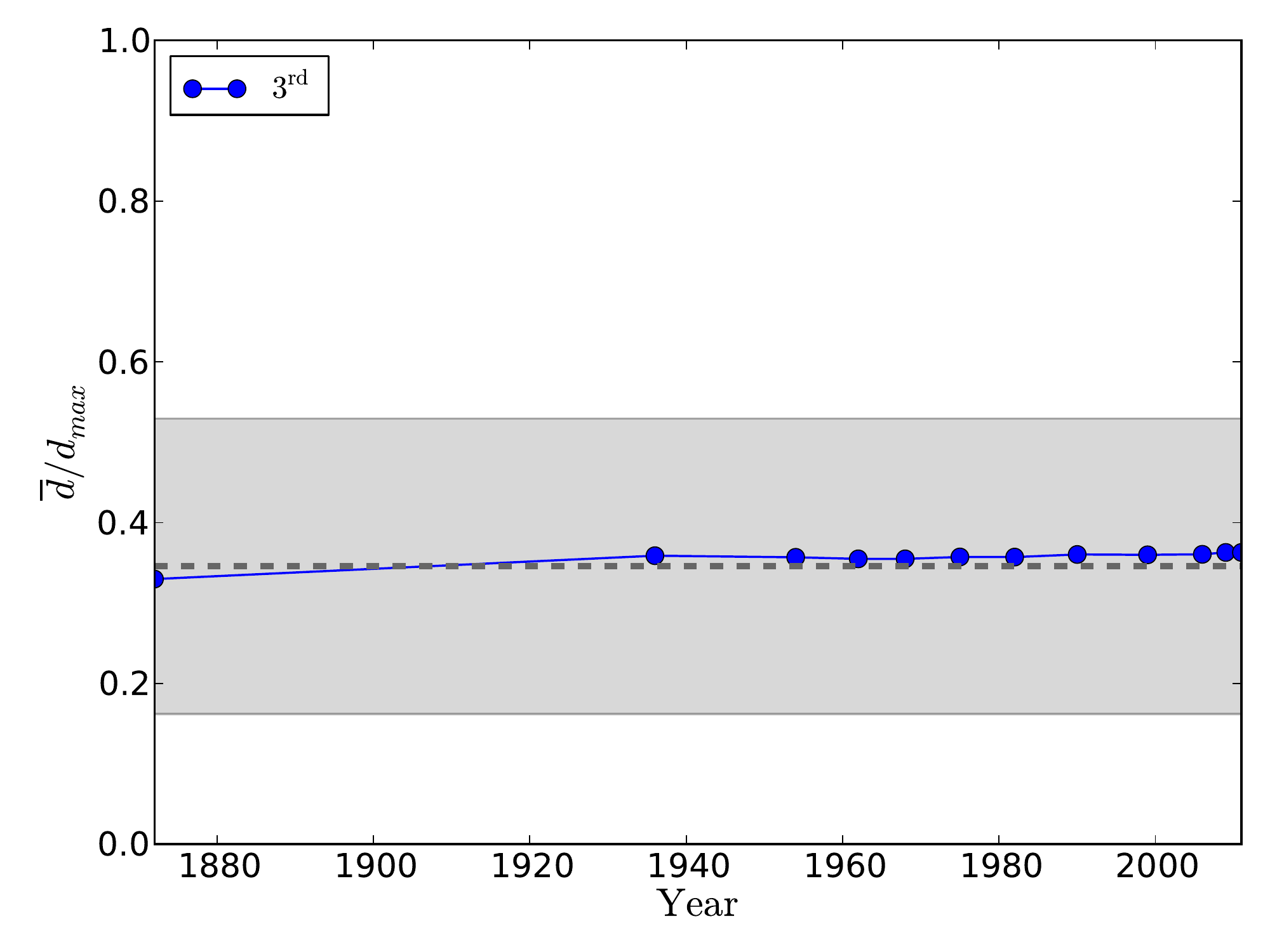}&
\includegraphics[angle=0, width=0.25\textwidth]{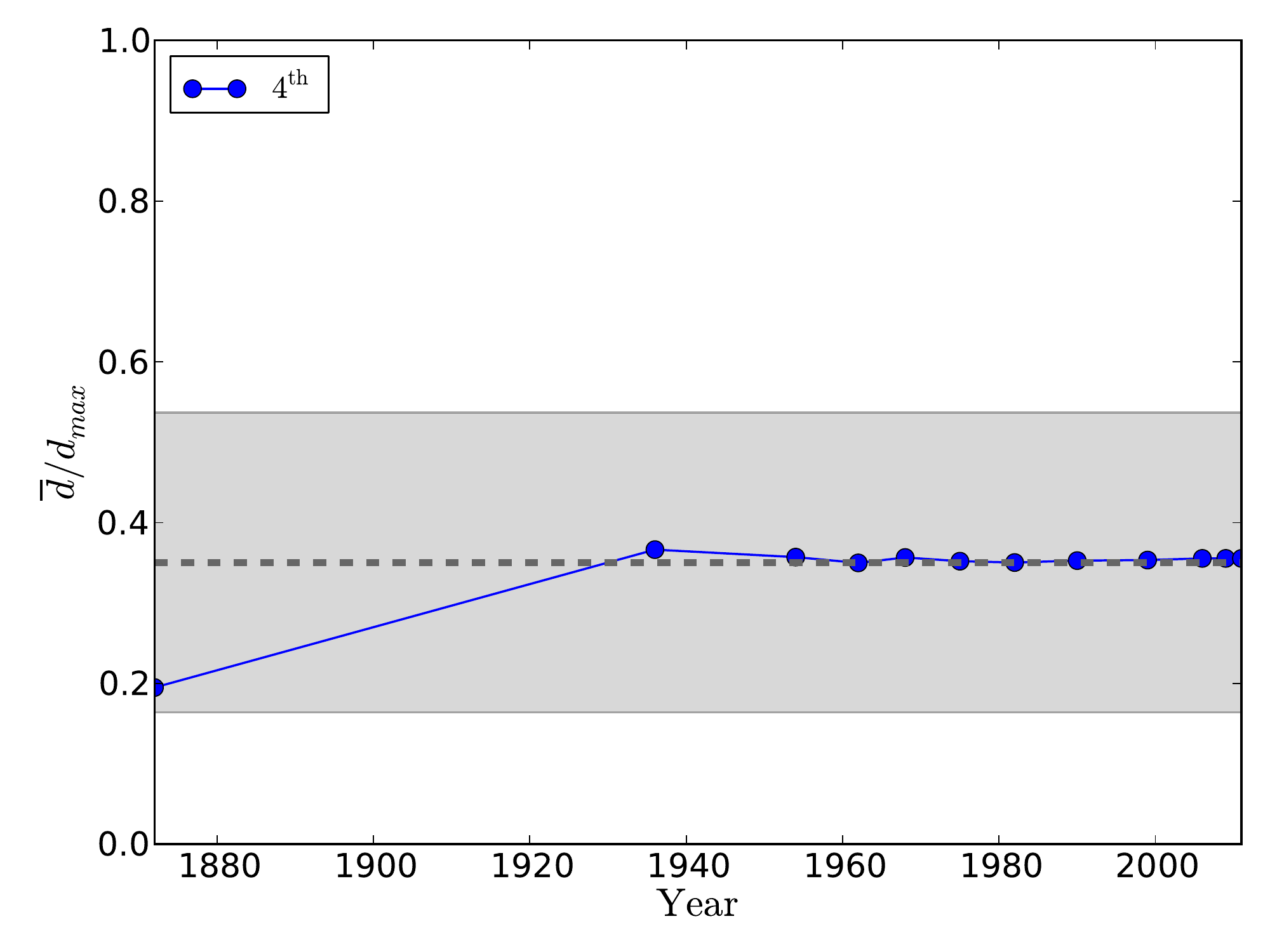} \\
\includegraphics[angle=0, width=0.25\textwidth]{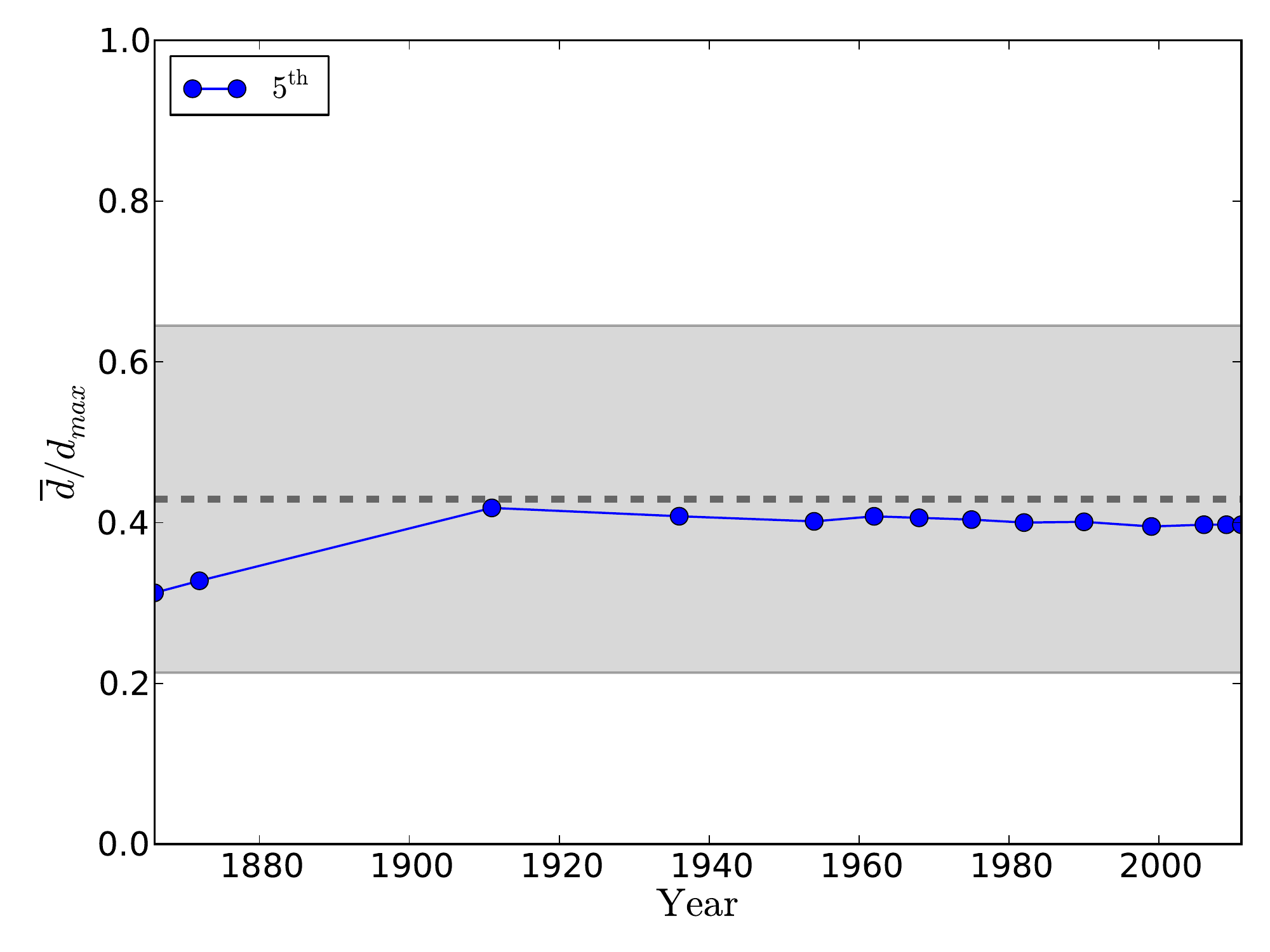} &
\includegraphics[angle=0, width=0.25\textwidth]{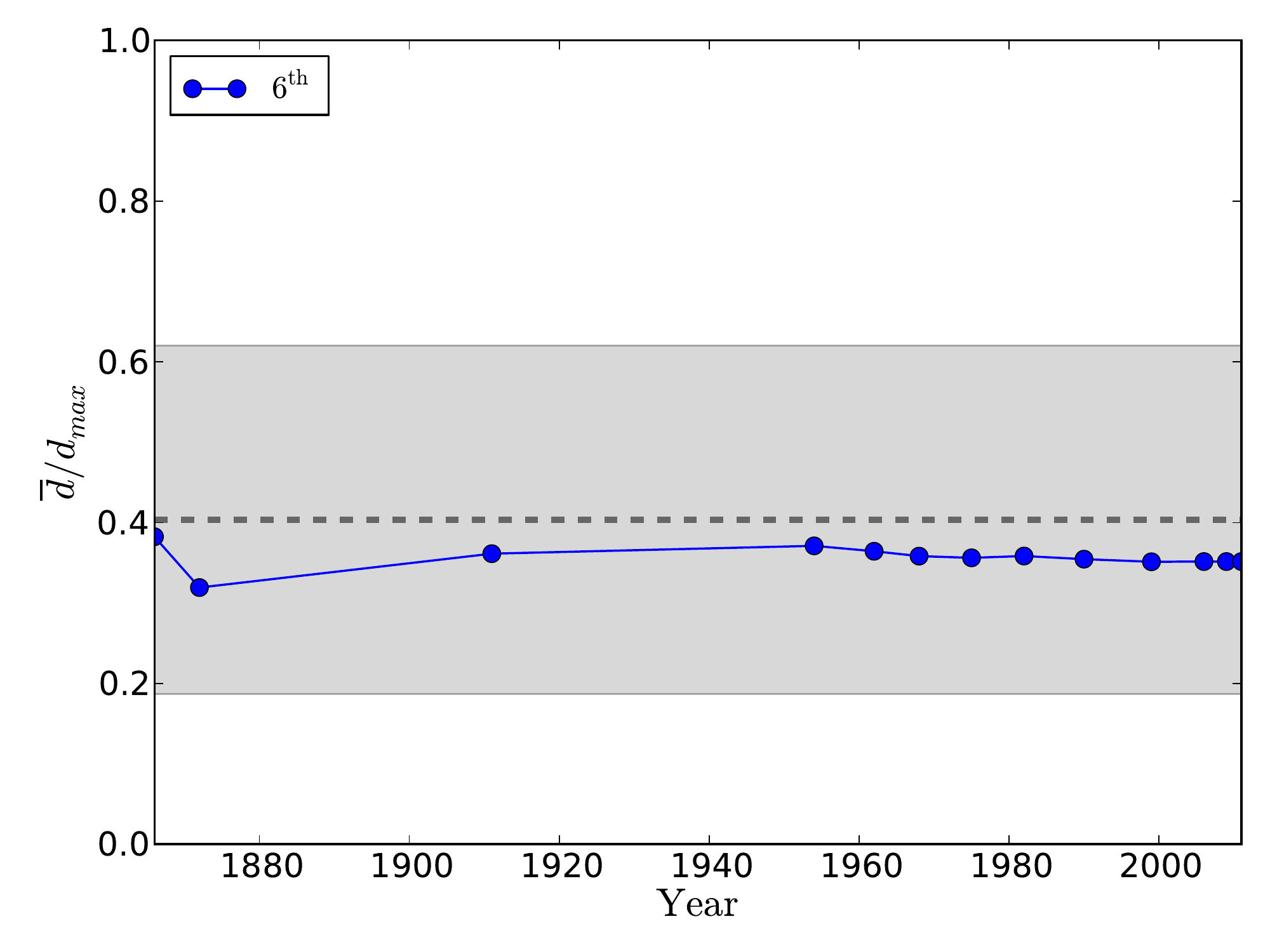} &
\includegraphics[angle=0, width=0.25\textwidth]{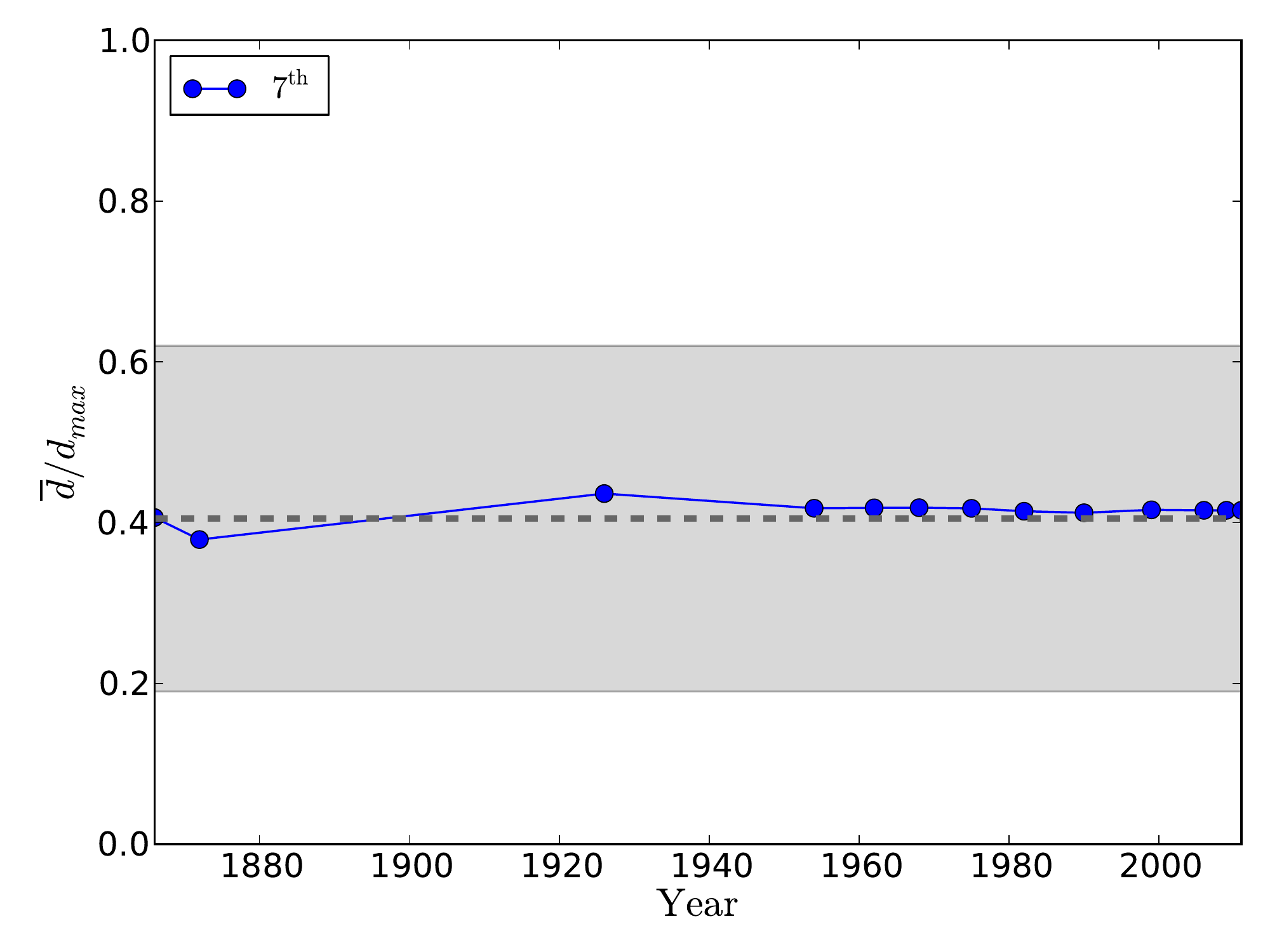}&
\includegraphics[angle=0, width=0.25\textwidth]{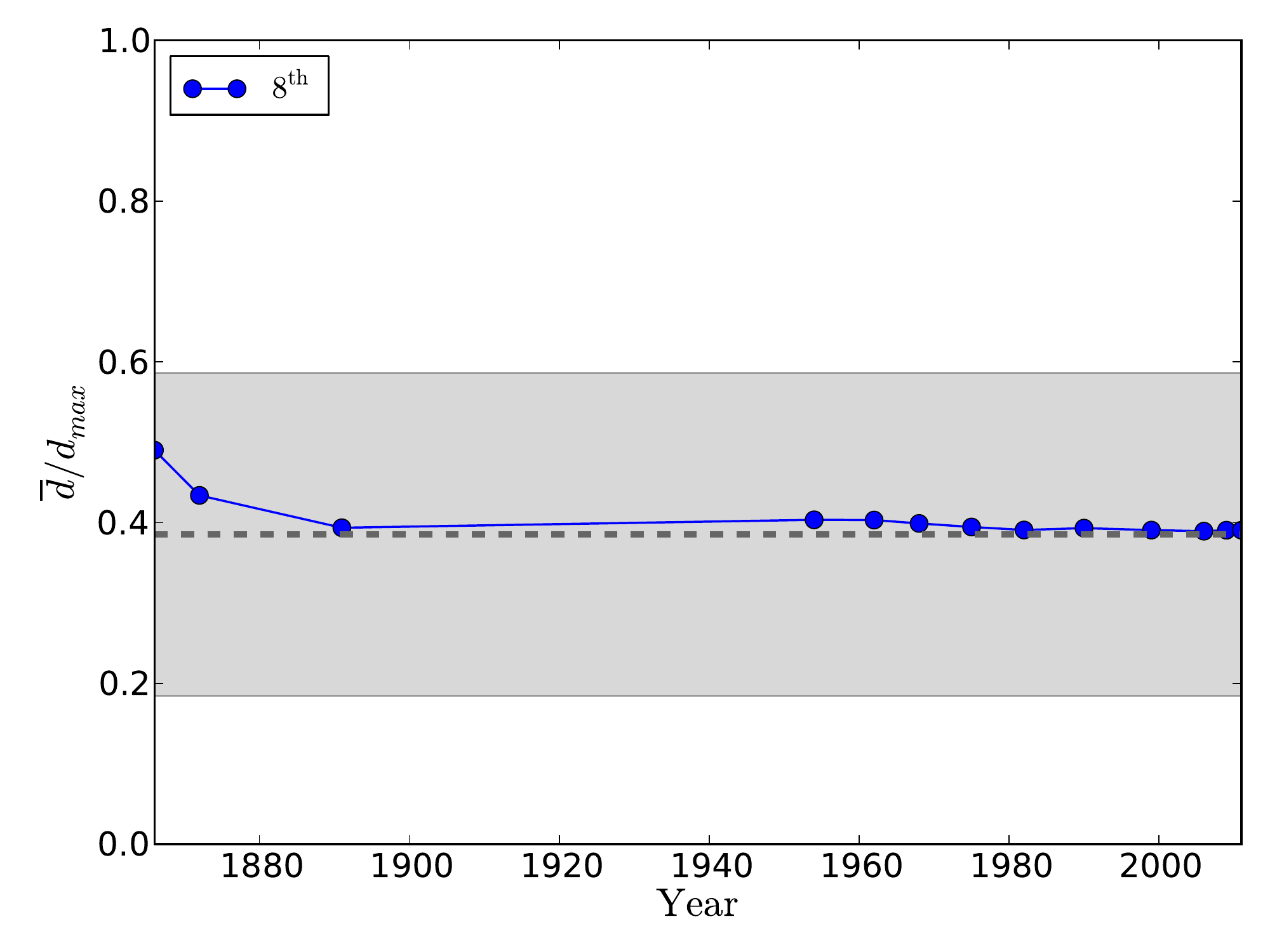} \\
\includegraphics[angle=0, width=0.25\textwidth]{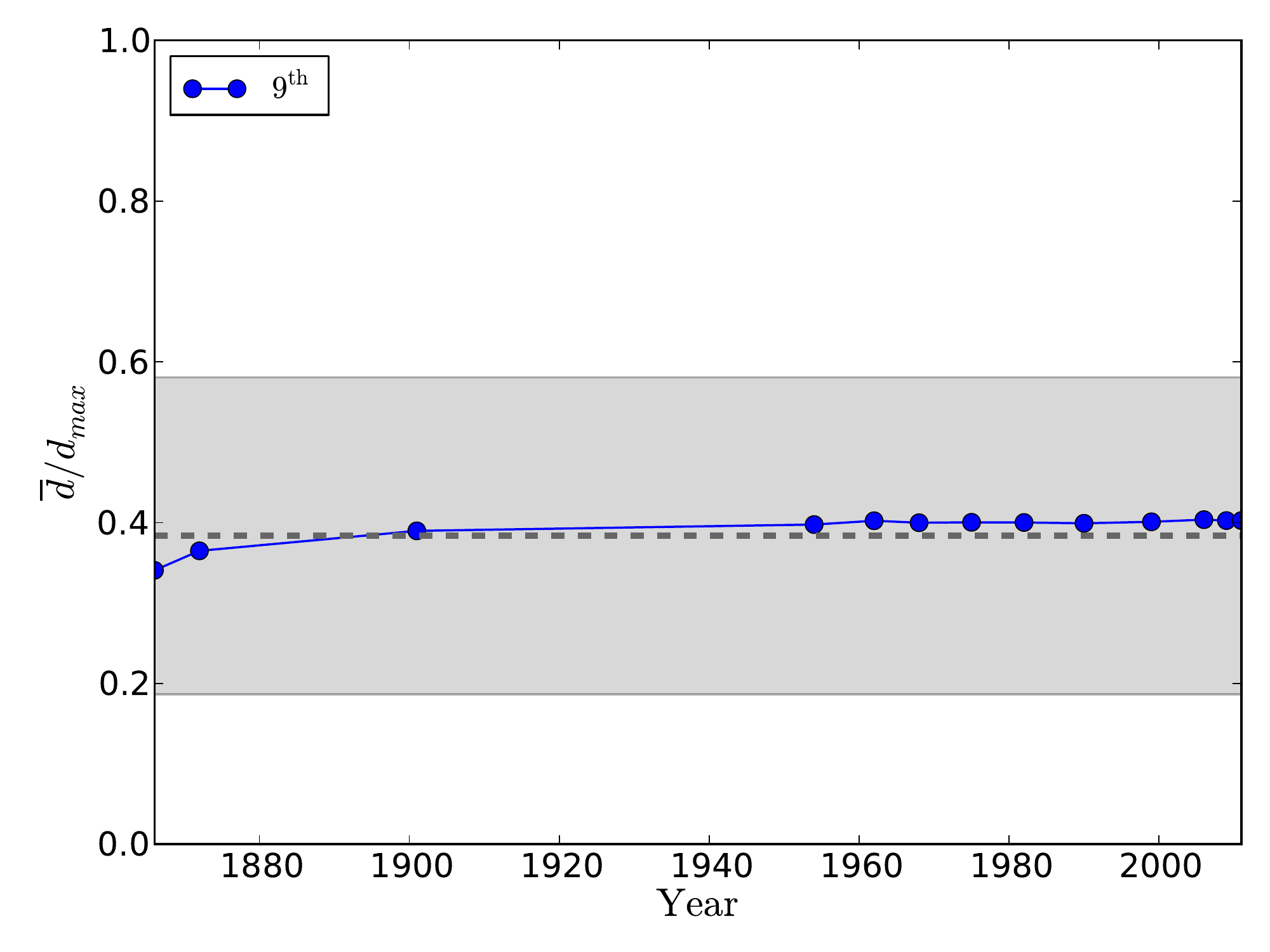} &
\includegraphics[angle=0, width=0.25\textwidth]{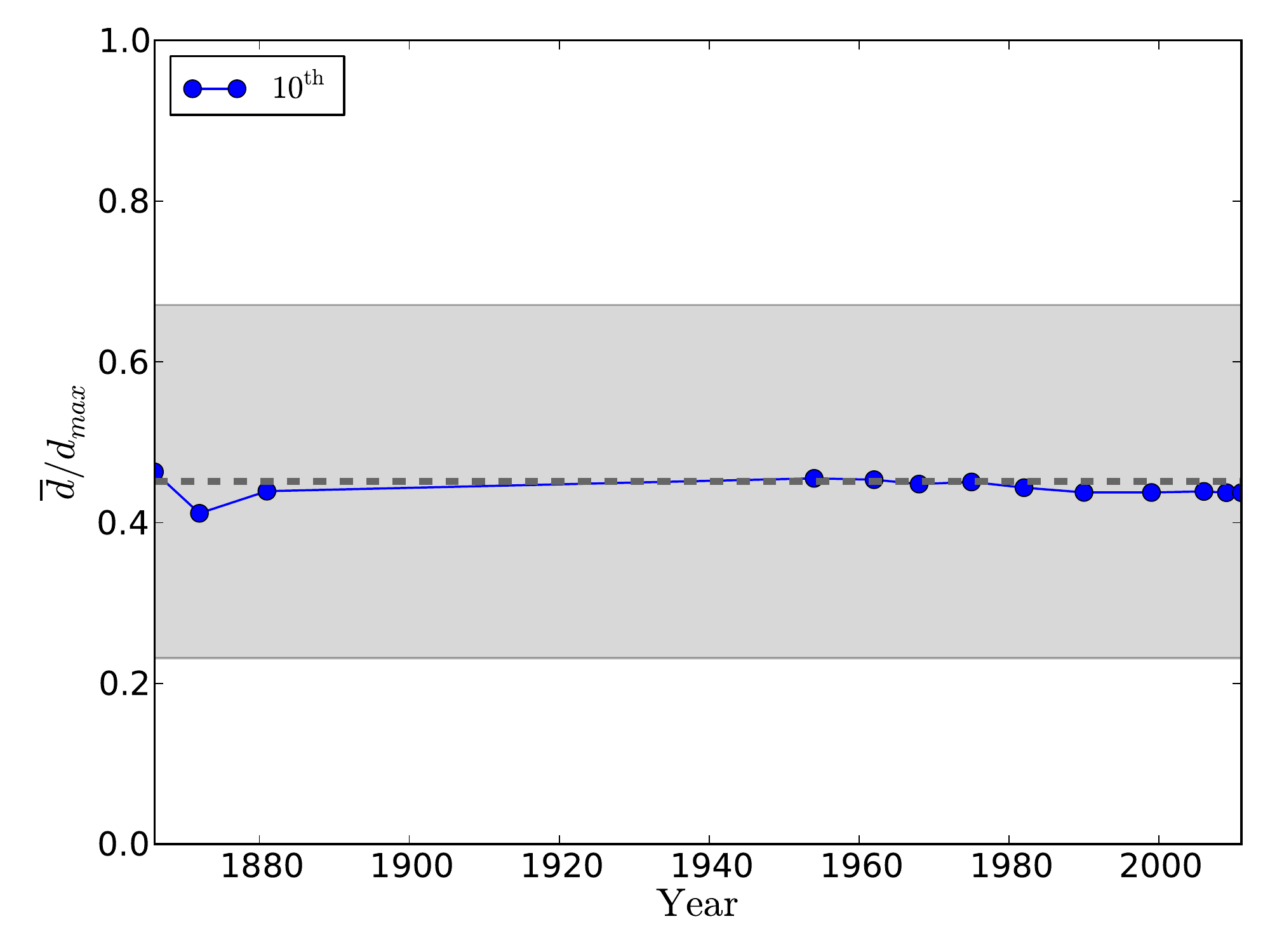} &
\includegraphics[angle=0, width=0.25\textwidth]{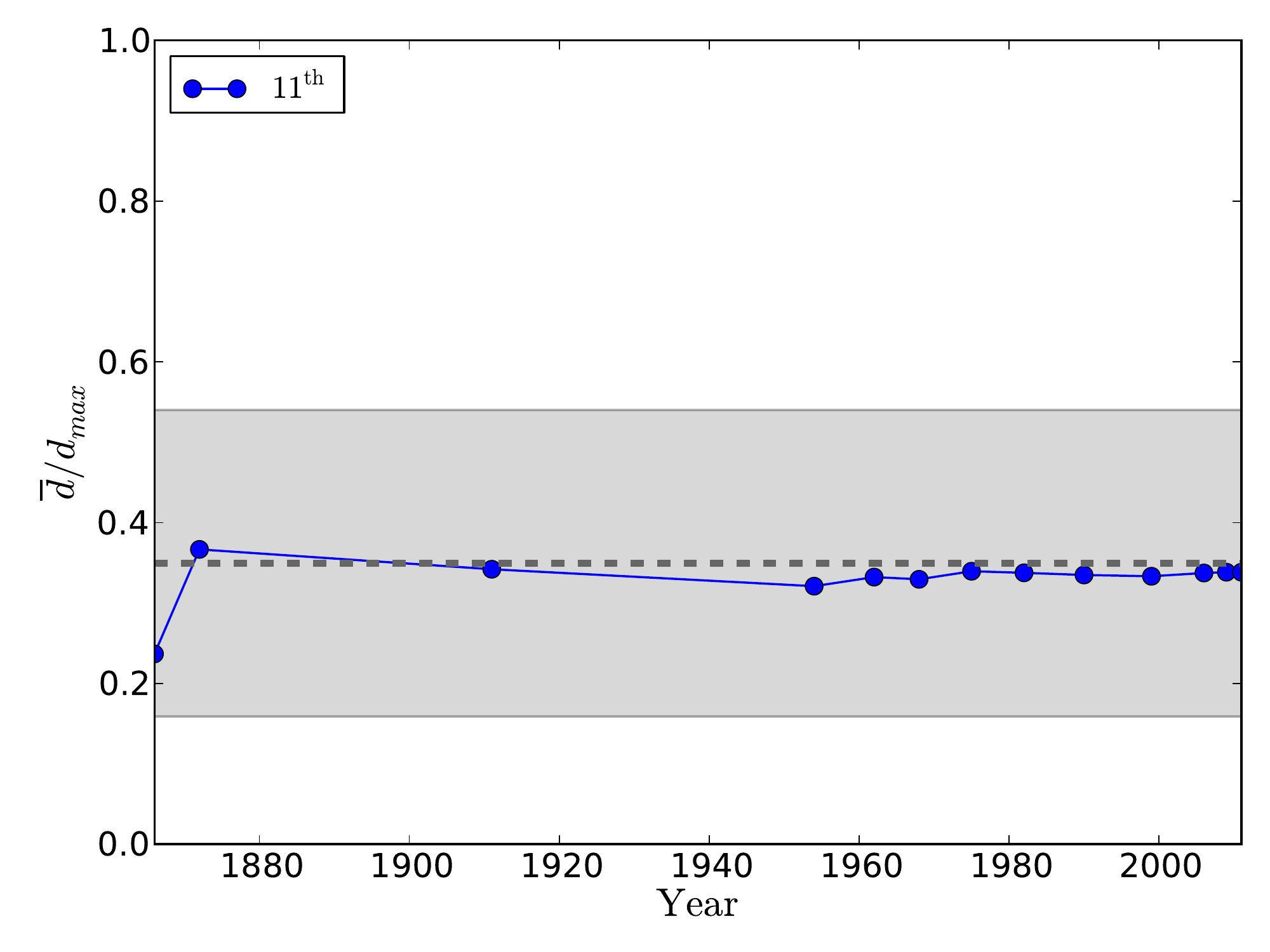}&
\includegraphics[angle=0, width=0.25\textwidth]{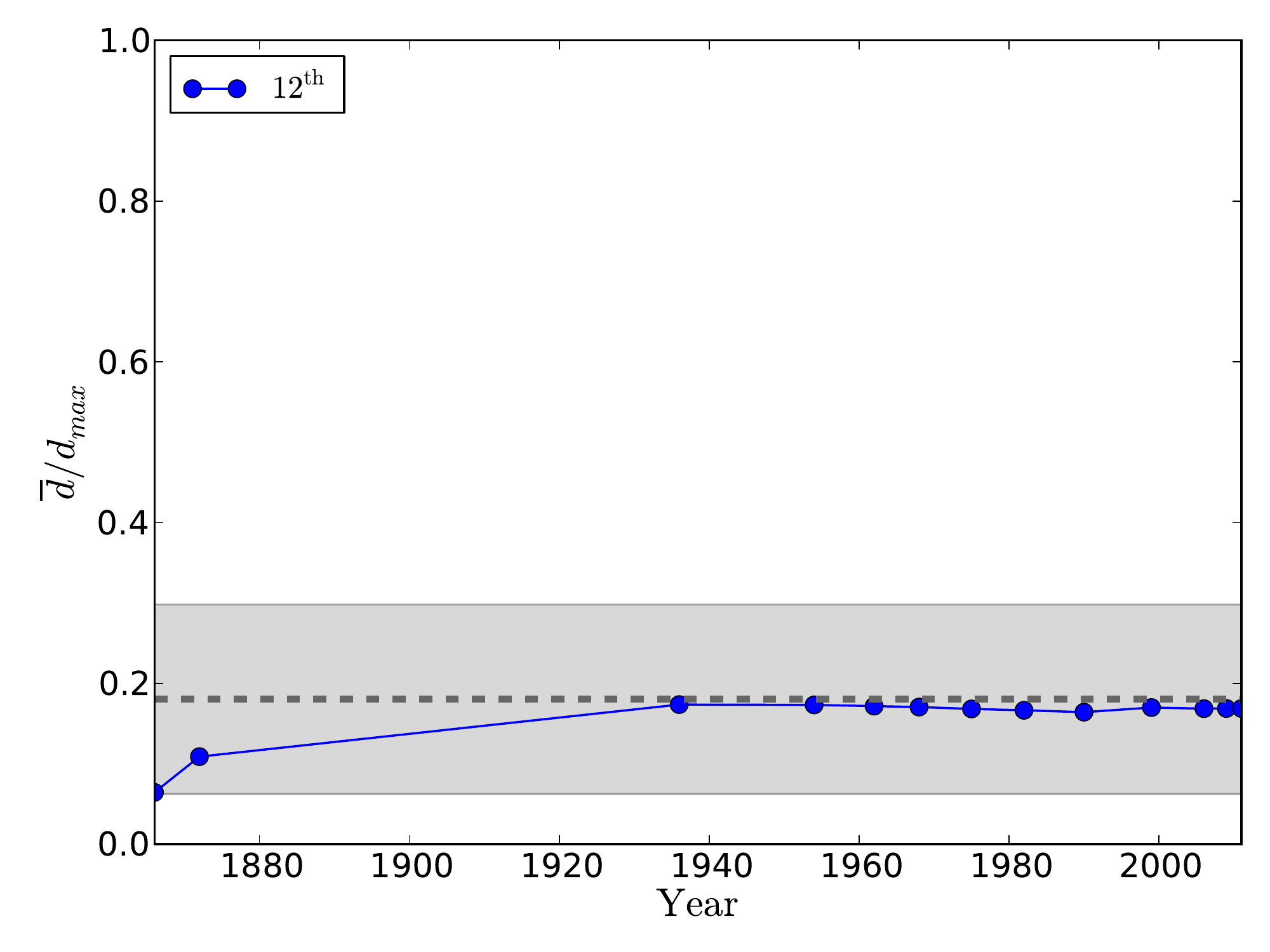} \\
\includegraphics[angle=0, width=0.25\textwidth]{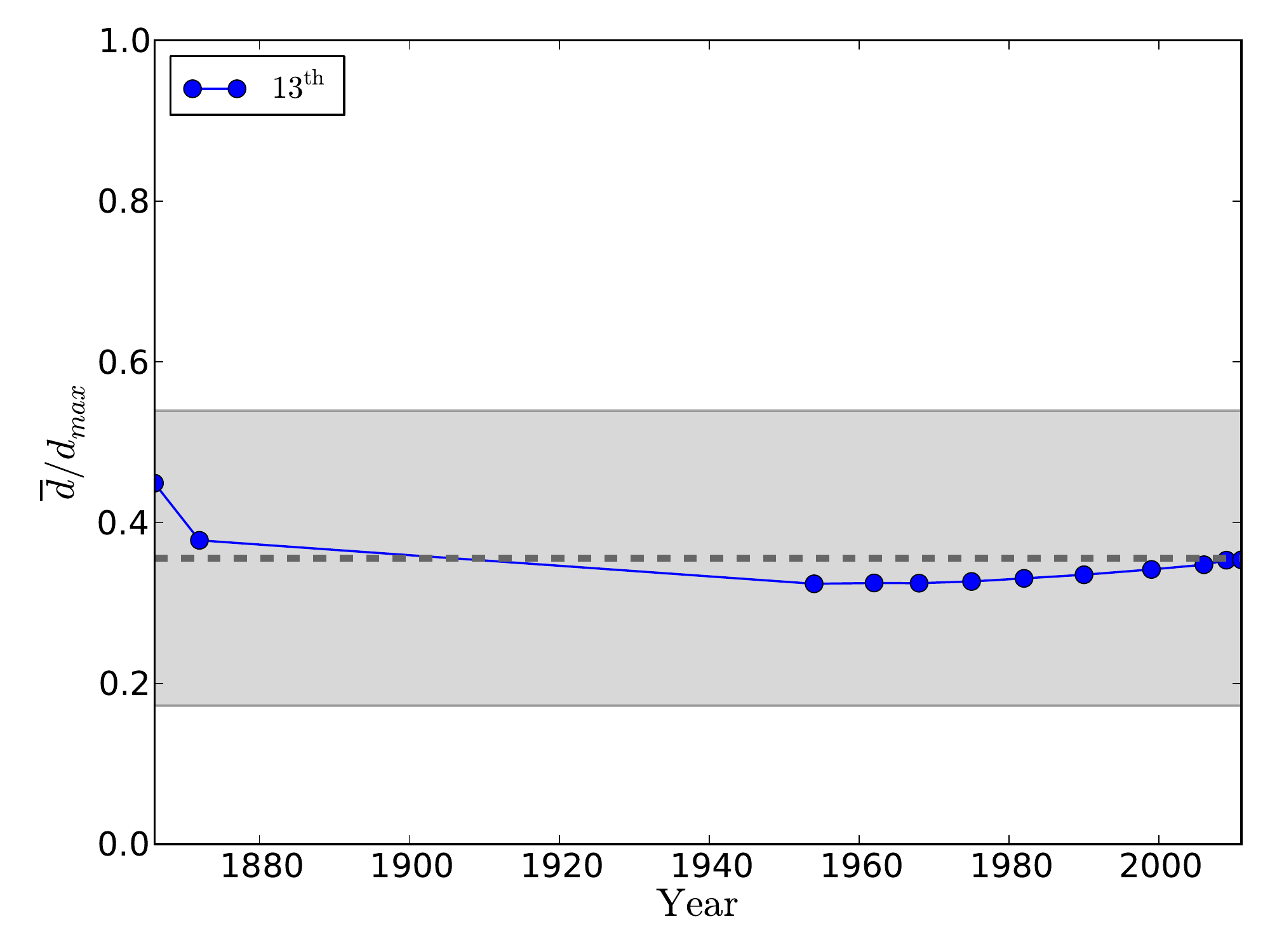} &
\includegraphics[angle=0, width=0.25\textwidth]{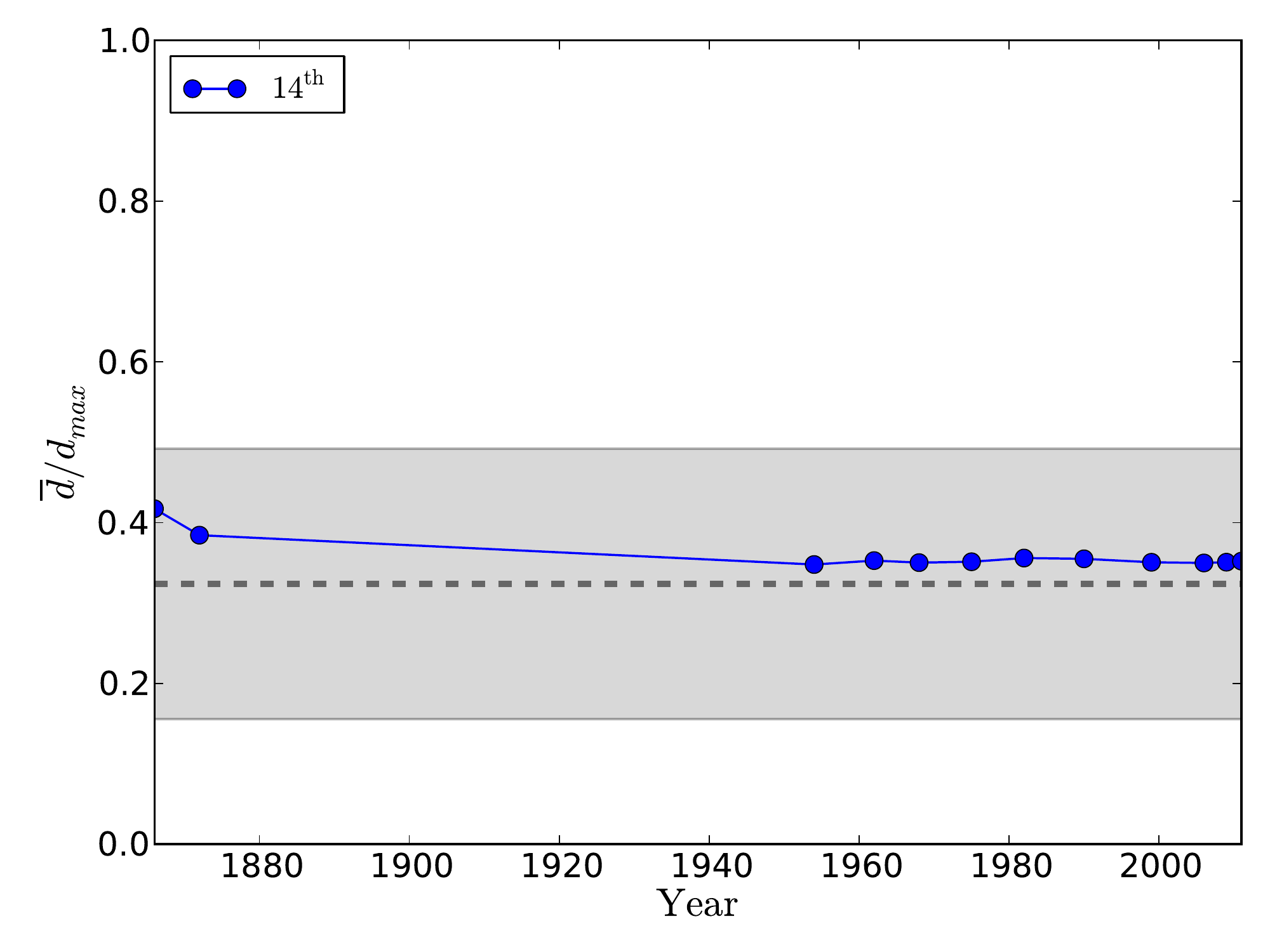} &
\includegraphics[angle=0, width=0.25\textwidth]{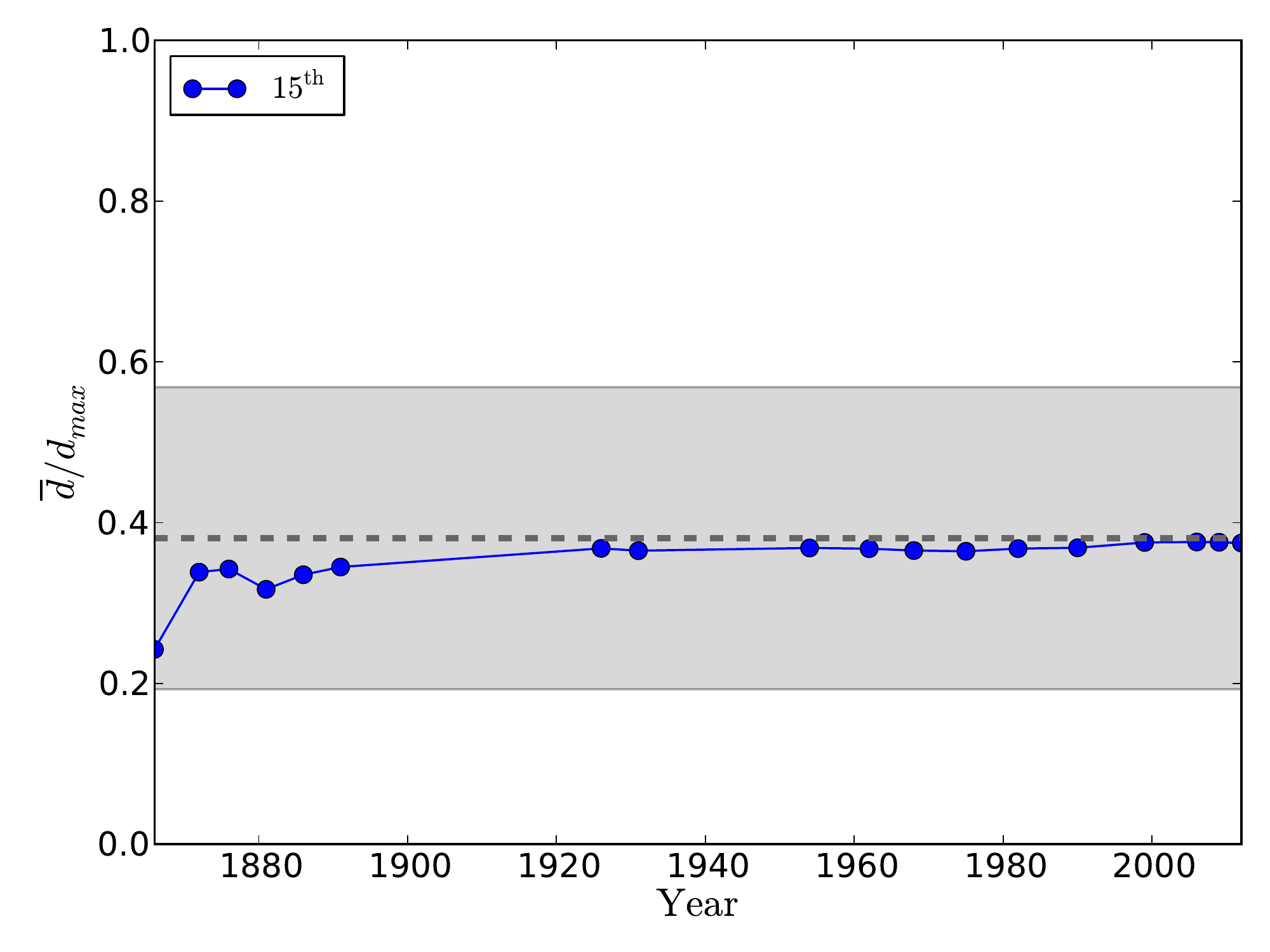}&
\includegraphics[angle=0, width=0.25\textwidth]{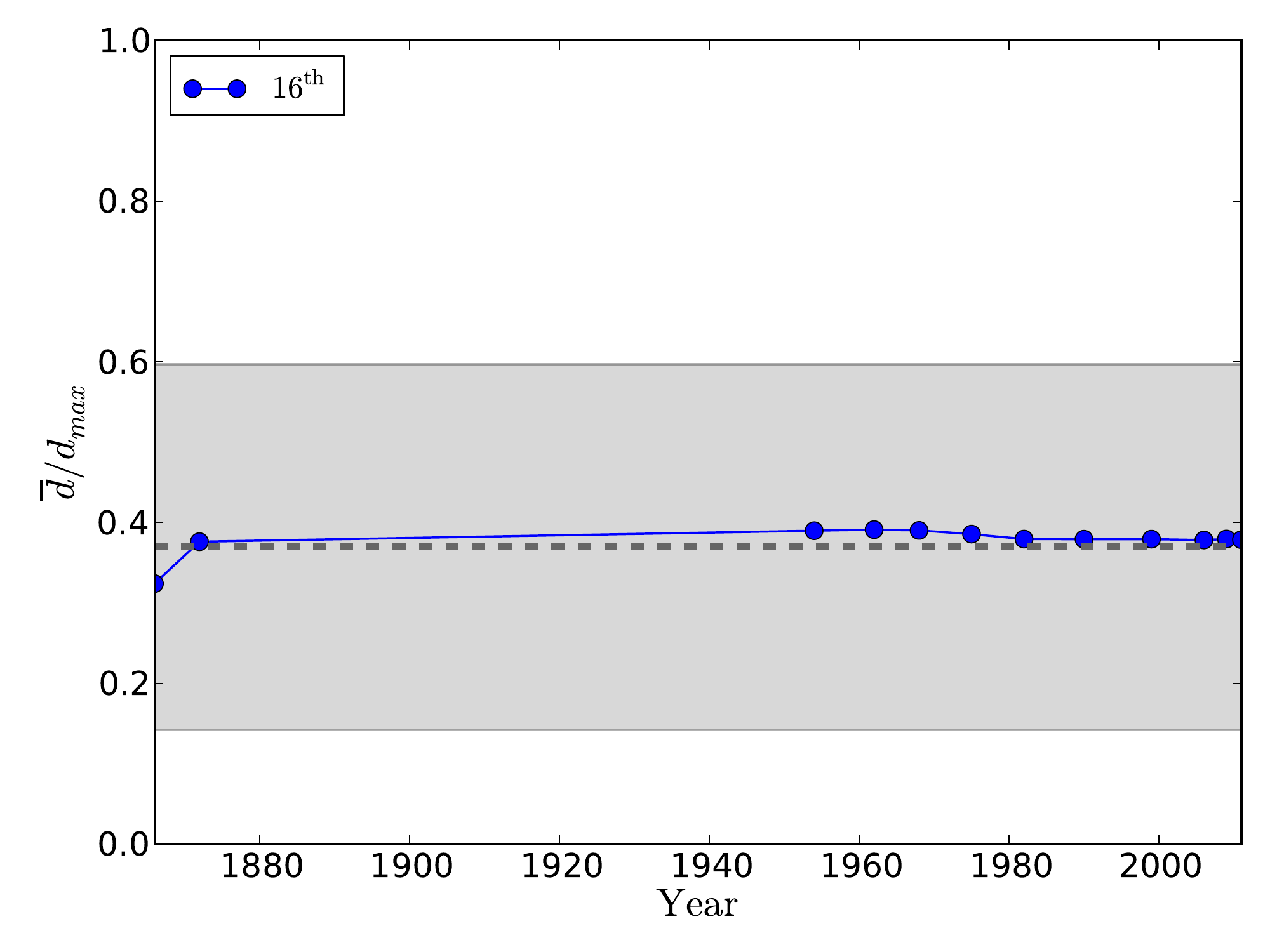} \\
\includegraphics[angle=0, width=0.25\textwidth]{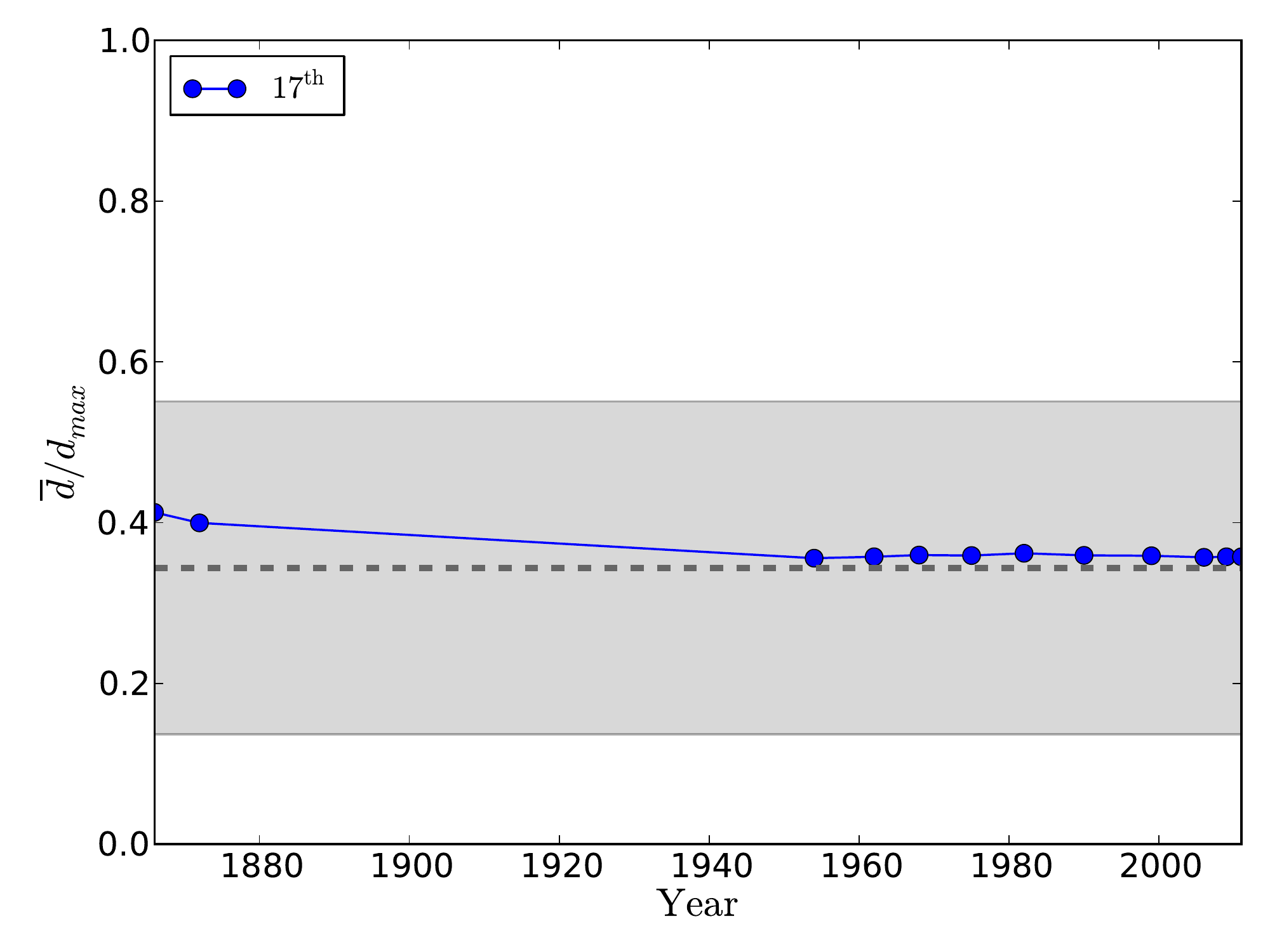} &
\includegraphics[angle=0, width=0.25\textwidth]{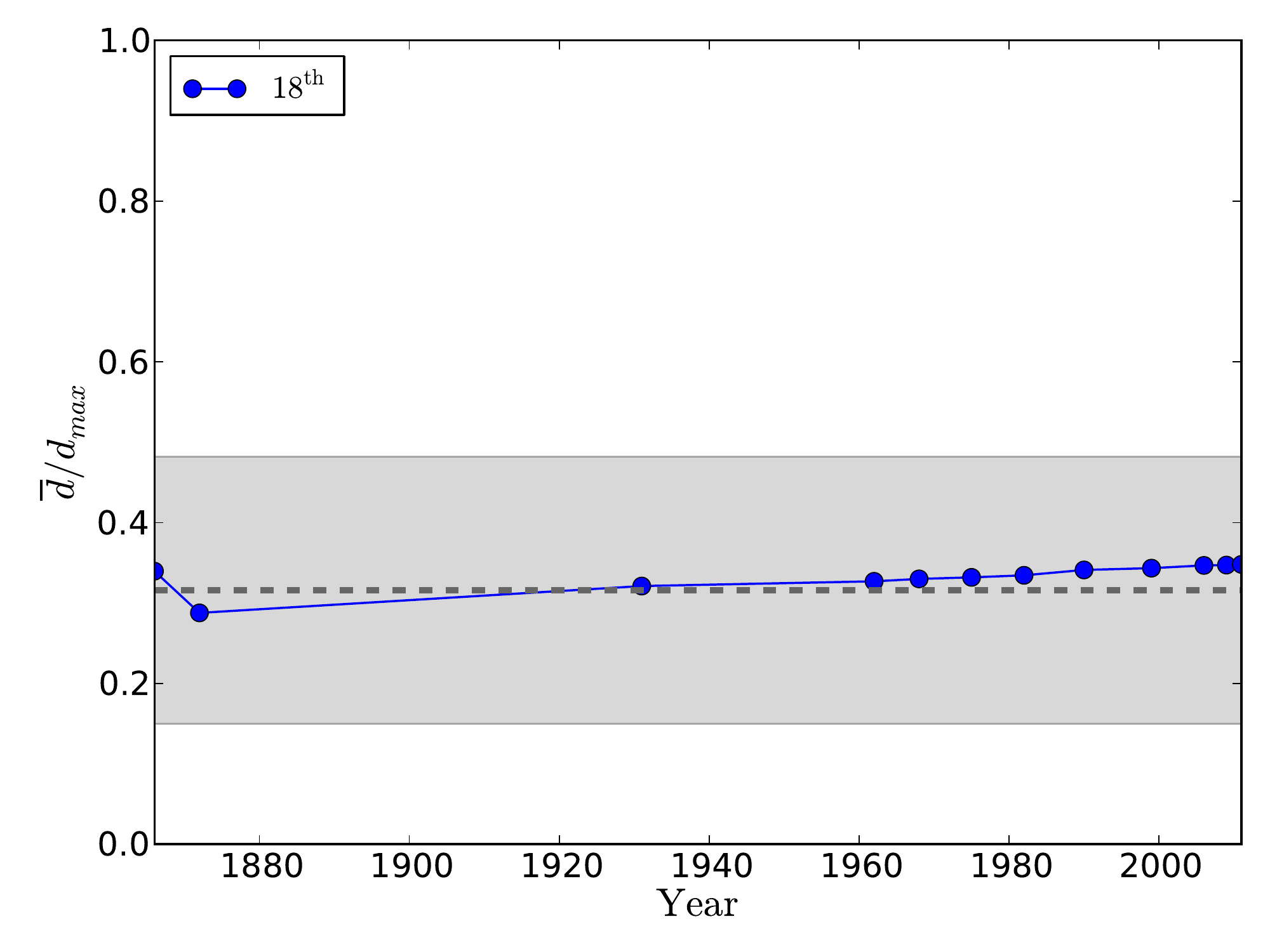} &
\includegraphics[angle=0, width=0.25\textwidth]{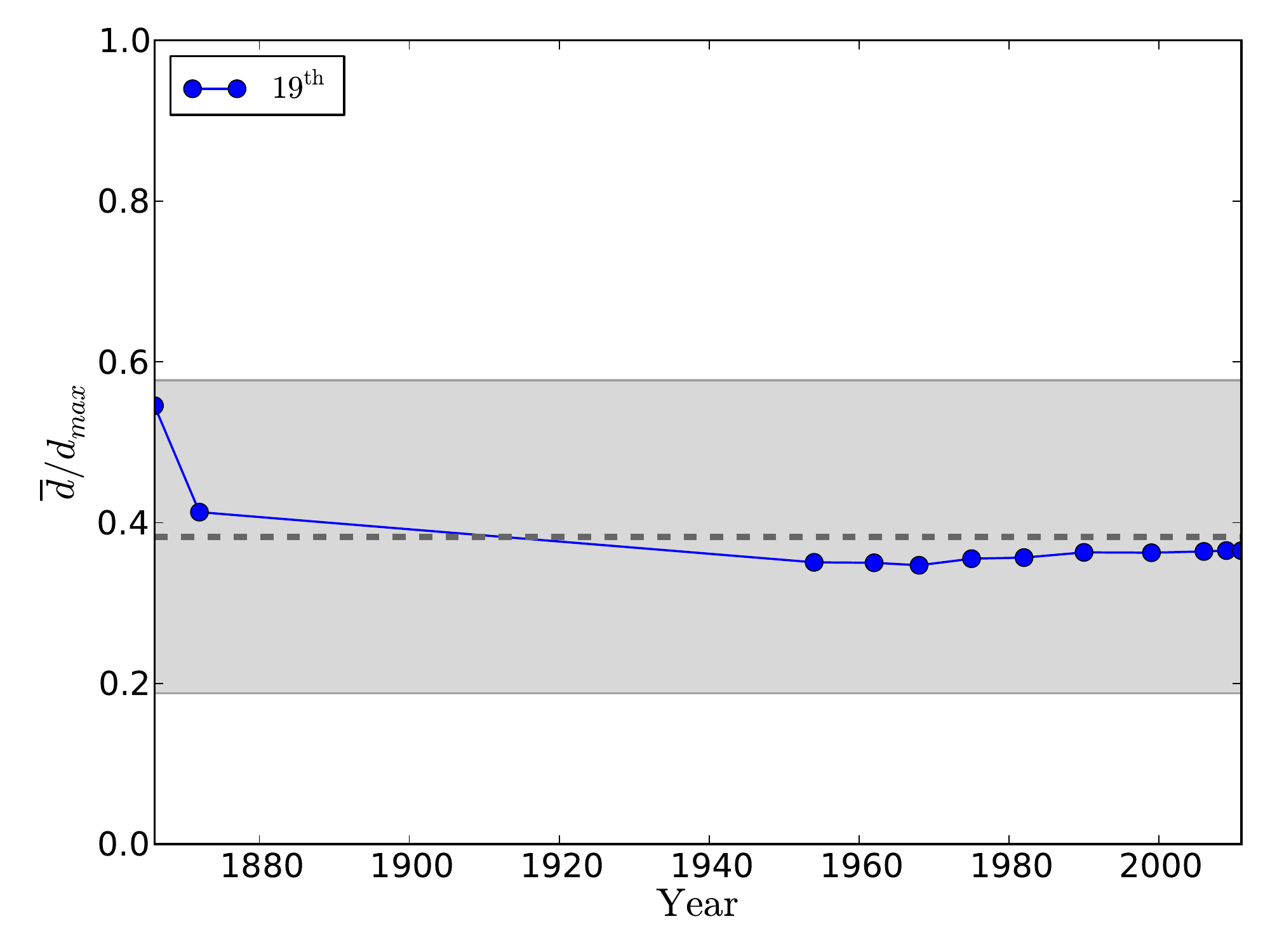}&
\includegraphics[angle=0, width=0.25\textwidth]{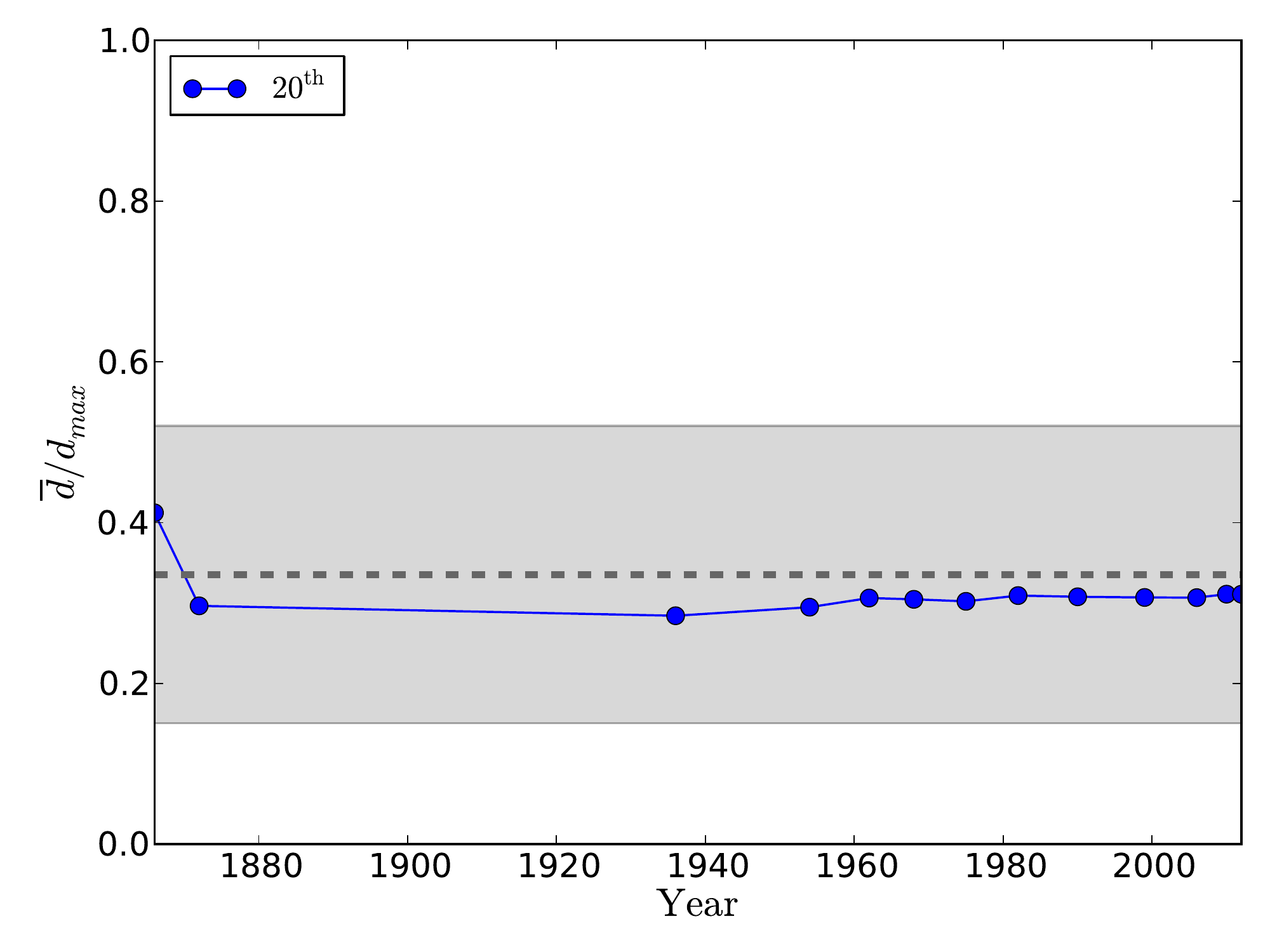} 
\end{tabular}%
\end{center}
\caption{\textbf{Paris arrondissements: homogeneity of growth in districts.} Average
distance between buildings at a given time (this distance is
normalized by the maximum distance found each district). The dotted line represents the average
value computed for a random uniform distribution and the grey
zone the dispersion computed with this null model.}
\label{figS8}
\end{figure}

\begin{figure}[!]
\begin{center}
\begin{tabular}{cc}
\includegraphics[angle=0, width=0.5\textwidth]{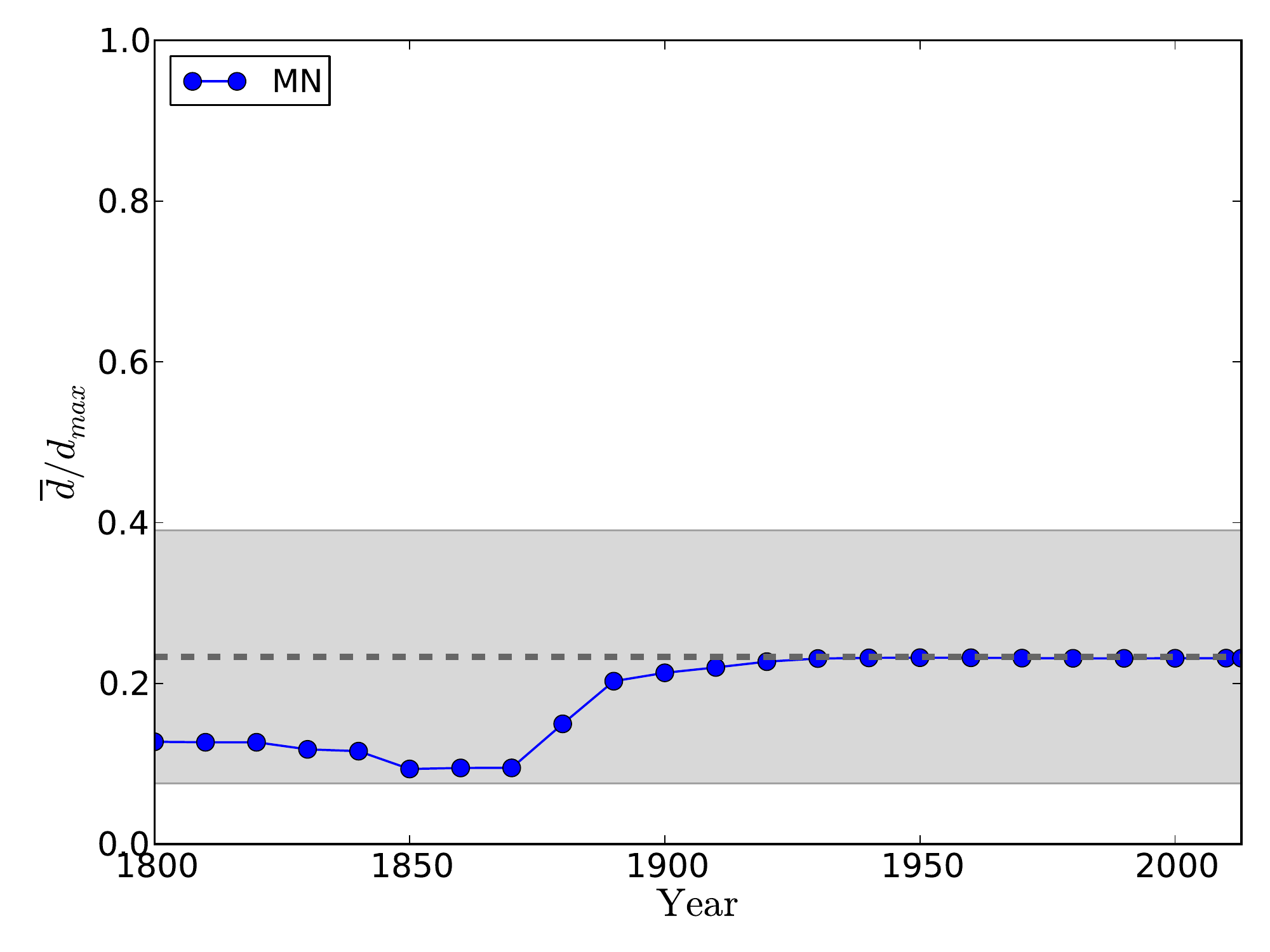} &
\includegraphics[angle=0, width=0.5\textwidth]{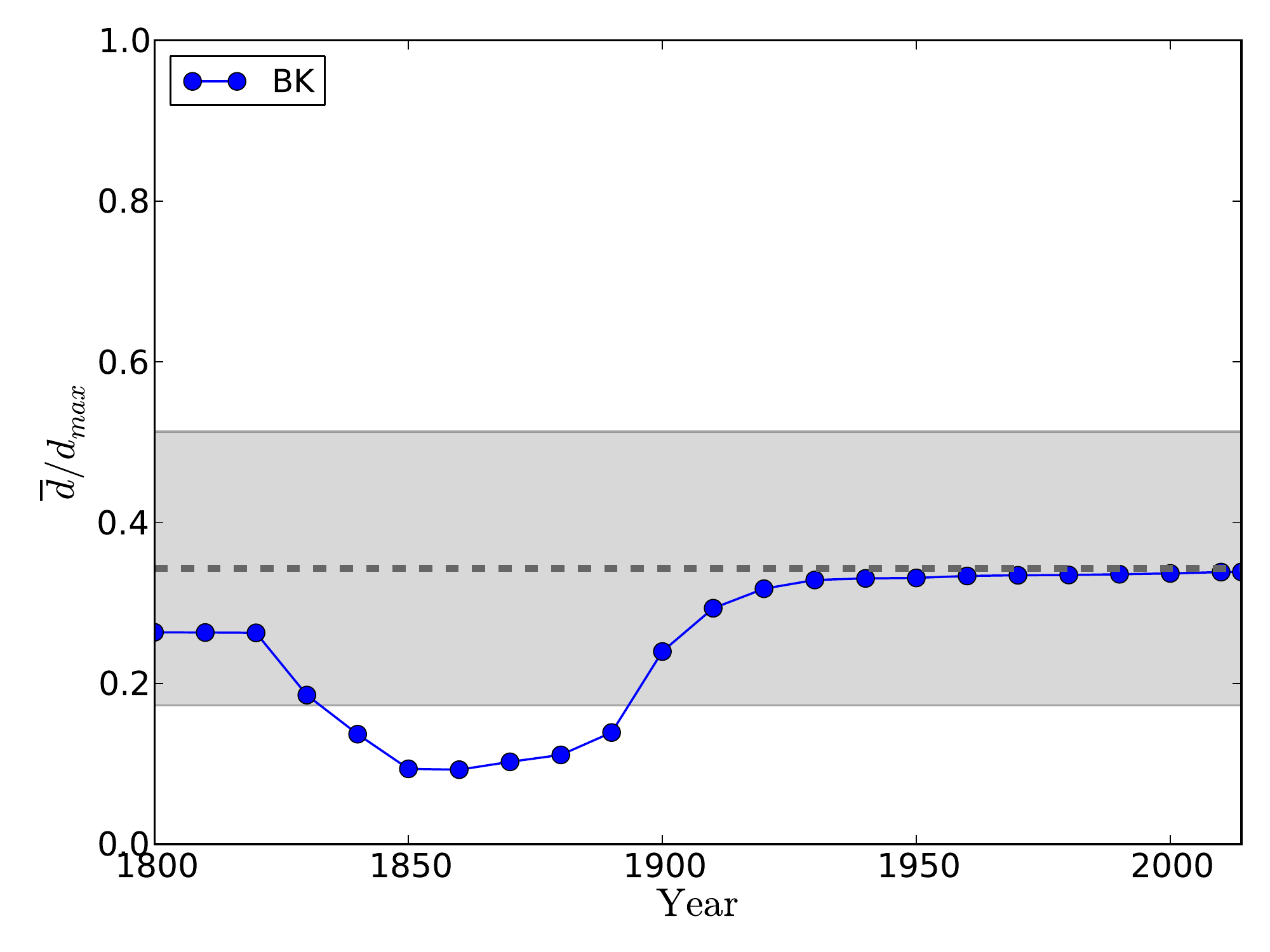} \\
\includegraphics[angle=0, width=0.5\textwidth]{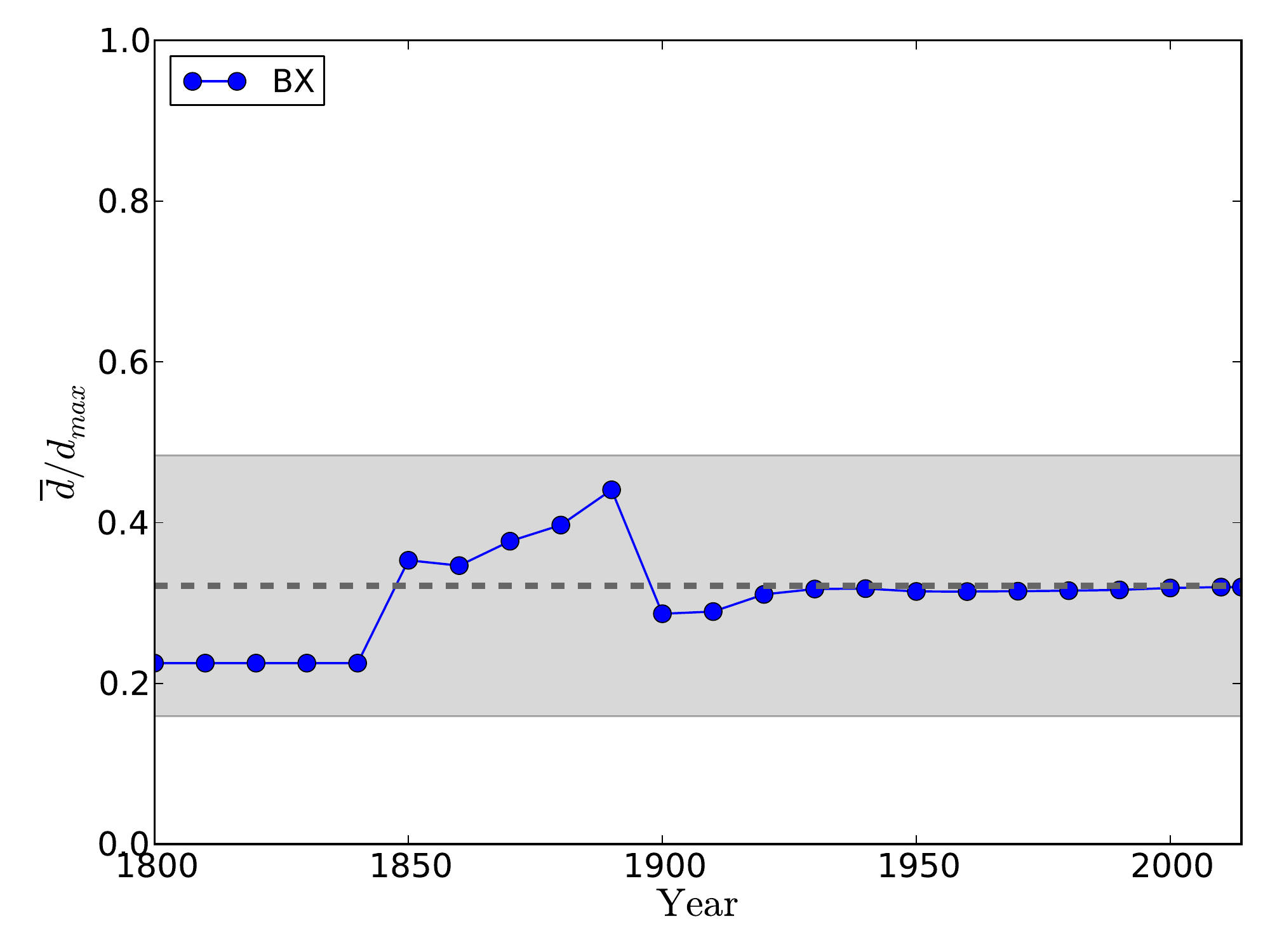} &
\includegraphics[angle=0, width=0.5\textwidth]{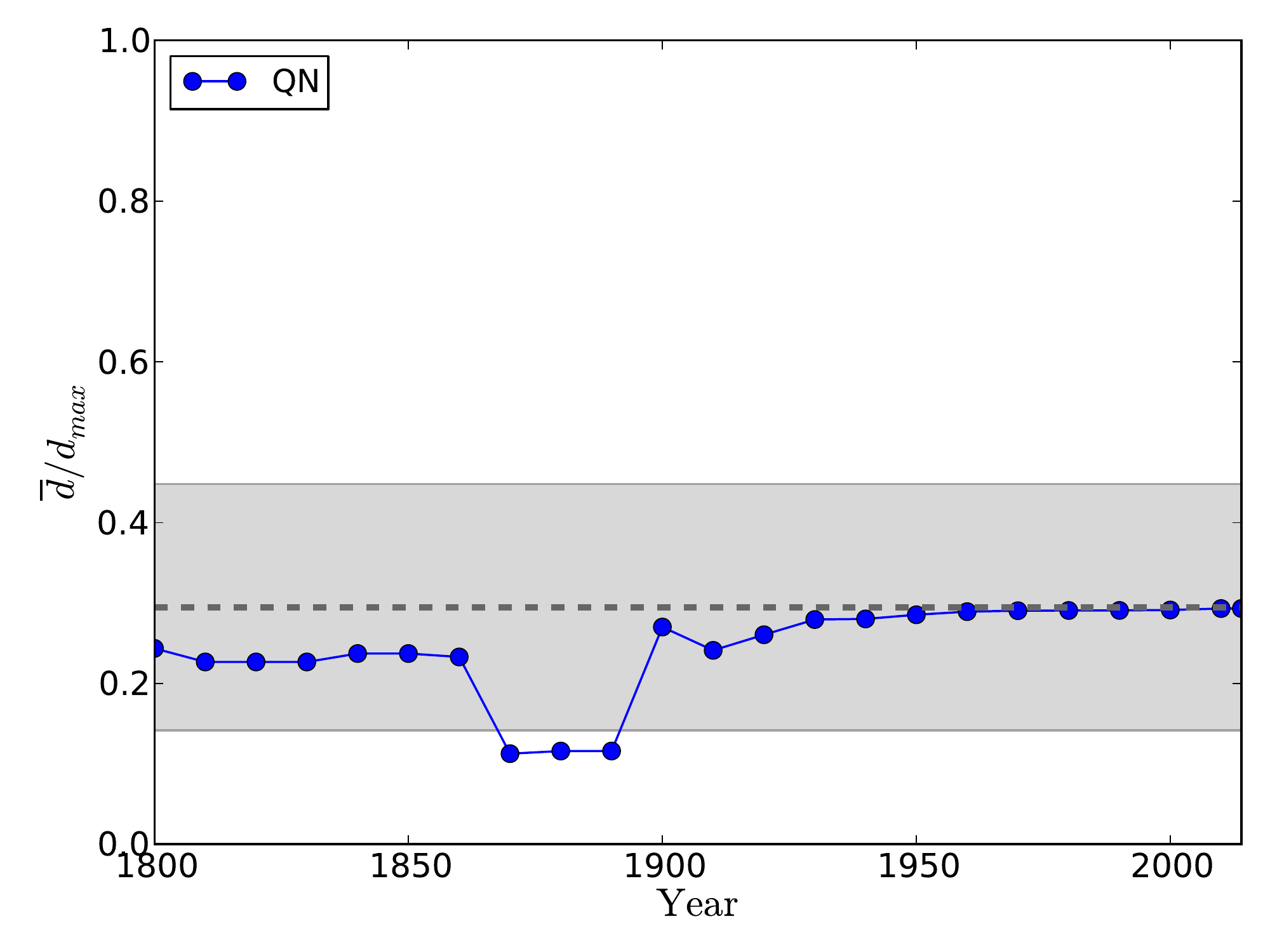} \\
\includegraphics[angle=0, width=0.5\textwidth]{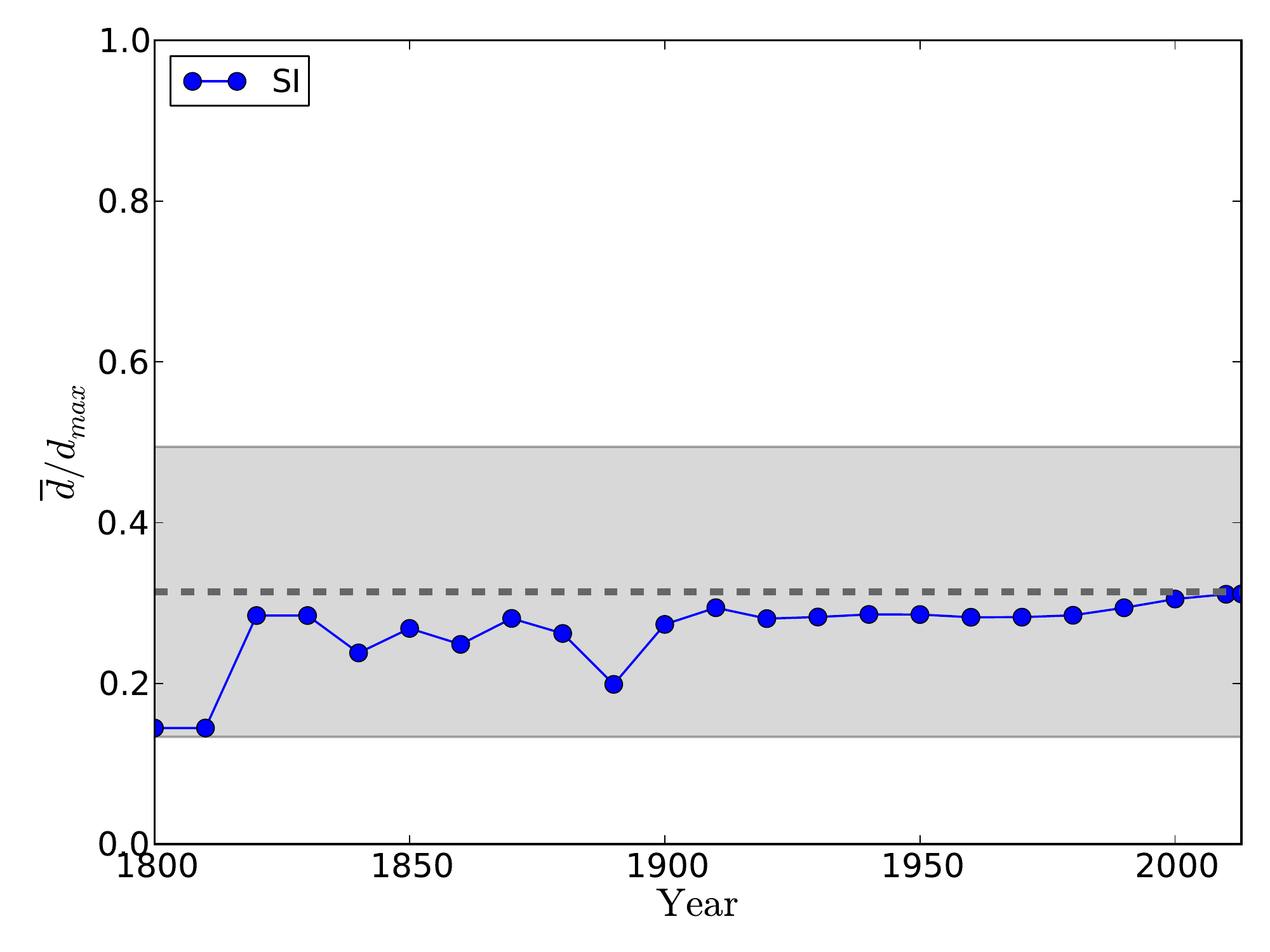} 
\end{tabular}%
\end{center}
\caption{\textbf{New York boroughs: homogeneity of growth in districts.} Average
distance between buildings at a given time (this distance is
normalized by the maximum distance found each district). The dotted line represents the average
value computed for a random uniform distribution and the grey
zone the dispersion computed with this null model.}
\label{figS9}
\end{figure}

\end{document}